\documentclass[aps,prl,reprint,superscriptaddress,longbibliography]{revtex4-2}

\usepackage{amsmath,amssymb,amsfonts,amsthm,mathrsfs}
\usepackage{bbm}
\usepackage{dsfont}
\usepackage{bm}
\usepackage{mathtools}
\usepackage{microtype}
\usepackage{tikz}
\usetikzlibrary{calc}
\usepackage{hyperref}
\hypersetup{colorlinks=true,citecolor=blue,linkcolor=blue,urlcolor=blue}
\usetikzlibrary{arrows.meta,positioning}
\usepackage{array}
\usepackage{needspace}

\definecolor{cobalt}{rgb}{0.0, 0.28, 0.67}

\newcommand{\id}{\mathbbm{1}}
\newcommand{\M}{\mathsf{M}}
\newcommand{\N}{\mathsf{N}}
\newcommand{\E}{\mathsf{E}}
\newcommand{\Tr}{\operatorname{tr}}
\newcommand{\F}{\mathbb{F}}
\newcommand{\cK}{\mathcal{K}}
\newcommand{\cT}{\mathcal{T}}

\newcommand{\cA}{\mathscr{A}}
\newcommand{\cR}{\mathscr{R}}
\newcommand{\cC}{\mathscr{C}}
\newcommand{\conv}{\operatorname{conv}}
\newtheorem{theorem}{Theorem}
\newtheorem{proposition}{Proposition}
\newtheorem{corollary}{Corollary}
\newtheorem{lemma}{Lemma}[section]

\begin{document}

\title{Compatibility of Pauli Observables}

\author{Daniel McNulty}
\author{Paolo Facchi}
\author{Saverio Pascazio}
\affiliation{Dipartimento di Fisica, Universit\`a di Bari, 70126 Bari, Italy}
\affiliation{INFN, Sezione di Bari, 70126 Bari, Italy}

\date{November 4, 2026}

\begin{abstract}
Given an arbitrary finite family of Pauli observables, how sharply can they be approximated by a single measurement? We derive an exact state-space correspondence that characterises the resulting sharpness tradeoffs, establishing a link between incompatibility and preparation uncertainty. In particular, suitable ensembles of quantum states determine the full compatibility region and generate explicit parent measurements through Pauli conjugation and classical randomisation. The compatibility region depends only on the anti-commutativity graph and determines all nonuniform tradeoffs under standard noise models, including depolarising, probabilistic and generalised noise. We show that optimal sharpness tradeoffs are determined by spectral problems for signed Pauli Hamiltonians, whose ground states supply the seeds for optimal parent measurements. The correspondence also shows that magic is necessary to attain the full compatibility region of every noncommuting Pauli family. We quantify the resulting sharpness advantage and relate it to the reduced robustness of magic. Finally, we apply the framework to several structured Pauli families, obtaining exact compatibility characterisations and explicit optimal parents.
\end{abstract}

\maketitle

\section{Introduction}\label{sec:introduction}

Measurement incompatibility is a fundamental manifestation of quantum complementarity~\cite{Bohr1928,Lahti1980Complementarity,BuschLahti1984}. A collection of measurements is incompatible if there is no common parent measurement from which all of them can be obtained by classical post-processing; otherwise the measurements are compatible, or jointly measurable~\cite{Heinosaari2008Joint,Heinosaari2016Invitation,Guhne2023}. Incompatibility, which for projective measurements is equivalent to noncommutativity, is also a quantum resource: it underpins quantum steering~\cite{Quintino2014,Uola2015}, is closely related to contextuality and Bell nonlocality~\cite{Wolf2009,TavakoliUola2020}, and guarantees an advantage in suitable state discrimination tasks~\cite{Skrzypczyk2019}.

A related but operationally distinct form of complementarity is preparation uncertainty~\cite{Robertson1929,MaassenUffink1988,BuschHeinonenLahti2007}, which constrains the joint predictability of the outcomes of several observables in a single quantum state. For arbitrary finite Pauli families, we establish a direct connection between these two notions: the sharpness tradeoffs allowed by joint measurability are completely determined by the geometry of absolute Pauli expectation values attainable by quantum states.

Consider a finite family of projective (sharp) binary observables $\{A_v\}_{v\in V}$, represented by Hermitian unitaries with outcomes $\pm1$. Assigning each observable an independent sharpness $s_v\in[0,1]$ gives the noisy (unsharp) measurement with effects $[\id\pm s_vA_v]/2$. We say that the sharpness vector $s=(s_v)_{v\in V}$ is compatible when these measurements are jointly measurable. The set of all such vectors forms the \emph{compatibility region}, which describes the full tradeoff between the noise added to the different observables. Its upper boundary describes the optimal sharpness tradeoffs. 

When all sharpnesses are equal, their largest compatible value is the \emph{optimal uniform sharpness}. For Pauli observables, this coincides with the \emph{incompatibility robustness} under depolarising noise~\cite{Heinosaari2015}. Alongside characterising the full region, we seek explicit parent measurements realising its points. Of particular interest are optimal parents along the boundary, which are relevant in applications such as simultaneous observable estimation~\cite{McNulty2023Estimation,McNulty2025Fermionic,Majsak2025Simple}.

Despite its simple definition, joint measurability can be difficult to characterise for large families. For $m$ binary measurements, the standard semidefinite formulation involves $2^m$ parent effects as optimisation variables~\cite{CavalcantiSkrzypczyk2017,Skrzypczyk2020Complexity}. Exact criteria are known in several special settings, including pairs and triples of qubit measurements~\cite{BuschHeinosaari2008,Yu2010Joint}, arbitrary finite families of unbiased binary qubit measurements~\cite{GrinkoUola2025}, and certain highly symmetric families in higher dimensions, such as complete sets of mutually unbiased bases~\cite{Designolle2019MUB,Nguyen2020Symmetries}.
%More generally, joint measurability can also be formulated as an inclusion problem for free spectrahedra~\cite{BluhmNechita2018,BluhmNechita2020}.

Pauli observables---tensor products of single-qubit Pauli and identity operators---are ubiquitous across quantum information~\cite{Gottesman1998,BravyiKitaev2005,Mermin1990,Huang2020ClassicalShadows,ChenGongYe2024,XuEtAl2026Shadows} and give the compatibility problem a useful discrete structure. Any two Pauli observables either commute or anti-commute, so their pairwise relations are encoded by an \emph{anti-commutativity graph}, whose vertices represent observables and whose edges join anti-commuting pairs. On the preparation side, this graph has been used to derive uncertainty relations~\cite{deGois2023Uncertainty} and to study the joint numerical range~\cite{Xu2024JointNumericalRange} of Pauli families. On the measurement side, existing results have largely concerned incompatibility robustness, with exact or asymptotic values known for a few structured families~\cite{KunjwalHeunenFritz2014,McNulty2025Fermionic}. The robustness has been shown to depend only on the anti-commutativity graph, with graph invariants providing general bounds~\cite{McNulty2025Graph}. Here we seek a characterisation of the full compatibility region, including all nonuniform sharpness tradeoffs, together with explicit parent measurements realising them.

The present work gives an exact state-space characterisation of the full compatibility region for arbitrary finite Pauli families. A state $\rho$ determines a \emph{predictability profile} $(|\Tr(\rho A_v)|)_{v\in V}$. We show that a sharpness vector $s=(s_v)_{v\in V}$ is compatible if and only if it is componentwise bounded by a convex combination of these profiles. The state-space description also gives an explicit construction of joint measurements: the same states that determine the compatibility region generate parent measurements through their Pauli-conjugation orbits. By mixing these orbit POVMs and applying classical post-processing, we realise every compatible sharpness vector. In a canonical irreducible representation, the binary rank of the adjacency matrix determines the number of labelled outcomes in each orbit POVM, while its binary nullity bounds the number of seed orbits needed to realise any point in the compatibility region.

The full compatibility region depends only on the anti-commutativity graph, extending the previously known graph invariance of the incompatibility robustness~\cite{McNulty2025Graph} to all nonuniform sharpness tradeoffs. It also characterises how accurately arbitrary jointly measurable binary measurements can approximate the sharp Pauli measurements, as quantified by their worst-case total-variation errors. For the noise models considered in~\cite{Designolle2019}, the same region determines the full nonuniform compatibility tradeoffs in \emph{all} cases. At uniform sharpness, all of the resulting robustness thresholds are determined by the depolarising noise, with the same parent attaining each optimum.

We show that optimal sharpness tradeoffs along the boundary of the compatibility region can be determined from the ground-state energies of Hamiltonians that are linear combinations of the Pauli observables with suitable weights and signs. Ensembles of the corresponding ground states then generate joint measurements attaining these tradeoffs. For vertex-transitive graphs, we show that symmetry removes the optimisation over the relative weights of the Hamiltonian terms, reducing the uniform sharpness problem to a spectral optimisation over sign choices.

The state-space characterisation also reveals how magic~\cite{BravyiKitaev2005} can enhance the sharpness attainable by joint measurements. Restricting the seed states in our construction to stabilizer states gives a corresponding class of \emph{stabilizer parents}. We show that their attainable sharpness vectors form exactly the stable-set polytope of the anti-commutativity graph, which is equivalent to the region obtained by classically randomising over sharp joint measurements of commuting subsets. For every noncommuting Pauli family, the full compatibility region is strictly larger. We quantify this advantage and relate it to the \emph{reduced robustness of magic}~\cite{VarelaEtAl2026Magic,LiuEtAl2026Nonstabilizerness}. When no commuting subset of the family has product proportional to the identity, this robustness equals the smallest factor by which a seed's sharpness vector must be reduced to become attainable by a stabilizer parent.

Finally, we apply the framework to structured graph families. For line graphs, quadratic Majorana realisations allow us to characterise the full compatibility region using Gaussian seed states. Their covariance matrices give the correlations that determine the sharpnesses attained by the corresponding orbit parents. For cycles and paths, we obtain exact compatibility characterisations and explicit optimal parents. In particular, removing one or two consecutive observables from an even cycle leaves the optimal uniform sharpness unchanged, and the same parent measurement remains optimal. For the complete quadratic Majorana family, skew-conference matrices provide Gaussian seeds whose single-orbit parents attain the optimal uniform sharpness. For the complete Pauli family, an optimal \emph{single-orbit} parent at uniform sharpness exists if and only if a Pauli-covariant symmetric informationally complete measurement (SIC) exists.

The paper is organised as follows. Sec.~\ref{sec:preliminaries} introduces the required preliminaries and notation. Sec.~\ref{sec:state-space} gives the state-space characterisation and parent construction, together with their consequences for measurement uncertainty and different noise models. Sec.~\ref{sec:boundary} characterises optimal sharpness tradeoffs and constructs parent measurements that attain them. In Sec.~\ref{sec:nonstabilizer} we treat stabilizer restrictions and magic, and Sec.~\ref{sec:structured-parents} presents exact results for several structured families. We conclude with a discussion of the main implications and open questions.

\section{Preliminaries}\label{sec:preliminaries}
% =============================================================================

\subsection{Measurements and joint measurability}

We work on a finite-dimensional Hilbert space $\mathcal H$ of dimension $d$, with identity operator $\id$. A measurement with finite outcome set $\Omega$ is a POVM $\E:=\{\E(a)\}_{a\in\Omega}$, whose effects satisfy $\E(a)\geq0$ and $\sum_a\E(a)=\id$. For a quantum state $\rho\geq0$ with $\Tr(\rho)=1$, the probability of outcome $a$ is $\Tr(\rho\E(a))$. We write $\|T\|_1:=\Tr(\sqrt{T^\dagger T})$ and $\|T\|_\infty$ for the trace and operator norms, respectively.

Let $\cA=\{A_v\}_{v\in V}$ be a finite family of sharp binary observables satisfying $A_v=A_v^\dagger$ and $A_v^2=\id$. We assign each observable a sharpness $s_v\in[0,1]$ and write $s=(s_v)_{v\in V}$ for the resulting sharpness vector of the family. The corresponding noisy (unsharp) binary POVMs are
\begin{equation}
 \M_v^{s_v}(b)=\frac12\bigl(\id+b\,s_vA_v\bigr),
 \label{eq:noisy}
\end{equation}
with outcomes $b\in\{\pm1\}$. Here $s_v=1$ gives the sharp measurement of $A_v$, while $s_v=0$ gives a uniformly random binary outcome. The noisy measurement can be obtained by flipping the sharp outcome with probability $(1-s_v)/2$.

A family of measurements is \emph{jointly measurable} if its members can be obtained as marginals of a single measurement, described by a \emph{parent} POVM. For the binary family $\{\M_v^{s_v}\}_{v\in V}$, this means that there exists a POVM $\{\E(a)\}_{a\in\{\pm1\}^V}$ satisfying
\[
 \sum_{a:a_v=b}\E(a)=\M_v^{s_v}(b),
 \qquad v\in V,
\]
for outcome $b\in\{\pm1\}$. Each parent outcome $a=(a_v)_{v\in V}$ records one binary result for each measurement. Measuring the parent and reporting the coordinate $a_v$ therefore reproduces the statistics of $\M_v^{s_v}$ for every input state. This definition can equivalently be formulated in terms of classical post-processing~\cite[Proposition~8]{AliEtAl2009}, with each measurement obtained from a common parent POVM by a classical stochastic map. Since the two effects of each binary marginal sum to $\id$, the marginal conditions are equivalent to
\begin{equation}
\sum_a a_v\E(a)=s_vA_v,
\qquad v\in V.
\label{eq:parent-condition}
\end{equation}
The largest $\eta\in[0,1]$ for which the family is jointly measurable with $s_v=\eta$ for every $v$ is its \emph{optimal uniform sharpness}. For Pauli observables, this coincides with the incompatibility robustness under depolarising noise~\cite{Heinosaari2015}. Sec.~\ref{subsec:noise-models} relates this threshold to other standard notions of incompatibility robustness.

Classical randomisation over measurements can itself be described by a single POVM. Choosing $\E_j=\{\E_j(x)\}$ with probability $p_j$ and recording both $j$ and $x$ gives the POVM with effects $p_j\E_j(x)$. We refer to this as \emph{mixing the POVMs}.

\subsection{Pauli families and conjugation symmetries}
\label{subsec:pauli-symmetries}

A Pauli observable on $n$ qubits is a tensor product $\pm P_1\otimes\cdots\otimes P_n$, where $P_j\in\{\id,X,Y,Z\}$. A finite collection of such observables is called a \emph{Pauli family}. Any two Pauli observables either commute or anti-commute, so their pairwise relations are encoded by the \emph{anti-commutativity graph} $G=(V,E)$, with
\[
 \{u,v\}\in E
 \quad\Longleftrightarrow\quad
 A_uA_v=-A_vA_u.
\]
Every finite simple graph admits a Pauli realisation~\cite{Abramsky24,Khovanova2010,Xu2024JointNumericalRange}. Writing $B_G$ for the adjacency matrix of $G$ over $\F_2$, the commutation relations take the form
\[
 A_uA_v=(-1)^{(B_G)_{uv}}A_vA_u.
\]
Together with $A_v=A_v^\dagger$ and $A_v^2=\id$, these are the \emph{graph-Clifford relations} associated with $G$.

The binary rank and nullity of $B_G$ are
\begin{equation}\label{eq:rank-nullity}
 \begin{aligned}
 r&:=\operatorname{rank}_{\F_2}B_G,\\
 \nu_2(G)&:=\dim_{\F_2}\ker B_G=|V|-r.
 \end{aligned}
\end{equation}
Since $B_G$ is alternating over $\F_2$, i.e., symmetric with zero diagonal, its rank $r$ is even~\cite{Khovanova2010}. Every irreducible realisation of the graph-Clifford relations has dimension
\[
 d_G=2^{r/2}.
\]
We call any such irreducible realisation a \emph{canonical representation}. The original physical representation need not be irreducible, and its dimension $d$ need not equal $d_G$. Appendix~\ref{app:clifford} gives the algebraic details. Appendix~\ref{app:representations} explains how parent measurements can be mapped between canonical and physical representations without changing their sharpness vectors.

The matrix $B_G$ also determines the sign changes induced by conjugation with products of the selected observables. Fix an ordering of $V$ and, for $z\in\F_2^V$, define the ordered monomial
\[
 A^z:=\prod_{v\in V}A_v^{z_v}.
\]
The graph-Clifford relations give
\[
 A^zA_v(A^z)^\dagger=(-1)^{(B_Gz)_v}A_v.
\]
Thus two labels $z,z'$ induce the same sign changes on the family exactly when $z-z'\in\ker B_G$. These distinct actions form the finite abelian group
\[
 \cK_G:=\F_2^V/\ker B_G,
 \qquad |\cK_G|=2^r=d_G^2.
\]
Thus $\cK_G$ may be viewed as the group of sign patterns on the family generated by Pauli conjugation. For each class $x\in\cK_G$, choose a representative $\widetilde x\in\F_2^V$ and set $U_x:=A^{\widetilde x}$. Define
\begin{equation}\label{eq:character}
 \chi_v(x):=(-1)^{(B_G\widetilde x)_v}.
\end{equation}
These functions are independent of the representative and are characters of $\cK_G$, satisfying $\chi_v(x+y)=\chi_v(x)\chi_v(y)$. In particular,
\begin{equation}
 U_xA_vU_x^\dagger=\chi_v(x)A_v.
 \label{eq:sign-change-action}
\end{equation}
If $k\in\ker B_G$, the product $A^k$ commutes with every observable $A_v$. In a canonical representation, irreducibility therefore forces $A^k=\lambda_k\id$, with $|\lambda_k|=1$. Changing the representative of $x\in\cK_G$ consequently multiplies $U_x$ only by a phase, which cancels in $U_x XU_x^\dagger$ for every operator $X$. The conjugation action is therefore independent of this choice. These symmetries underlie the orbit parent construction below; their algebraic properties are collected in Appendix~\ref{app:clifford}.

Majorana monomials (see e.g. Sec.~\ref{sec:structured-parents}) fit the same framework through standard fermion-to-qubit mappings. Such monomials can be identified, up to phase, with Pauli strings that preserve their commutation relations~\cite{JordanWigner1928,BravyiKitaev2002}.

% 
% =============================================================================

\section{State-space characterisation and parent measurements}
\label{sec:state-space}
% =============================================================================

We now formulate joint measurability of Pauli observables in terms of the absolute expectation values attainable by quantum states. The same states generate explicit parent measurements through Pauli-conjugation orbits. We first develop this construction and bound the number of seeds needed, then derive its consequences for measurement uncertainty and compatibility under different noise models.

\subsection{State-space characterisation}
\label{subsec:state-space-characterisation}

We first present the construction for a canonical Pauli family $\cA=\{A_v\}_{v\in V}$ on $\mathbb C^{d_G}$, and then extend the result to arbitrary physical Pauli families.

For a state $\rho$, define its \emph{absolute expectation-value vector}
\begin{equation}
 c_{\cA}(\rho):=\bigl(|\Tr(\rho A_v)|\bigr)_{v\in V}.
 \label{eq:absolute-expectation-vector}
\end{equation}
The correspondence below follows from two complementary constructions: normalising parent effects produces an ensemble of states, while Pauli conjugation turns suitable seed states into a parent measurement.

\paragraph{From parent measurements to states.} For every nonzero effect $\E(a)$ of a parent satisfying Eq.~\eqref{eq:parent-condition}, set
\begin{equation}
 p_a=\frac{\Tr(\E(a))}{d_G},
 \qquad
 \rho_a=\frac{\E(a)}{\Tr(\E(a))}.
 \label{eq:effect-state}
\end{equation}
These define an ensemble with $\sum_a p_a=1$ and $\sum_a p_a\rho_a=\id/d_G$. Multiplying Eq.~\eqref{eq:parent-condition} by $A_v$, taking the trace, and using $A_v^2=\id$ gives
\begin{equation}
 s_v
 =\sum_a p_a a_v\Tr(\rho_aA_v)
 \leq\sum_a p_a|\Tr(\rho_aA_v)|.
 \label{eq:necessary-state-bound}
\end{equation}
Thus $s$ lies componentwise below a convex combination of the vectors $c_{\cA}(\rho_a)$. This necessary condition holds for any family of binary observables of the form~\eqref{eq:noisy}. The Pauli-specific result is that it is also sufficient.

\paragraph{From states to parent measurements.} In the canonical representation, averaging over Pauli conjugations removes every operator component except the identity component. Weighting the same average by $\chi_v$ (defined in Eq.~\eqref{eq:character}) instead selects the component along $A_v$. Thus, for every operator $X$,
\begin{align}
 \frac1{|\cK_G|}\sum_{x\in\cK_G}U_xXU_x^\dagger
 &=\frac{\Tr(X)}{d_G}\id,\nonumber\\
 \frac1{|\cK_G|}\sum_{x\in\cK_G}\chi_v(x)U_xXU_x^\dagger
 &=\frac{\Tr(A_vX)}{d_G}A_v.
 \label{eq:character-projection}
\end{align}
Appendix~\ref{app:clifford} derives these identities from the conjugation action~\eqref{eq:sign-change-action}.

Given a state $\rho$, define
\begin{equation}
 \E_\rho(x)=\frac1{d_G}U_x\rho U_x^\dagger,
 \qquad x\in\cK_G.
 \label{eq:orbit-povm}
\end{equation}
The first identity in Eq.~\eqref{eq:character-projection}, together with $|\cK_G|=d_G^2$, shows that these effects sum to $\id$. We call $\rho$ the \emph{seed state}, the family $\{U_x\rho U_x^\dagger\}_{x\in\cK_G}$ its \emph{Pauli orbit}, and $\E_\rho$ the corresponding \emph{orbit POVM}.

For each $v$, choose $\alpha_v\in\{\pm1\}$ such that $\alpha_v\Tr(\rho A_v)=|\Tr(\rho A_v)|$, with either sign allowed when the expectation vanishes. We then assign the binary outcome $\alpha_v\chi_v(x)$ to each orbit outcome $x$. The second character identity gives
\begin{equation}
 \sum_{x\in\cK_G}\alpha_v\chi_v(x)\E_\rho(x)
 =|\Tr(\rho A_v)|A_v.
 \label{eq:orbit-marginal}
\end{equation}
Since the binary marginal effects sum to $\id$, this is precisely the measurement $\M_v^{|\Tr(\rho A_v)|}$. Thus one seed state generates a parent with sharpness vector $c_{\cA}(\rho)$.

Now consider a finite ensemble $\{p_j,\rho_j\}$. Choose $j$ with probability $p_j$, perform the orbit POVM $\E_{\rho_j}$ on the input system, and record both $j$ and the orbit outcome $x$. For each $j$ and $v$, choose $\alpha_{jv}\in\{\pm1\}$ such that $\alpha_{jv}\Tr(\rho_jA_v)=|\Tr(\rho_jA_v)|$, and report $\alpha_{jv}\chi_v(x)$ as the result for observable $v$. Averaging Eq.~\eqref{eq:orbit-marginal} over $j$ gives sharpnesses
\[
 t_v:=\sum_jp_j|\Tr(\rho_jA_v)|.
\]
Whenever $s\leq t$, independently flipping the reported result for $v$ with probability $(1-s_v/t_v)/2$ reduces its sharpness to $s_v$. If $t_v=0$, then $s_v=0$ and no additional noise is needed.

Appendix~\ref{app:representations} shows that the ensemble condition and parent construction extend beyond the canonical representation to arbitrary physical Pauli families. In the following definitions and theorem, $\cA$ may therefore act on its original Hilbert space $\mathcal H$, and the states in the ensemble are states on $\mathcal H$.

Define the state region and compatibility region by
\begin{align}
 \cR_{\cA}
 &:=\{c_{\cA}(\rho):\rho\ \text{a state}\},
 \label{eq:state-region}\\
 \cC_{\cA}
 &:=\bigl\{s\in[0,1]^V:
 \{\M_v^{s_v}\}_{v\in V}\ \text{is jointly measurable}\bigr\}.
 \label{eq:compatibility-region}
\end{align}
The two constructions above establish the following characterisation.

\begin{theorem}
\label{thm:state-space}
Let $\cA=\{A_v\}_{v\in V}$ be a finite Pauli family on $\mathcal H$, and let $s\in[0,1]^V$. The measurements $\{\M_v^{s_v}\}_{v\in V}$ are jointly measurable if and only if there is a finite ensemble $\{p_j,\rho_j\}$ of states on $\mathcal H$, with $p_j\geq0$ and $\sum_jp_j=1$, satisfying
\begin{equation}
 s_v\leq\sum_jp_j|\Tr(\rho_jA_v)|,
 \qquad v\in V.
 \label{eq:main-theorem}
\end{equation}
In a canonical representation, every such ensemble gives the parent POVM
\begin{equation}
 \E(j,x)=p_j\E_{\rho_j}(x)
 =\frac{p_j}{d_G}U_x\rho_jU_x^\dagger,
 \qquad x\in\cK_G.
 \label{eq:ensemble-orbit-parent}
\end{equation}
For an arbitrary physical representation, Appendix~\ref{app:representations} supplies a corresponding parent with the same target sharpness vector $s$.
\end{theorem}

\begin{proof}
Necessity follows from Eq.~\eqref{eq:necessary-state-bound}, with $d_G$ replaced by $\dim\mathcal H$ for an arbitrary physical representation. The preceding orbit construction, followed by classical output noise, proves sufficiency in a canonical representation and gives Eq.~\eqref{eq:ensemble-orbit-parent}. Lemmas~\ref{lem:C-statevectors} and~\ref{lem:C-parent-transfer} in Appendix~\ref{app:representations} extend this construction to the original Hilbert space.
\end{proof}

Theorem~\ref{thm:state-space} implies that the compatible sharpness vectors are those lying componentwise below a convex combination of state profiles:
\begin{equation}
 \cC_{\cA}
 =\bigl\{s\in[0,1]^V:
 s\leq t\ \text{for some }t\in\conv\cR_{\cA}\bigr\}.
 \label{eq:compatibility-region-characterisation}
\end{equation}

The seeds may always be chosen pure: decomposing each mixed seed and applying the triangle inequality can only increase the averaged absolute expectations. Any excess sharpness is removed by classical noise.

As mentioned above, the converse in Theorem~\ref{thm:state-space} is Pauli-specific. For example, for $A_1=X$ and $A_2=(X+Z)/\sqrt2$, an $X$ eigenstate has absolute expectation-value vector $(1,1/\sqrt2)$. However, the measurements with sharpness vector $s=(1,1/\sqrt2)$ are incompatible since any measurement jointly measurable with sharp $X$ must commute with it.

Representation independence also implies that two Pauli families with the same anti-commutativity graph have the same compatibility region. This extends the graph invariance of the uniform threshold~\cite{McNulty2025Graph} to all nonuniform sharpness tradeoffs.

\begin{corollary}
\label{cor:graph-invariance}
\emph{Let $\cA=\{A_v\}_{v\in V}$ and $\cA'=\{A'_v\}_{v\in V}$ be finite-dimensional Pauli families with the same anti-commutativity graph $G$. Then}
\[
 \cC_{\cA}=\cC_{\cA'}.
\]
\emph{Hence the compatibility region depends only on $G$, and we denote it by $\cC_G$.}
\end{corollary}

\begin{proof}
Appendix~\ref{app:representations} maps parent measurements between any realisation of $G$ and a fixed canonical family $\widehat{\cA}_G$ without changing their sharpness vectors. Thus $\cC_{\cA}=\cC_{\widehat{\cA}_G}=\cC_{\cA'}$.
\end{proof}

The corresponding \emph{optimal uniform sharpness} is
\begin{equation}
 \eta(G):=\max\{\eta\in[0,1]:\eta\mathbf1\in\cC_G\},
 \label{eq:uniform-sharpness-definition}
\end{equation}
where $\mathbf1\in\mathbb R^V$ is the all-ones vector.

\subsection{Example: the single-qubit Pauli family}
\label{subsec:single-qubit}

We illustrate Theorem~\ref{thm:state-space} for $\cA=\{X,Y,Z\}$, whose anti-commutativity graph is $K_3$. Here the Bloch ball determines the compatibility region and supplies a seed for every compatible sharpness vector.

For a qubit state $\rho$, the coordinates of $c_{\cA}(\rho)$ are the absolute values of its Bloch components. The state region is therefore the nonnegative octant of the Bloch ball,
\begin{equation}
 \cR_{XYZ}
 =\{(x,y,z)\in\mathbb R_+^3:x^2+y^2+z^2\leq1\}.
 \label{eq:qubit-state-region}
\end{equation}
This region is already convex and closed under coordinatewise decrease within the nonnegative orthant. Thus neither averaging profiles nor reducing their components enlarges it, and Theorem~\ref{thm:state-space} gives $\cC_{XYZ}=\cR_{XYZ}$. Equivalently, $s=(s_X,s_Y,s_Z)\in[0,1]^3$ is compatible if and only if
\begin{equation}
 s_X^2+s_Y^2+s_Z^2\leq1,
 \label{eq:qubit-ball}
\end{equation}
recovering the known criterion~\cite{Busch1986,YuOh2013,GrinkoUola2025}.

To construct a parent for such a vector, choose the seed whose Bloch vector is $s$, namely
\begin{equation}
 \rho_s=\frac12(\id+s_XX+s_YY+s_ZZ).
 \label{eq:qubit-seed}
\end{equation}
Equation~\eqref{eq:qubit-ball} ensures positivity, and $c_{\cA}(\rho_s)=s$. The Pauli-orbit construction gives the four effects $\frac12U\rho_sU^\dagger$, with $U\in\{\id,X,Y,Z\}$. Each nonidentity Pauli conjugation flips the other two Bloch components, so the resulting sign triples satisfy $a_Xa_Ya_Z=1$. Using these triples as outcome labels, the parent is
\begin{equation}
 \E(a)=\frac14\bigl(
 \id+a_Xs_XX+a_Ys_YY+a_Zs_ZZ\bigr),
 \label{eq:qubit-parent}
\end{equation}
with $a\in\{\pm1\}^3$ and $a_Xa_Ya_Z=1$.

The classical readout for observable $A_v$ is simply $a_v$. For fixed $v\in\{X,Y,Z\}$ and $b\in\{\pm1\}$, the two effects with $a_v=b$ have opposite coefficients for the other two Paulis. Those terms cancel, giving
\begin{equation}
 \sum_{\substack{a_Xa_Ya_Z=1\\ a_v=b}}\E(a)
 =\frac12(\id+b\,s_vA_v)
 =\M_v^{s_v}(b).
 \label{eq:qubit-marginal}
\end{equation}

On the spherical boundary $s_X^2+s_Y^2+s_Z^2=1$, the seed $\rho_s$ is pure and is the ground state of $-(s_XX+s_YY+s_ZZ)$, anticipating the general construction in Sec.~\ref{sec:boundary}. At the uniform optimum $s_X=s_Y=s_Z=1/\sqrt3$, this parent is the tetrahedral qubit SIC~\cite{RenesEtAl2004SIC}; the higher-dimensional connection is developed in Sec.~\ref{subsec:complete-pauli}.

Thus every compatible vector is realised exactly by a single, possibly mixed, seed, without mixing orbit POVMs or adding further output noise. Whether one seed suffices for a general Pauli family is the subject of the next subsection.

\subsection{Single seeds and mixtures}
\label{subsec:orbit-structure}

The qubit example above uses one seed for every compatible sharpness vector. For a general Pauli family, the ensemble in Theorem~\ref{thm:state-space} need not be replaceable by a single state. We first characterise when one seed suffices, and then bound the number required in general.

Throughout this subsection, we fix a canonical irreducible representation of $G$. A parent obtained from one orbit POVM by classical post-processing is called a \emph{single-orbit parent}.

\begin{proposition}
\label{prop:single-orbit}
A sharpness vector $s$ admits a single-orbit parent if and only if there exists a state $\rho$ such that
\begin{equation}
 s\leq c_{\cA}(\rho),
 \label{eq:general-single-orbit}
\end{equation}
componentwise.
\end{proposition}

\begin{proof}
If $s\leq c_{\cA}(\rho)$, the orbit POVM generated by $\rho$ has sharpness vector $c_{\cA}(\rho)$, and classical output noise reduces it to $s$.

Conversely, let $f_v(x)\in[-1,1]$ be the conditional mean of the binary result reported for observable $v$ after obtaining orbit outcome $x$. Then $s_vA_v=\sum_{x\in\cK_G}f_v(x)\E_\rho(x)$. Multiplying by $A_v$, taking the trace, and using Eq.~\eqref{eq:sign-change-action} gives
\[
 s_v=\Tr(\rho A_v)\,
 \frac1{|\cK_G|}\sum_{x\in\cK_G}f_v(x)\chi_v(x).
\]
Since the absolute value of the average is at most one, $s_v\leq|\Tr(\rho A_v)|$ for every $v$.
\end{proof}

If the set of sharpness vectors admitting a single-orbit parent is convex, it already coincides with $\mathscr C_G$. Mixing orbit POVMs generated by different seeds then yields no additional sharpness vectors: every compatible vector admits a single-orbit parent.

The ensemble size also admits a general bound in terms of the nullity $\nu_2(G)$, defined in Eq.~\eqref{eq:rank-nullity}.
\begin{corollary}
\label{cor:orbit-bound}
In a canonical irreducible representation, every $s\in\cC_G$ admits a parent obtained by mixing $N_{\mathrm{orb}}$ orbit POVMs with possibly mixed seeds, where
\begin{equation}
 N_{\mathrm{orb}}\leq\min\bigl\{|V|+1,\,2^{\nu_2(G)}\bigr\}.
 \label{eq:orbit-number-bound}
\end{equation}
A single orbit suffices whenever $B_G$ is nonsingular over $\F_2$ or $G$ is an odd cycle.
\end{corollary}

The bound $|V|+1$ follows from Carath\'eodory's theorem applied to $\cR_{\cA}\subset\mathbb R^{|V|}$~\cite{Boyd2004}. Appendix~\ref{app:stateproof} proves the nullity bound and gives the corresponding bound on parent outcomes. For odd cycles, an additional symmetry under complex conjugation reduces the bound from two orbits to one. This extends the single-seed construction for $K_3$ in Sec.~\ref{subsec:single-qubit} to all odd cycles, despite their nonzero binary nullity.

\subsection{Preparation and measurement uncertainty}

The compatibility region also determines how accurately arbitrary jointly measurable binary measurements can approximate the sharp Pauli measurements. We show that noisy Pauli marginals already attain all optimal error tradeoffs, and relate these tradeoffs to preparation uncertainty in ensembles of seed states.

For binary distributions, the total-variation distance is
\begin{equation}
 D_{\mathrm{TV}}(p,q)
 :=\frac12\sum_{a=\pm1}|p(a)-q(a)|
 =|p(+1)-q(+1)|.
 \label{eq:total-variation-distance}
\end{equation}
We measure the error of a binary POVM $\mathsf F_v$ by the largest distance between its outcome distribution and that of the sharp measurement, over all input states $\xi$~\cite{Werner2004,BuschHeinonenLahti2007,SchwonnekReebWerner2016}:
\begin{align}
 e(\mathsf F_v,A_v)
 &:=\sup_{\xi}
 \left|\Tr\left(
 \xi\bigl(\mathsf F_v(+1)-\M_v^1(+1)\bigr)
 \right)\right|
 \nonumber\\
 &=\bigl\|\mathsf F_v(+1)-\M_v^1(+1)\bigr\|_\infty.
 \label{eq:measurement-error}
\end{align}
Here, $\M_v^1$ is the POVM in Eq.~\eqref{eq:noisy} with $s_v=1$. This compares outcome statistics, not the disturbance caused by performing a measurement. For the noisy Pauli marginals,
\begin{equation}
 e(\M_v^{s_v},A_v)=\frac12(1-s_v).
 \label{eq:bitflip-measurement-error}
\end{equation}

Preparation uncertainty concerns a different task: predicting an outcome when the state is known~\cite{MaassenUffink1988,Catani2022Nonclassical}. For a state $\rho$, choosing the more probable sign minimises the probability of an incorrect prediction, giving
\begin{equation}
 u_v(\rho):=\frac12\bigl(1-c_v(\rho)\bigr),
 \qquad c_v(\rho)=|\Tr(\rho A_v)|.
 \label{eq:prediction-error}
\end{equation}
We write $u(\rho)=(u_v(\rho))_{v\in V}$. The following result expresses the attainable measurement-error bounds in terms of both $\mathscr C_G$ and averages of these prediction errors.

\begin{corollary}
\label{cor:uncertainty-compatibility}
Let $\cA=\{A_v\}_{v\in V}$ be a finite Pauli family with anti-commutativity graph $G$, and let $\epsilon\in[0,1/2]^V$. There exist jointly measurable binary POVMs $\{\mathsf F_v\}_{v\in V}$ with $e(\mathsf F_v,A_v)\leq\epsilon_v$ for every $v$ if and only if $(1-2\epsilon_v)_{v\in V}\in\mathscr C_G$. Equivalently, there is a finite ensemble of states $\{p_j,\rho_j\}$ such that
\begin{equation}
 \epsilon\geq\sum_jp_j u(\rho_j)
 \label{eq:uncertainty-compatibility}
\end{equation}
componentwise. Whenever these conditions hold, the approximations can be chosen as $\mathsf F_v=\M_v^{1-2\epsilon_v}$.
\end{corollary}

\begin{proof}
Let $\E(a)$, with $a\in\{\pm1\}^V$, be a joint POVM for the approximations. For its nonzero effects, set $p_a=\Tr(\E(a))/d$ and $\rho_a=\E(a)/\Tr(\E(a))$, where $d=\dim\mathcal H$, and write $B_v:=\mathsf F_v(+1)-\mathsf F_v(-1)=\sum_a a_v\E(a)$. Then
\begin{align*}
 \sum_a p_a u_v(\rho_a)
 &\leq\frac{1}{2d}\Tr(A_v(A_v-B_v))\\
 &\leq\frac12\|A_v-B_v\|_\infty
 =e(\mathsf F_v,A_v)\leq\epsilon_v.
\end{align*}
The first inequality uses $a_v\Tr(\rho_aA_v)\leq|\Tr(\rho_aA_v)|$ and $A_v^2=\id$; the second uses $\|A_v\|_1=d$. Thus the ensemble condition is necessary.

Conversely, since $c_v(\rho)=1-2u_v(\rho)$, the ensemble condition is equivalent to $1-2\epsilon_v\leq\sum_jp_jc_v(\rho_j)$ for every $v$. By Theorem~\ref{thm:state-space}, this is the condition $(1-2\epsilon_v)_{v\in V}\in\mathscr C_G$. The noisy Pauli marginals $\M_v^{1-2\epsilon_v}$ then attain the prescribed errors by Eq.~\eqref{eq:bitflip-measurement-error}.
\end{proof}

Thus allowing biased or otherwise more general binary approximations does not improve the error tradeoffs. The connection to preparation uncertainty is also constructive: a seed $\rho$ generates a parent with marginals satisfying
\begin{equation}
 e(\M_v^{c_v(\rho)},A_v)=u_v(\rho),
 \qquad v\in V.
 \label{eq:preparation-noise-transfer}
\end{equation}
Here $\rho$ specifies the parent measurement; it is not the unknown input $\xi$ over which the measurement error is maximised. Mixing the seed parents with probabilities $p_j$ and retaining the seed label gives marginal errors $\sum_jp_j u_v(\rho_j)$. On the preparation side, this average assumes that the label $j$ is known, so the best prediction is chosen separately for each $\rho_j$, rather than from the averaged state.

Related correspondences are known for covariant position--momentum and number--angle measurements~\cite{Werner2004,BuschKiukasWerner2018}, and for finite Fourier-conjugate pairs~\cite{Carmeli2012Informationally,SchwonnekReebWerner2016}. Here the correspondence holds for arbitrary finite Pauli families with total-variation error, allowing ensembles when a single seed does not suffice.

\subsection{Compatibility under different noise models}
\label{subsec:noise-models}

The preceding subsection compares approximate measurements through their worst-case statistical error. We now ask a different question: how much of each sharp Pauli measurement can be retained in a compatible mixture when the replacement measurements are restricted to a specified class? For a family $\cA=\{A_v\}_{v\in V}$ of nontrivial Pauli observables, consider
\[
\mathsf F_v=t_v\M_v^1+(1-t_v)\N_v,
\qquad t=(t_v)_{v\in V}\in[0,1]^V,
\]
where each $\N_v$ is a binary POVM describing the measurement noise. Write $\mathsf F:=\{\mathsf F_v\}_{v\in V}$ and $\N:=\{\N_v\}_{v\in V}$ for the resulting measurement family and the noise family, respectively. Different noise models restrict $\N$ in different ways. We ask which vectors $t$ admit an allowed noise family for which $\mathsf F$ is jointly measurable.

\emph{Depolarising noise} takes $\N_v(b)=\Tr(\M_v^1(b))\id/d$ \cite{Carmeli2012Informationally,Heinosaari2015}, while \emph{random noise} fixes $\N_v(b)=\id/2$~\cite{Heinosaari2015}. These coincide here because $\Tr(\M_v^1(b))=d/2$, recovering the noisy Pauli measurements considered throughout this work, with $t_v=s_v$. \emph{Probabilistic noise} allows $\N_v(b)=q_v(b)\id$ for arbitrary binary probability distributions $q_v$~\cite{Busch2013}. \emph{Jointly measurable noise} requires the noise family $\N$ to be jointly measurable~\cite{Cavalcanti2016}, whereas \emph{generalised noise} allows arbitrary binary noise POVMs~\cite{Haapasalo2015}. These are the noise models considered in~\cite{Designolle2019}.

The following proposition characterises all five cases in terms of the original compatibility region.

\begin{proposition}
\label{prop:pauli-nonuniform-noise}
Let $\cA=\{A_v\}_{v\in V}$ be a finite family of nontrivial Pauli observables with anti-commutativity graph $G$, and let $t\in[0,1]^V$. There exists a noise family $\N$ of the specified type such that the measurements
\[
\mathsf F_v=t_v\M_v^1+(1-t_v)\N_v,
\qquad v\in V,
\]
are jointly measurable if and only if the corresponding condition holds:
\begin{enumerate}
\item For depolarising, random, or probabilistic noise, $t\in\mathscr C_G$.
\item For jointly measurable noise, there exist $s,\gamma\in\mathscr C_G$ such that $t_v=s_v+(1-t_v)\gamma_v$ for every $v\in V$.
\item For generalised noise, $\bigl(\max\{0,2t_v-1\}\bigr)_{v\in V}\in\mathscr C_G$.
\end{enumerate}
In these three cases, the resulting compatible measurements can be chosen, respectively, as $\mathsf F_v=\M_v^{t_v}$, $\mathsf F_v=\M_v^{s_v}$, and $\mathsf F_v=\M_v^{\max\{0,2t_v-1\}}$.
\end{proposition}

\begin{proof}
For any binary measurement family $\mathsf F$, define $x(\mathsf F)=(x_v(\mathsf F))_{v\in V}$ by
\[
x_v(\mathsf F):=\frac1d\Tr\left(
A_v\bigl(\mathsf F_v(+1)-\mathsf F_v(-1)\bigr)
\right).
\]
If $\mathsf F$ is jointly measurable, normalising the nonzero effects of a parent gives an ensemble $\{p_\alpha,\rho_\alpha\}$ with
\[
|x_v(\mathsf F)|
\leq\sum_\alpha p_\alpha|\Tr(\rho_\alpha A_v)|.
\]
The state-space characterisation therefore implies $|x(\mathsf F)|\in\mathscr C_G$, where the absolute value is taken componentwise.

For the mixture, $x_v(\mathsf F)=t_v+(1-t_v)x_v(\N)$. In the first three models, the noise effects are proportional to the identity and $\Tr(A_v)=0$, so $x_v(\N)=0$ and hence $t\in\mathscr C_G$. If $\N$ is jointly measurable, then also $|x(\N)|\in\mathscr C_G$, while
\[
t_v\leq |x_v(\mathsf F)|+(1-t_v)|x_v(\N)|.
\]
Reducing the two vectors componentwise until equality holds gives $s,\gamma\in\mathscr C_G$ satisfying the stated condition, since decreasing compatible sharpnesses preserves compatibility. For generalised noise, $x_v(\N)\geq-1$, so $\max\{0,2t_v-1\}\leq|x_v(\mathsf F)|$. The same componentwise-decrease property gives the generalised-noise condition.

Conversely, in the first case choose $\N_v(b)=\id/2$, giving $\mathsf F_v=\M_v^{t_v}$. In the second, choose $\N_v(b)=\M_v^{\gamma_v}(-b)$. This noise family is jointly measurable, since swapping outcome labels preserves compatibility, and the mixture gives $\mathsf F_v=\M_v^{s_v}$. For generalised noise, choose
\[
\N_v(b)=
\begin{cases}
\M_v^1(-b), & t_v\geq 1/2,\\[2pt]
\M_v^{t_v/(1-t_v)}(-b), & t_v<1/2.
\end{cases}
\]
Both choices are valid binary POVMs and give $\mathsf F_v=\M_v^{\max\{0,2t_v-1\}}$, completing the proof.
\end{proof}

In the jointly measurable-noise case, $s$ gives the sharpnesses of the resulting compatible measurements, while $\gamma$ specifies the noise family $\N_v(b)=\M_v^{\gamma_v}(-b)$. These two compatible families need not have the same parent.

Thus the compatibility regions for all five models are determined entirely by $\mathscr C_G$ and depend only on the anti-commutativity graph, not on the particular Pauli realisation.

The usual robustness thresholds are obtained by requiring $t_v=\tau$ for every $v$ and maximising $\tau\in[0,1]$. Following the notation of~\cite{Designolle2019}, we denote them by $\eta^{\mathrm d}(\cA)$, $\eta^{\mathrm r}(\cA)$, $\eta^{\mathrm p}(\cA)$, $\eta^{\mathrm{jm}}(\cA)$, and $\eta^{\mathrm g}(\cA)$ for the five models listed above.

\begin{corollary}
\label{prop:pauli-noise-robustness}
For a finite family $\cA$ of nontrivial Pauli observables with anti-commutativity graph $G$,
\[
\eta^{\mathrm d}(\cA)
=\eta^{\mathrm r}(\cA)
=\eta^{\mathrm p}(\cA)
=\eta(G),
\]
\[
\eta^{\mathrm{jm}}(\cA)
=\frac{2\eta(G)}{1+\eta(G)},
\qquad
\eta^{\mathrm g}(\cA)
=\frac{1+\eta(G)}{2}.
\]
Any parent realising $\{\M_v^{\eta(G)}\}_{v\in V}$ attains all five optima.
\end{corollary}

\begin{proof}
Write $\eta=\eta(G)$. The first three identities and the generalised-robustness formula follow directly from Proposition~\ref{prop:pauli-nonuniform-noise}. For jointly measurable noise, its condition becomes $\tau\mathbf{1}=s+(1-\tau)\gamma$. Convexity gives
\[
\frac{\tau}{2-\tau}\mathbf{1}
=
\frac{s}{2-\tau}
+\frac{1-\tau}{2-\tau}\gamma
\in\mathscr C_G.
\]
Thus $\tau/(2-\tau)\leq\eta$, or $\tau\leq2\eta/(1+\eta)$. Equality is attained by taking $s=\gamma=\eta\mathbf{1}$. At the respective optimal values of $\tau$, the constructions in the proposition all give the same compatible family $\{\M_v^\eta\}_{v\in V}$.
\end{proof}

All five thresholds therefore give the same ordering of sharp Pauli families by incompatibility. For the single-qubit Pauli family $\{X,Y,Z\}$, the tetrahedral parent of Sec.~\ref{subsec:single-qubit} attains all five uniform robustness thresholds.

\section{Optimal tradeoffs and parent measurements}
\label{sec:boundary}
% =============================================================================

Since $\cC_G$ also determines the tradeoffs for the noise models in Sec.~\ref{subsec:noise-models}, we now focus on computing this sharpness region and constructing parents at its boundary. The key object is the support function, which records the largest attainable weighted sum of sharpnesses. For a compact convex region, these weighted optima determine the region itself. We show that, for Pauli families, they can be computed from the spectra of signed Pauli Hamiltonians, whose ground states provide seeds for optimal parent measurements. Finally, we use graph symmetry to obtain parents with optimal uniform sharpness.

\subsection{Support functions and signed Pauli Hamiltonians}
\label{subsec:boundary-geometry}

The support function records the best weighted sharpness attainable by a parent measurement. Assign nonnegative weights $w_v$ to the observables, with $\sum_v w_v=1$, and let $\Delta(V)$ denote the simplex of these weight vectors. The support function of $\cC_G$, restricted to weights $w\in \Delta(V)$, is
\begin{equation}
 h_G(w):=\max_{s\in\cC_G}w\cdot s.
 \label{eq:support-function}
\end{equation}
Different weights select different tradeoffs: increasing $w_v$ gives more importance to the sharpness of observable $A_v$. Equivalently, since the marginal measurement error is $(1-s_v)/2$, the minimum weighted sum of these errors is $(1-h_G(w))/2$.

For each $w$, the hyperplane $w\cdot s=h_G(w)$ supports $\cC_G$, with normal direction $w$. The separating-hyperplane theorem implies that support bounds over all directions determine a compact convex region~\cite{Schneider2014ConvexBodies}. 	It is enough to consider nonnegative coefficients. When a coefficient is negative, setting the corresponding sharpness to zero cannot decrease the objective and preserves compatibility. Replacing the negative coefficients by zero therefore leaves the maximum unchanged. The resulting inequality is at least as restrictive for any nonnegative vector $s$. Normalising the coefficients then gives
\begin{equation}
 \cC_G=\left\{s\in[0,1]^V:
 \begin{array}{l}
 w\cdot s\leq h_G(w)\quad \forall w\in\Delta(V)
 \end{array}\right\}.
 \label{eq:support-halfspaces}
\end{equation}
Thus the support bounds give necessary and sufficient conditions for joint measurability, including nonuniform tradeoffs: a vector $s$ is incompatible if and only if $w\cdot s>h_G(w)$ for some $w\in\Delta(V)$.

It remains to calculate these support values. For a Pauli family, this reduces to the spectra of the \emph{signed Pauli Hamiltonians}
\begin{equation}
 H_{w,a}:=\sum_{v\in V}w_va_vA_v,
 \qquad a\in\{\pm1\}^V.
 \label{eq:signed-hamiltonian}
\end{equation}
The state-space characterisation expresses these support values in terms of the largest eigenvalues of signed Pauli Hamiltonians and identifies pure seed states that attain them.

\begin{proposition}
\label{prop:boundary-variational}
For every $w\in\Delta(V)$,
\begin{align}
 h_G(w)
 &=\max_\rho\sum_{v\in V}w_v|\Tr(\rho A_v)|
 \label{eq:support-state}\\
 &=\max_{a\in\{\pm1\}^V}\lambda_{\max}(H_{w,a})
 \label{eq:support-spectral}\\
 &=-\min_{a\in\{\pm1\}^V}\lambda_{\min}(H_{w,a}).
 \label{eq:support-ground}
\end{align}
In a canonical representation, each support value is attained by a single orbit POVM with a pure seed.
\end{proposition}

\begin{proof}
By Theorem~\ref{thm:state-space}, every compatible vector is bounded componentwise by an average of seed profiles, and every seed profile is itself compatible. Since $w\geq0$, averaging profiles and reducing their coordinates cannot increase the maximum of $w\cdot s$. Hence
\begin{align*}
h_G(w)
&=\max_\rho\sum_{v\in V}w_v|\Tr(\rho A_v)|\\
&=\max_\rho\max_{a\in\{\pm1\}^V}\Tr(\rho H_{w,a})\\
&=\max_{a\in\{\pm1\}^V}\lambda_{\max}(H_{w,a}),
\end{align*}
where in the second line each sign $a_v$ is chosen to agree with $\Tr(\rho A_v)$. This proves Eqs.~\eqref{eq:support-state} and~\eqref{eq:support-spectral}.

If $a_\star$ attains the maximum, any pure eigenstate $\rho_\star$ corresponding to the largest eigenvalue of $H_{w,a_\star}$ satisfies $\sum_v w_v|\Tr(\rho_\star A_v)|=h_G(w)$. Thus its profile attains the support value and, in a canonical representation, its orbit POVM does so as well.

Finally, $H_{w,-a}=-H_{w,a}$ and both signings are allowed, so $\max_a\lambda_{\max}(H_{w,a}) =-\min_a\lambda_{\min}(H_{w,a})$, proving Eq.~\eqref{eq:support-ground}.
\end{proof}

For each choice of weights, one pure seed suffices to attain the maximum weighted sharpness. However, several sharpness vectors may attain the same maximum, and realising a specified one of them can require a mixture of seeds.

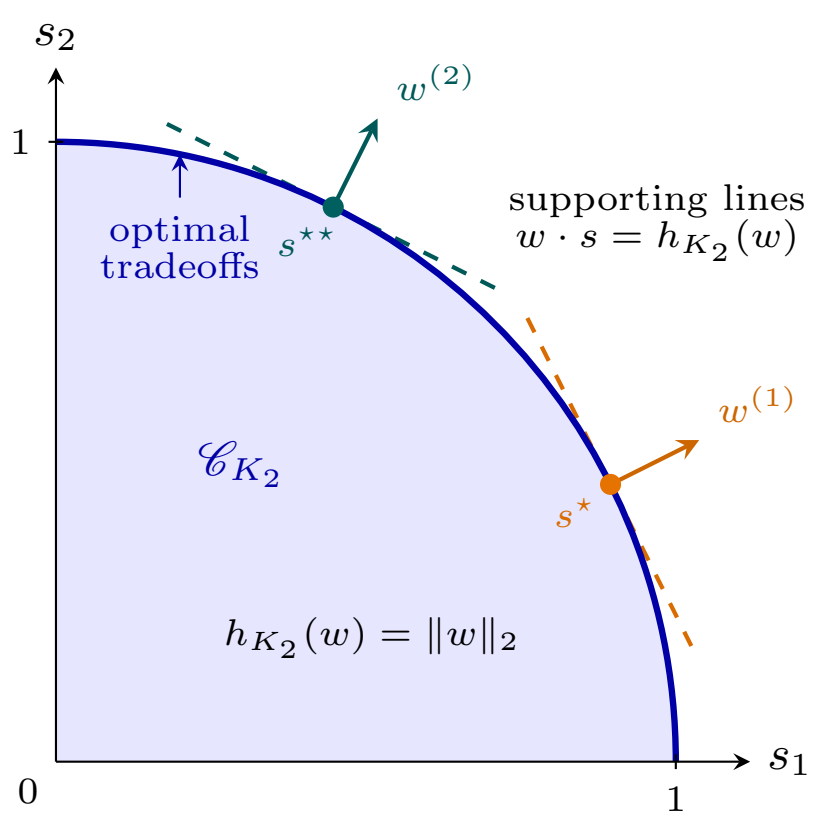
\begin{figure}[tbp]
\centering
\begin{tikzpicture}[scale=4.1,>=stealth]
  \pgfmathsetmacro{\q}{1/sqrt(5)}
  % Compatibility region.
  \fill[blue!10] (0,0)--(1,0)
    arc[start angle=0,end angle=90,radius=1]--cycle;
  % Weight choices (2/3,1/3) and (1/3,2/3).
  \coordinate (sone) at ({2*\q},\q);
  \coordinate (stwo) at (\q,{2*\q});
  % Tight weighted bounds.
  \draw[thick,dashed,orange!85!black]
    (sone) ++({-.30*\q},{.60*\q})
    --++({.60*\q},{-1.20*\q});
  \draw[thick,dashed,teal!70!black]
    (stwo) ++({-.60*\q},{.30*\q})
    --++({1.20*\q},{-.60*\q});
  % Optimal tradeoffs.
  \draw[very thick,blue!65!black]
    (1,0) arc[start angle=0,end angle=90,radius=1];
  \draw[->] (0,0)--(1.12,0) node[right] {$s_1$};
  \draw[->] (0,0)--(0,1.12) node[above] {$s_2$};
  \draw (1,.012)--(1,-.012)
    node[below,font=\scriptsize] {$1$};
  \draw (.012,1)--(-.012,1)
    node[left,font=\scriptsize] {$1$};
  \node[below left,font=\scriptsize] at (0,0) {$0$};
  % Weight directions.
  \draw[->,thick,orange!80!black]
    (sone)--++({.32*\q},{.16*\q})
    node[above right,font=\scriptsize] {$w^{(1)}$};
  \draw[->,thick,teal!70!black]
    (stwo)--++({.16*\q},{.32*\q})
    node[above right,font=\scriptsize] {$w^{(2)}$};
  \fill[orange!90!black] (sone) circle (.017);
  \fill[teal!75!black] (stwo) circle (.017);
  \node[anchor=north east,font=\scriptsize,orange!85!black]
    at (.90,.453) {$s^{\star}$};
  \node[anchor=north east,font=\scriptsize,teal!70!black]
    at (.484,.89) {$s^{\star\star}$};
  \draw[->,thin,blue!65!black]
    (.20,.91)--(.20,{sqrt(1-.20^2)});
  \node[anchor=north,font=\scriptsize,blue!65!black,align=center]
    at (.20,.91) {optimal\\[-1pt]tradeoffs};
  \node[blue!65!black] at (.30,.48) {$\cC_{K_2}$};
  \node[font=\scriptsize,align=center] at (.97,.87)
    {supporting lines\\[-1pt]$w\cdot s=h_{K_2}(w)$};
  \node[font=\scriptsize,align=center] at (.51,.20)
    {$h_{K_2}(w)=\lVert w\rVert_2$};
\end{tikzpicture}
\caption{The support function describes the compatibility region through its tight linear bounds. For two anti-commuting Pauli observables, $h_{K_2}(w)=\lVert w\rVert_2$, and the supporting line $w\cdot s=h_{K_2}(w)$ touches the region at $s(w)=w/\lVert w\rVert_2$. The weights $w^{(1)}=(2/3,1/3)$ and $w^{(2)}=(1/3,2/3)$ select the two illustrated tradeoffs, each attained by an orbit parent generated by a pure seed. The arrows show the normal directions. In the nonnegative quadrant, intersecting the half-spaces $w\cdot s\leq h_{K_2}(w)$ for all weights recovers exactly the shaded quarter disk.}
\label{fig:pareto-boundary}
\end{figure}

For example, consider $A_1=X$ and $A_2=Z$. Anti-commutation gives
\[
 H_{w,a}^2=(w_1^2+w_2^2)\id,
 \qquad h_{K_2}(w)=\sqrt{w_1^2+w_2^2}.
\]
Writing $q_w=\lVert w\rVert_2$, the pure seed
\[
 \rho_w=\frac12\left(\id+\frac{w_1X+w_2Z}{q_w}\right)
\]
is a ground state of $-(w_1X+w_2Z)$ and has profile $s(w)=w/q_w$. Its orbit POVM therefore attains the support bound $w\cdot s\leq\lVert w\rVert_2$. Taken over all nonnegative weights, these inequalities describe exactly the quarter unit disk $s_1^2+s_2^2\leq1$: one direction follows from Cauchy--Schwarz, and the converse follows by taking $w$ proportional to a nonzero candidate $s$. Figure~\ref{fig:pareto-boundary} illustrates how the supporting lines select individual tradeoffs while together determining the whole region.

Although Eq.~\eqref{eq:support-spectral} involves $2^{|V|}$ signings, Pauli conjugation makes many of them spectrally equivalent. Equation~\eqref{eq:sign-change-action} gives
\[
 U_xH_{w,a}U_x^\dagger=H_{w,a\chi(x)},
 \qquad (a\chi(x))_v=a_v\chi_v(x).
\]
The action partitions the signings into $2^{\nu_2(G)}$ equivalence classes, and only one representative from each class is needed; see Appendix~\ref{app:stateproof}. When $B_G$ is nonsingular, all signings have the same spectrum and the sign optimisation is unnecessary.

The state optimisation in Eq.~\eqref{eq:support-state} is the $k=1$ case of the generalised beta number $\beta(G,w,k)$ introduced in~\cite{XuEtAl2025Hbar}. Thus $\beta(G,w,1)=h_G(w)$ has a joint measurement interpretation: it is the largest weighted sum of sharpnesses attainable by jointly measurable noisy Pauli observables. Equation~\eqref{eq:support-halfspaces} further shows that these values characterise the entire compatibility region.

\subsection{Ground-state parent measurements}
\label{subsec:parent-seeds}

For fixed weights, several sharpness vectors may attain the same support value. These vectors form the \emph{exposed face}
\[
 F_G(w):=\{s\in\cC_G:w\cdot s=h_G(w)\},
\]
namely the part of the compatibility region that touches the supporting hyperplane. In the two-observable example, each such face is a single point; in general it may contain a continuum of tradeoffs. Proposition~\ref{prop:boundary-variational} supplies a parent generated by a pure seed attaining one point of $F_G(w)$. We now show that every point of the face can be obtained by mixing ground-state orbit POVMs, with classical post-processing when necessary. The proof is a direct consequence of the state-space characterisation: if an averaged profile attains the support value, then every seed appearing with positive probability must attain it as well.

\begin{theorem}
\label{thm:boundary-parents}
Let $w\in\Delta(V)$ and $s\in F_G(w)$. There exist a finite ensemble of pure states $\rho_j=|\psi_j\rangle\langle\psi_j|$, with probabilities $p_j$, and signings $a^{(j)}\in\{\pm1\}^V$ such that
\begin{equation}
 \begin{aligned}
 H_{w,a^{(j)}}|\psi_j\rangle
 &=-h_G(w)|\psi_j\rangle,\\
 s&\leq\sum_jp_jc_{\cA}(\rho_j).
 \end{aligned}
 \label{eq:ground-state-boundary}
\end{equation}
Consequently, $s$ admits a parent obtained from these pure ground-state seeds by the orbit construction, mixing, and classical output noise when necessary.
\end{theorem}

\begin{proof}
By Theorem~\ref{thm:state-space}, refining the seeds into pure states if necessary, choose an ensemble $\{p_j,\rho_j\}$ with $p_j>0$ such that $t:=\sum_jp_jc_{\cA}(\rho_j)\geq s$. Since $s\in F_G(w)$, Proposition~\ref{prop:boundary-variational} gives
\[
 \begin{aligned}
 h_G(w)=w\cdot s
 &\leq \sum_jp_j\,w\cdot c_{\cA}(\rho_j)\\
 &\leq \sum_jp_j h_G(w)=h_G(w).
 \end{aligned}
\]
Hence equality holds throughout. Since every $p_j>0$, each seed satisfies $w\cdot c_{\cA}(\rho_j)=h_G(w)$.

For each $j$, choose signs so that $a_v^{(j)}\Tr(\rho_jA_v)=-|\Tr(\rho_jA_v)|$ for every $v$, with either sign allowed when the expectation vanishes. Then $\Tr(\rho_jH_{w,a^{(j)}})=-h_G(w)$. By Eq.~\eqref{eq:support-ground}, $\lambda_{\min}(H_{w,a})\geq-h_G(w)$ for every signing $a$. Thus each $\rho_j$ is a ground state of its corresponding signed Hamiltonian, proving Eq.~\eqref{eq:ground-state-boundary}. The orbit construction of Theorem~\ref{thm:state-space}, followed by classical output noise when necessary, gives the stated parent.
\end{proof}

The same conclusion can also be seen directly at the parent level. If $\E(a)$ is any parent for $s\in F_G(w)$, then
\[
 0=h_G(w)-w\cdot s
 =\frac1d\sum_a\Tr\left(\E(a)\bigl(h_G(w)\id-H_{w,a}\bigr)\right).
\]
Each term is nonnegative, so every nonzero effect $\E(a)$ is supported on the $h_G(w)$-eigenspace of $H_{w,a}$. Equivalently, $E(a)$ is supported on the ground-state space of $-H_{w,a}$.

In a canonical representation, the ensemble gives the parent in Eq.~\eqref{eq:ensemble-orbit-parent}. Upon observing $(j,x)$, report $-a_v^{(j)}\chi_v(x)$ for observable $A_v$. Equation~\eqref{eq:orbit-marginal} then gives sharpness $t_v=\sum_jp_jc_v(\rho_j)$. The sign choice also includes observables with $w_v=0$, since the parent must reproduce their marginals as well. Since $t\geq s$ and $w\cdot(t-s)=0$, we have $t_v=s_v$ whenever $w_v>0$. Additional output noise is needed only on zero-weight coordinates where $t_v>s_v$; flipping the reported result with probability $(1-s_v/t_v)/2$ then gives the target.

To relate this construction to optimal tradeoffs, call a compatible vector $s$ \emph{maximal} if there is no $t\in\cC_G$ with $t\geq s$ and $t\neq s$. Their set is the \emph{upper}, or \emph{Pareto}, boundary~\cite{Boyd2004},
\[
 \partial\cC_G:=\{s\in\cC_G:s\text{ is maximal}\}.
\]
At such a point, no sharpness can be increased without decreasing another or leaving the region. Every maximal point belongs to $F_G(w)$ for some $w\in\Delta(V)$, so the theorem covers every optimal tradeoff. Conversely, if all weights are positive, every point of $F_G(w)$ is maximal; zero weights can also select nonmaximal points. If the desired sharpness vector $s$ is maximal, then the compatible averaged profile $t\geq s$ must equal $s$, so the ground-state construction requires no additional output noise.

Finally, all seed profiles in the proof belong to $F_G(w)$. Carath\'eodory's theorem represents the same averaged profile $t$ using at most $\dim F_G(w)+1\leq|V|$ pure ground-state seeds, since the face lies in the hyperplane $w\cdot t=h_G(w)$. If $F_G(w)$ consists of a single point, one pure ground-state orbit POVM suffices.

The optimal sharpness tradeoffs also determine those for the other noise models of Sec.~\ref{subsec:noise-models}. For depolarising, random, and probabilistic noise, the corresponding compatibility region in the variables $t$ is exactly $\mathscr C_G$, while for generalised noise its Pareto boundary is obtained from that of $\mathscr C_G$ by $t_v=(1+s_v)/2$. Hence, for $w\in\Delta(V)$,
\[
h_G^{\mathrm d}(w)=h_G^{\mathrm r}(w)=h_G^{\mathrm p}(w)=h_G(w),
\]
and
\[
h_G^{\mathrm g}(w)=\frac{1+h_G(w)}{2}\,.
\]
The same compatible measurements, and therefore the same parent, attain the corresponding optima. For jointly measurable noise one instead has
\[
h_G^{\mathrm{jm}}(w)
=
\max_{s,\gamma\in\mathscr C_G}
\sum_v w_v\frac{s_v+\gamma_v}{1+\gamma_v},
\]
so an additional optimisation over the compatible noise profile $\gamma$ is required.

\subsection{Uniform sharpness and graph symmetry}
\label{subsec:uniform-sharpness}
\label{subsec:graph-symmetry}

We now determine the optimal uniform sharpness $\eta(G)$, defined in Eq.~\eqref{eq:uniform-sharpness-definition}. Setting $s=\eta\mathbf1$ in a support inequality gives $\eta\leq h_G(w)$, since the weights sum to one. Because these inequalities determine the entire region, the strongest such bound is the exact threshold. Uniform sharpness need not correspond to uniform weights: without vertex transitivity, nonuniform weights can give a tighter bound.

\begin{corollary}
\label{cor:uniform-sharpness}
For a Pauli family $\cA=\{A_v\}_{v\in V}$ with anti-commutativity graph $G$, the optimal uniform sharpness is
\begin{align}
 \eta(G)
 &=\min_{w\in\Delta(V)}h_G(w)
 \label{eq:eta-support}\\
 &=\min_{w\in\Delta(V)}
   \max_{a\in\{\pm1\}^V}\lambda_{\max}(H_{w,a}).
 \label{eq:eta-minimax}
\end{align}
For every minimiser $w_\star$, an optimal parent is obtained from a suitable ensemble of pure ground states of signed Hamiltonians $H_{w_\star,a}$ with energy $-\eta(G)$, using the orbit construction and classical output noise when necessary.
\end{corollary}

\begin{proof}
Since $w\cdot\mathbf1=1$, Eq.~\eqref{eq:support-halfspaces} gives $\eta\mathbf1\in\cC_G$ if and only if $\eta\leq h_G(w)$ for every $w\in\Delta(V)$. This proves Eq.~\eqref{eq:eta-support}, and Proposition~\ref{prop:boundary-variational} gives Eq.~\eqref{eq:eta-minimax}. The minimum exists because $h_G$ is continuous and $\Delta(V)$ is compact. For any minimiser $w_\star$, we have $w_\star\cdot\bigl(\eta(G)\mathbf1\bigr) =\eta(G)=h_G(w_\star)$. Theorem~\ref{thm:boundary-parents} supplies the ground-state ensemble and parent.
\end{proof}

For vertex-transitive graphs, symmetry allows us to use equal weights and construct a parent whose marginals all have the same sharpness. Let $\mathfrak G=\operatorname{Aut}(G)$ be the group of vertex permutations preserving adjacency. In a fixed canonical representation, Appendix~\ref{app:symmetry} shows that for every seed $\rho$ and $\pi\in\mathfrak G$ there is a transformed seed $\rho^\pi$ with
\[
 c_v(\rho^\pi)=c_{\pi^{-1}(v)}(\rho).
\]
If $G$ is \emph{vertex-transitive}, every vertex can be mapped to every other by an automorphism. Averaging these permuted profiles therefore gives the same value in every coordinate.

\begin{theorem}
\label{thm:vertex-transitive}
If $G$ is vertex-transitive, the uniform weight vector $\bar w=\mathbf1/|V|$ minimises $h_G$, and
\begin{equation}
 \begin{aligned}
 \eta(G)
 &=\frac1{|V|}
   \max_{a\in\{\pm1\}^V}
   \lambda_{\max}\!\left(\sum_{v\in V}a_vA_v\right)\\
 &=\frac1{|V|}\max_\rho
   \sum_{v\in V}|\Tr(\rho A_v)|.
 \end{aligned}
 \label{eq:vertex-transitive}
\end{equation}
In a canonical representation, for any state $\rho_\star$ attaining the state maximum, an optimal parent is
\begin{equation}
 \E(\pi,x)=\frac{1}{|\mathfrak G|d_G}
 U_x\rho_\star^\pi U_x^\dagger,
 \quad (\pi,x)\in\mathfrak G\times\cK_G.
 \label{eq:vertex-transitive-parent}
\end{equation}
\end{theorem}

\begin{proof}
Let $m=|V|$, $\bar w=\mathbf1/m$, and $b=h_G(\bar w)$. The support bound gives $\eta(G)\leq b$. Choose a state $\rho_\star$ attaining $b=m^{-1}\sum_v c_v(\rho_\star)$. Vertex transitivity implies
\[
 \frac1{|\mathfrak G|}\sum_{\pi\in\mathfrak G}c_v(\rho_\star^\pi)
 =\frac1m\sum_u c_u(\rho_\star)=b
 \qquad\text{for every }v.
\]
Mixing the orbit POVMs of the transformed seeds uniformly over $\mathfrak G$ gives Eq.~\eqref{eq:vertex-transitive-parent}. For outcome $(\pi,x)$, report $\alpha_{\pi,v}\chi_v(x)$, where $\alpha_{\pi,v}\Tr(\rho_\star^\pi A_v)=|\Tr(\rho_\star^\pi A_v)|$, with either sign allowed at zero. The resulting marginals all have sharpness $b$, so $\eta(G)\geq b$. Thus $\eta(G)=h_G(\bar w)$; Corollary~\ref{cor:uniform-sharpness} shows that uniform weights minimise $h_G$, and Proposition~\ref{prop:boundary-variational} gives Eq.~\eqref{eq:vertex-transitive}.
\end{proof}

The averaged profile has equal components, although the individual seed profiles need not. The permutation label must remain part of the measurement outcome because the binary outputs assigned to the observables can depend on it. Averaging the density operators before constructing the orbit POVM need not give the same sharpnesses: expectation values of opposite signs can cancel.

A seed is \emph{balanced} when its absolute expectation values are equal for all selected observables. If a balanced seed attains the optimal uniform sharpness, its orbit POVM is already optimal, so averaging over graph automorphisms is unnecessary.

\begin{corollary}
\label{cor:balanced-orbit}
For vertex-transitive $G$, an optimal single-orbit parent at uniform sharpness exists if and only if there is a state $\rho_\star$ satisfying
\begin{equation}
 |\Tr(\rho_\star A_v)|=\eta(G),
 \qquad v\in V.
 \label{eq:balanced-optimal-seed}
\end{equation}
\end{corollary}

\begin{proof}
Such a state generates an optimal parent by the orbit construction. Conversely, a seed whose orbit parent is optimal satisfies $c_{\cA}(\rho_\star)\geq\eta(G)\mathbf1$ by Proposition~\ref{prop:single-orbit}, while Eq.~\eqref{eq:vertex-transitive} gives $\sum_{v\in V}|\Tr(\rho_\star A_v)|\leq|V|\eta(G)$. Every coordinate must therefore equal $\eta(G)$.
\end{proof}

For example, consider a mutually anti-commuting family $\{A_v\}_{v=1}^m$, with $m\geq2$, whose graph is $K_m$. For every signing $a$, set $H_a=\sum_{v=1}^m a_vA_v$. Anti-commutation gives $H_a^2=m\id$ and $\Tr(H_a)=0$, so its eigenvalues are $\pm\sqrt m$. Theorem~\ref{thm:vertex-transitive} therefore yields
\begin{equation}
 \eta(K_m)=\frac1{\sqrt m},
 \label{eq:complete-anticommuting}
\end{equation}
recovering the known threshold~\cite{KunjwalHeunenFritz2014}.

Every ground-state seed of $H_a$ is balanced. Indeed, $\{H_a,A_v\}=2a_v\id$, so a state $\rho$ supported on the ground eigenspace satisfies
\[
 \Tr(\rho A_v)
 =-\frac{\Tr\bigl(\rho\{H_a,A_v\}\bigr)}{2\sqrt m}
 =-\frac{a_v}{\sqrt m}.
\]
Thus $c_{\cA}(\rho)=\mathbf1/\sqrt m$, and its orbit POVM attains the optimal uniform sharpness without mixing seed orbits, even when the ground eigenspace is degenerate.

\section{Magic as a resource for joint measurements}
\label{sec:nonstabilizer}
% =============================================================================

\subsection{The stabilizer compatibility region}

We now ask how restricting the seeds in Theorem~\ref{thm:state-space} to stabilizer mixtures limits the attainable sharpnesses. We take convex mixtures of pure stabilizer states as the resource-free states; states outside this convex hull are nonstabilizer, or \emph{magic}, states~\cite{Gottesman1998,BravyiKitaev2005}. We call a parent a \emph{stabilizer parent} if every nonzero effect $\E(a)$ has $\E(a)/\Tr(\E(a))$ in the convex hull of pure stabilizer states. We denote by $\mathscr C_G^{\mathrm{stab}}$ the set of sharpness vectors $s\in[0,1]^V$ for which $\{\M_v^{s_v}\}_{v\in V}$ admits such a parent.

For a pure stabilizer state $\sigma$, every Pauli expectation is $0$ or $\pm1$, and the observables with nonzero expectation mutually commute. Thus $I_\sigma:=\{v\in V:|\Tr(\sigma A_v)|=1\}$ is an independent set of $G$, and
\begin{equation}
 c_{\mathscr A}(\sigma)=\mathbf1_{I_\sigma},
 \label{eq:stabilizer-profile}
\end{equation}
where $\mathbf1_I$ denotes the incidence vector of $I$. The orbit POVM generated by $\sigma$ is a stabilizer parent, since its normalised effects are Pauli conjugates of $\sigma$. Mixing these parents gives convex combinations of their profiles, which therefore lie in the stable-set polytope~\cite{Chvatal1975},
\begin{equation}
 \operatorname{STAB}(G)
 :=\operatorname{conv}\{\mathbf1_I:I\in\mathcal I(G)\},
 \label{eq:stable-set-polytope}
\end{equation}
where $\mathcal I(G)$ is the family of independent sets of $G$. Here ``stable set'' is another name for ``independent set''; $\operatorname{STAB}(G)$ denotes a graph polytope, not a set of stabilizer states.

The following theorem shows that every point of $\operatorname{STAB}(G)$ is attainable and that no stabilizer parent, whether or not it is constructed from seed orbits, can achieve sharpnesses outside this region.

\begin{figure*}[t]
\centering
\begin{minipage}[t]{0.47\textwidth}
\centering
\begin{tikzpicture}[scale=1.1]
  \definecolor{fcA}{RGB}{0,114,178}
  \definecolor{fcB}{RGB}{230,159,0}
  \definecolor{fcC}{RGB}{0,158,115}
  \definecolor{fcD}{RGB}{204,121,167}
  \definecolor{fcE}{RGB}{213,94,0}

  % Cycle edges: anti-commuting pairs
  \foreach \k in {1,...,5}{
    \coordinate (v\k) at
      ({-.8+1.3*cos(90-72*(\k-1))},
       {1.3*sin(90-72*(\k-1))});
  }
  \foreach \a/\b in {1/2,2/3,3/4,4/5,5/1}{
    \draw[thick,black!65] (v\a)--(v\b);
  }

  % Each colour represents an independent set of weight 1/2.
  % Every vertex belongs to two such sets.
  \foreach \k/\leftcol/\rightcol in
    {1/fcA/fcE,2/fcC/fcD,3/fcA/fcB,
     4/fcD/fcE,5/fcB/fcC}{
    \begin{scope}[shift={(v\k)}]
      \fill[\leftcol!80]
        (0,0)--(90:.24) arc (90:270:.24)--cycle;
      \fill[\rightcol!80]
        (0,0)--(-90:.24) arc (-90:90:.24)--cycle;

      \draw[white,line width=.7pt]
        (0,-.24)--(0,.24);
      \draw[black!65,thin] (0,0) circle (.24);

      % Vertex number
      \fill[white] (0,0) circle (.115);
      \node[font=\scriptsize,inner sep=0pt]
        at (0,0) {$\k$};
    \end{scope}
  }

  \node[font=\scriptsize,align=center] at (-.8,0)
    {At each vertex\\[2pt]
     $2\times\frac12=1$};

  % Colour key
  \node[font=\scriptsize,anchor=west] at (0.60,1.30)
    {Independent sets};

  \foreach \col/\a/\b/\y in
    {fcA/1/3/.80,fcB/3/5/.35,fcC/2/5/-.10,
     fcD/2/4/-.55,fcE/1/4/-1.00}{
    \fill[\col!80]
      (1.05,{\y-.085}) rectangle (1.22,{\y+.085});
    \node[font=\scriptsize,anchor=west] at (1.34,\y)
      {$\{\a,\b\}$};
  }

  % Fractional colouring weights
  \node[font=\scriptsize] at (.10,-1.69)
    {$w_I=\frac12,\qquad
      \chi_f(C_5)=5\times\frac12=\frac52$};

  % Corresponding measurement probabilities and sharpness
  \node[font=\scriptsize] at (.10,-2.18)
    {$p_I=\frac{w_I}{\chi_f(C_5)}=\frac15,
      \qquad \eta_{\rm stab}=\frac25$};
\end{tikzpicture}

\smallskip
\textbf{(a) Fractional colouring of $C_5$.}
\end{minipage}
\hfill
\begin{minipage}[t]{0.47\textwidth}
\centering
\begin{tikzpicture}[scale=4.0,>=stealth]
  \fill[blue!9] (0,0)--(1,0)
    arc[start angle=0,end angle=90,radius=1]--cycle;
  \fill[orange!28,opacity=.85] (0,0)--(1,0)--(0,1)--cycle;
  \draw[very thick,blue!65!black]
    (1,0) arc[start angle=0,end angle=90,radius=1];
  \draw[very thick,orange!75!black] (1,0)--(0,1);
  \draw[->] (0,0)--(1.08,0) node[right] {$s_X$};
  \draw[->] (0,0)--(0,1.08) node[above] {$s_Z$};
  \draw[dashed,gray!70] (0,0)--(.76,.76);
  \fill (0.5,0.5) circle (0.014);
  \fill (0.7071,0.7071) circle (0.014);
  \draw[<->,thick,purple!70!black]
    (.525,.525)--(.682,.682)
    node[midway,sloped,above,font=\scriptsize] {$\times\sqrt2$};
  \node[anchor=north east,font=\scriptsize] at (.49,.49)
    {$\eta_{\rm stab}=1/2$};
  \node[anchor=south west,font=\scriptsize] at (.71,.71)
    {$\eta=1/\sqrt2$};
  \node[font=\tiny,orange!75!black] at (.55,.13)
    {$\operatorname{STAB}(K_2)$};
  \node[font=\tiny,blue!65!black] at (.20,.89) {$\cC_{K_2}$};
  \node[below,font=\scriptsize] at (.5,-.08) {$X\;\text{---}\;Z$};
\end{tikzpicture}

\smallskip
\textbf{(b) One anti-commuting pair.}
\end{minipage}
\caption{Stabilizer-restricted and unrestricted joint measurements. (a) Each colour identifies one of the five maximum independent sets of $C_5$, corresponding to a commuting pair of observables. Every vertex receives two colours of weight $1/2$, giving total coverage $1$ and an optimal fractional colouring of weight $\chi_f(C_5)=5/2$. Normalising these weights gives probability $1/5$ for each commuting-pair measurement and uniform sharpness $\eta_{\rm stab}(C_5)=2/5=1/\chi_f(C_5)$. (b) For two anti-commuting Paulis $X$ and $Z$, stabilizer parents fill the triangle $\operatorname{STAB}(K_2)$, whereas arbitrary parents fill the quarter-disc $\mathscr C_{K_2}$. Along the uniform sharpness direction, allowing nonstabilizer effects increases the optimal sharpness from $\eta_{\rm stab}=1/2$ to $\eta=1/\sqrt2$, a factor of $\sqrt2$ equal to the reduced robustness of magic of the corresponding optimal seed state.}
\label{fig:stab-geometry}
\end{figure*}

\begin{theorem}
\label{thm:stabilizer-region}
For every Pauli family with anti-commutativity graph $G$,
\begin{equation}
 \mathscr C_G^{\mathrm{stab}}
 =\operatorname{STAB}(G).
 \label{eq:stab-region}
\end{equation}
Moreover, if $s\in\mathscr C_G\setminus\operatorname{STAB}(G)$, then every parent realising $s$ has a nonzero effect $\E(a)$ for which $\E(a)/\Tr(\E(a))$ is a nonstabilizer state.
\end{theorem}

\begin{proof}
Suppose $s$ is realised by a stabilizer parent. Normalising its effects as in Eq.~\eqref{eq:effect-state}, the parent-to-state bound Eq.~\eqref{eq:necessary-state-bound}, the triangle inequality, and Eq.~\eqref{eq:stabilizer-profile} give
\[
 s\leq\sum_\alpha p_\alpha c_{\mathscr A}(\rho_\alpha)
 \leq\sum_I q_I\mathbf1_I
\]
for some probability distribution $\{q_I\}$ over independent sets, obtained by decomposing the $\rho_\alpha$ into pure stabilizer states. Since $\operatorname{STAB}(G)$ is closed under coordinatewise decrease in the nonnegative orthant, $s\in\operatorname{STAB}(G)$.

Conversely, for any independent set $I$, the commuting Pauli observables $\{A_v:v\in I\}$ admit a common stabilizer eigenbasis. Every nonzero joint spectral projector, divided by its trace, is therefore a convex mixture of pure stabilizer states. This sharp joint measurement, with uniformly random outputs for $v\notin I$, is a stabilizer parent with sharpness vector $\mathbf1_I$. Mixing these parents gives every point of $\operatorname{STAB}(G)$, proving the region equality. The final assertion follows immediately.
\end{proof}

The theorem has a direct operational interpretation. Every $s\in\operatorname{STAB}(G)$ can be written as $s=\sum_I p_I\mathbf1_I$, where $p_I\geq0$ and $\sum_I p_I=1$. To realise $s$, choose the commuting subset $I$ with probability $p_I$, jointly measure its observables sharply, and assign uniformly random outputs to the remaining observables. The sharpness of observable $v$ is then $\sum_{I\ni v}p_I=s_v$. Thus classical randomisation over commuting-subset measurements attains every sharpness vector achievable by a stabilizer parent.

The optimal uniform sharpness for stabilizer parents is governed by the fractional chromatic number,
\begin{equation}
 \begin{aligned}
 \chi_f(G)=\min\biggl\{
 &\sum_{I\in\mathcal I(G)}w_I:\ w_I\geq0,\\
 &\sum_{I\ni v}w_I\geq1\quad\forall v\in V
 \biggr\}.
 \end{aligned}
 \label{eq:fractional-colouring-lp}
\end{equation}
Define
\[
 \eta_{\mathrm{stab}}(G)
 :=\max\{\eta\in[0,1]:
 \eta\mathbf1\in\mathscr C_G^{\mathrm{stab}}\}.
\]

\begin{corollary}
\label{cor:stabilizer-uniform}
For every Pauli family with anti-commutativity graph $G$,
\begin{equation}
 \eta_{\mathrm{stab}}(G)=\frac1{\chi_f(G)}.
 \label{eq:stab-uniform}
\end{equation}
\end{corollary}

\begin{proof}
By Theorem~\ref{thm:stabilizer-region} and the possibility of reducing excess sharpness by classical noise, $\eta\mathbf1\in\mathscr C_G^{\mathrm{stab}}$ if and only if there are probabilities $p_I$ with $\sum_{I\ni v}p_I\geq\eta$ for every $v$. For $\eta>0$, the weights $w_I=p_I/\eta$ form a fractional colouring of total weight $1/\eta$. Hence $\eta\leq1/\chi_f(G)$.

Conversely, an optimal fractional colouring $\{w_I\}$ gives probabilities $p_I=w_I/\chi_f(G)$ satisfying $\sum_{I\ni v}p_I\geq1/\chi_f(G)$ for every $v$. The same equivalence therefore proves achievability.
\end{proof}

An optimal fractional colouring therefore specifies the probabilities $p_I=w_I/\chi_f(G)$ in the commuting-subset construction above. Each observable is measured sharply with total probability at least $1/\chi_f(G)$; classical noise removes any excess sharpness to give the optimal uniform value $\eta_{\mathrm{stab}}(G)$.

The cycle $C_5$, for example, has five maximum independent sets, each corresponding to a commuting pair of observables; see Fig.~\ref{fig:stab-geometry}(a). Choosing one pair uniformly, measuring its observables sharply, and reporting fair random outcomes for the remaining three gives each observable sharpness $2/5$, since each belongs to two pairs. The resulting sharpness vector is the average of the five incidence vectors in $\operatorname{STAB}(C_5)$. This strategy corresponds to assigning fractional-colouring weight $1/2$ to each pair: every vertex receives total weight one, and normalising the weights gives the sampling probability $1/5$. Theorem~\ref{thm:stabilizer-region} guarantees optimality among all stabilizer parents, since every independent set has at most two vertices and hence $\sum_v s_v\leq2$ throughout $\operatorname{STAB}(C_5)$. Thus
\begin{equation}
 \eta_{\mathrm{stab}}(C_5)=\frac25,
 \qquad
 \chi_f(C_5)=\frac52.
 \label{eq:C5-stabilizer}
\end{equation}
By contrast, sampling the three colour classes of a fixed proper $3$-colouring uniformly gives sharpness only $1/3$: randomising over overlapping commuting pairs improves on using a fixed partition into colour classes.

The simplest strict separation of arbitrary and stabilizer parents occurs for two anti-commuting observables, such as $X$ and $Z$:
\begin{align}
 \operatorname{STAB}(K_2)
 &=\{(s_X,s_Z)\geq0:s_X+s_Z\leq1\},\notag\\
 \mathscr C_{K_2}
 &=\{(s_X,s_Z)\geq0:s_X^2+s_Z^2\leq1\}.
 \label{eq:K2-stabilizer-v-full}
\end{align}
The stabilizer triangle is strictly contained in the compatibility quarter-disc, as shown in Fig.~\ref{fig:stab-geometry}(b), with
\[
 \eta_{\mathrm{stab}}(K_2)=\frac12,
 \qquad
 \eta(K_2)=\frac1{\sqrt2}.
\]

More generally, if $(u,v)\in E(G)$, an optimal parent for this anti-commuting pair, with uniformly random outputs for all other observables, realises $s_u=s_v=1/\sqrt2$ and $s_w=0$ for $w\neq u,v$. This violates the stable-set inequality $s_u+s_v\leq1$. Consequently,
\[
 \mathscr C_G^{\mathrm{stab}}=\mathscr C_G
 \quad\Longleftrightarrow\quad
 G\ \text{is edgeless}.
\]
We next quantify this separation and determine its correspondence to the reduced robustness of magic.

\subsection{Maximal sharpness advantage}
\label{subsec:sv-sharpness-advantage}

We now quantify how much unrestricted parents can improve on the stabilizer benchmark. For a fixed choice of weights $w\in\Delta(V)$, Sec.~\ref{sec:boundary} gives the unrestricted optimum $h_G(w)$. Since $\cC_G^{\mathrm{stab}}=\operatorname{STAB}(G)$, the corresponding stabilizer optimum is
\begin{equation}
 \begin{aligned}
 h_G^{\mathrm{stab}}(w)
 &:=\max_{s\in\cC_G^{\mathrm{stab}}}w\cdot s\\
 &=\max_{I\in\mathcal I(G)}\sum_{v\in I}w_v
 =\alpha(G,w),
 \end{aligned}
 \label{eq:sv-stabilizer-support}
\end{equation}
where $\alpha(G,w)$ is the weighted independence number. A commuting subset of maximum total weight already attains this value; randomising between subsets cannot improve a fixed linear objective. Thus $h_G(w)/h_G^{\mathrm{stab}}(w)$ measures the multiplicative advantage for the same weighted sharpness objective.

There is also a geometric way to compare all tradeoffs at once. Define
\begin{equation}
 \operatorname{Adv}(G):=
 \inf\{\lambda\geq1:\cC_G\subseteq \lambda\operatorname{STAB}(G)\}.
 \label{eq:sv-advantage-definition}
\end{equation}
This is the smallest factor by which every compatible sharpness vector can be divided to make it attainable by a stabilizer parent. For the profile generated by a particular seed $\rho$, define similarly
\begin{equation}
 R_G(\rho):=\min\left\{\lambda\geq1:
 \frac{c_{\cA}(\rho)}{\lambda}\in\operatorname{STAB}(G)\right\}.
 \label{eq:seed-sharpness-advantage}
\end{equation}
Here $R_G(\rho)=1$ means that the seed's sharpness profile is already attainable by a stabilizer parent, not necessarily that $\rho$ is a stabilizer mixture.

\begin{proposition}
\label{prop:measurement-advantage}
For every Pauli family with anti-commutativity graph $G$,
\begin{equation}
 \operatorname{Adv}(G)
 =\max_\rho R_G(\rho)
 =\max_{w\in\Delta(V)}\frac{h_G(w)}{h_G^{\mathrm{stab}}(w)}.
 \label{eq:sv-advantage-equivalences}
\end{equation}
\end{proposition}

\begin{proof}
By Theorem~\ref{thm:state-space}, $\cC_G$ is the downward closure of the convex hull of the profiles $c_{\cA}(\rho)$. Since $\lambda\operatorname{STAB}(G)$ is convex and downward closed, it contains $\cC_G$ precisely when it contains every such profile. This proves the first equality. By the support-function characterisation of Sec.~\ref{sec:boundary}, the same containment holds precisely when $h_G(w)\leq \lambda h_G^{\mathrm{stab}}(w)$ for every $w\in\Delta(V)$. The smallest such $\lambda$ is the largest support-function ratio.
\end{proof}

Thus maximising the sharpness advantage over the weights gives the same result as maximising $R_G(\rho)$ over seed states. In a canonical representation, the orbit POVM of one pure seed attains this maximum, although realising a prescribed compatible sharpness vector may require several seeds. To construct such a seed, choose weights maximising the ratio and a pure ground state of an optimal signed Hamiltonian from Proposition~\ref{prop:boundary-variational}. Its orbit parent attains $h_G(w)$ for those weights.

The $X,Z$ example in Fig.~\ref{fig:stab-geometry}(b) illustrates both comparisons. The seed
\begin{equation}
 \rho_\star=\frac12\left(\id+\frac{X+Z}{\sqrt2}\right)
 \label{eq:sv-qubit-seed}
\end{equation}
has profile $(1,1)/\sqrt2$, which must be divided by $\sqrt2$ to reach the stabilizer triangle $s_X+s_Z\leq1$. Thus $R_{K_2}(\rho_\star)=\sqrt2$. For equal weights, the unrestricted and stabilizer optima are $1/\sqrt2$ and $1/2$, respectively, giving the same ratio.

For general families, the gain at uniform sharpness need not attain $\operatorname{Adv}(G)$. Since $\eta(G)\mathbf1\in\cC_G$, the definition gives $\eta(G)/\eta_{\mathrm{stab}}(G)\leq\operatorname{Adv}(G)$. The weights may be zero on some observables, so the maximal weighted advantage can be determined by a proper subfamily.

For perfect graphs, the maximal gain is determined by a largest mutually anti-commuting subfamily. Recall that a graph is perfect when every induced subgraph has chromatic number equal to its clique number.

\begin{corollary}
\label{cor:sv-perfect-advantage}
For a perfect graph $G$ with clique number $\omega(G)$,
\begin{equation}
 \operatorname{Adv}(G)=\sqrt{\omega(G)}.
 \label{eq:sv-perfect-advantage}
\end{equation}
\end{corollary}

\begin{proof}
For a perfect graph, $\operatorname{STAB}(G)$ is described by $s\geq0$ together with the clique inequalities $\sum_{v\in Q}s_v\leq1$ for every clique $Q$ of $G$~\cite{Chvatal1975}. Since the observables indexed by $Q$ mutually anti-commute, $\sum_{v\in Q}c_v(\rho)^2\leq1$, and Cauchy--Schwarz gives $\sum_{v\in Q}c_v(\rho)\leq\sqrt{|Q|}\leq\sqrt{\omega(G)}$. Thus $c_{\cA}(\rho)/\sqrt{\omega(G)}$ satisfies every clique inequality and therefore lies in $\operatorname{STAB}(G)$. Hence $R_G(\rho)\leq\sqrt{\omega(G)}$ for every $\rho$.

For the reverse inequality, let $Q_\star$ be a maximum clique, so $|Q_\star|=\omega(G)$. The balanced-seed construction for a mutually anti-commuting family gives a state $\rho_\star$ with $c_v(\rho_\star)=1/\sqrt{\omega(G)}$ for every $v\in Q_\star$ (for $\omega(G)=1$, choose an eigenstate of the selected observable). If $c_{\cA}(\rho_\star)/\lambda\in\operatorname{STAB}(G)$, then the clique inequality for $Q_\star$ gives $\lambda\geq\sum_{v\in Q_\star}c_v(\rho_\star)=\sqrt{\omega(G)}$. Therefore $R_G(\rho_\star)=\sqrt{\omega(G)}$, and Proposition~\ref{prop:measurement-advantage} completes the proof.
\end{proof}

The four-vertex path $P_4$, with edges $\{1,2\}$, $\{2,3\}$ and $\{3,4\}$, shows that the uniform-sharpness ratio can be strictly smaller than the maximal weighted advantage. Mixing sharp joint measurements of the commuting pairs $\{A_1,A_3\}$ and $\{A_2,A_4\}$ with equal probabilities attains uniform sharpness $1/2$. The stabilizer constraint $s_1+s_2\leq1$ shows that this is optimal, so $\eta_{\mathrm{stab}}(P_4)=1/2$. We will prove in Proposition~\ref{prop:paths-cycles} that unrestricted parents attain $\eta(P_4)=2/3$. Since $P_4$ is perfect with clique number two, Corollary~\ref{cor:sv-perfect-advantage} gives
\[
 \frac{\eta(P_4)}{\eta_{\mathrm{stab}}(P_4)}
 =\frac43<\sqrt2=\operatorname{Adv}(P_4).
\]
The maximal weighted advantage is attained by assigning weight $1/2$ to each endpoint of any one edge and zero weight to the remaining vertices.

\subsection{Connection to reduced magic}
\label{subsec:sv-reduced-magic}

Robustness of magic is the minimum sum of the absolute values of the coefficients in a decomposition of a state into pure stabilizer states, with real coefficients that sum to one and may be negative~\cite{HowardCampbell2017}. The reduced robustness applies the same construction to selected Pauli expectations, allowing magic to be certified without full state tomography~\cite{VarelaEtAl2026Magic}.

Unlike the sharpness advantage $R_G(\rho)$, reduced robustness retains the signs of the expectation values. Writing $b_{\cA}(\rho)=(\Tr(\rho A_v))_{v\in V}$, define
\[
 \begin{aligned}
 \operatorname{RoM}_{\cA}(\rho):=\min\Bigl\{&\sum_j|q_j|:
 \sum_jq_j=1,\\[-2pt]
 &b_{\cA}(\rho)=\sum_jq_jb_{\cA}(\sigma_j)\Bigr\},
 \end{aligned}
\]
where the $\sigma_j$ are pure stabilizer states and $q_j\in\mathbb R$. A value of one means that a stabilizer mixture reproduces these expectations. A larger value certifies magic from the selected observables, because negative coefficients are then unavoidable.

Signed expectations can obey constraints absent from the graph. For example, $(Z\otimes\id)(\id\otimes Z)(Z\otimes Z)=\id$ constrains the simultaneous eigenvalues of these commuting observables. Following~\cite{LiuEtAl2026Nonstabilizerness}, we say that $\cA$ has \emph{no active dependencies} if no nonempty mutually commuting subset has product proportional to the identity.

\begin{proposition}
\label{prop:reduced-magic-advantage}
Let $\cA$ be a Pauli family with anti-commutativity graph $G$. For every state $\rho$,
\begin{align}
 R_G(\rho)
 &\leq \operatorname{RoM}_{\cA}(\rho),
 \label{eq:rom-radial-bound}\\
 \operatorname{Adv}(G)
 &\leq \max_\rho\operatorname{RoM}_{\cA}(\rho).
 \label{eq:rom-advantage-bound}
\end{align}
Both bounds are equalities if $\cA$ has no active dependencies.
\end{proposition}

\begin{proof}
Any feasible decomposition has total absolute weight $t=\sum_j|q_j|\geq1$ and satisfies $c_{\cA}(\rho)\leq\sum_j|q_j|c_{\cA}(\sigma_j)$. Thus $c_{\cA}(\rho)/t$ lies componentwise below a convex combination of stabilizer profiles. This combination belongs to $\operatorname{STAB}(G)$, as does $c_{\cA}(\rho)/t$, since reducing nonnegative coordinates preserves membership. Hence $R_G(\rho)\leq t$, and minimising over decompositions proves the first bound. Appendix~\ref{app:magic} proves pointwise equality when $\cA$ has no active dependencies. The bound and equality for $\operatorname{Adv}(G)$ then follow from $\operatorname{Adv}(G)=\max_\rho R_G(\rho)$ in Proposition~\ref{prop:measurement-advantage}.
\end{proof}

Thus, in the absence of active dependencies, the reduced robustness of magic equals the smallest factor by which a seed's sharpness vector must be divided to become attainable by a stabilizer parent. Its maximum, the \emph{witness capacity} of the selected family~\cite{LiuEtAl2026Nonstabilizerness}, then equals the maximal joint measurement advantage. For perfect graphs, Corollary~\ref{cor:sv-perfect-advantage} therefore recovers the witness-capacity formula $\sqrt{\omega(G)}$ of that reference. The corollary's measurement-advantage formula itself holds even when active dependencies are present.

% =============================================================================

\section{Exact results for structured graph families}
\label{sec:structured-parents}

We now apply the general framework to structured graph families, obtaining exact compatibility characterisations and explicit parent measurements. A quadratic Majorana realisation provides a common description of line graphs, including paths and cycles. We then treat the complete quadratic Majorana and complete Pauli families, for which symmetry gives parents attaining the optimal uniform sharpness and balanced seeds sometimes make a single orbit sufficient.

\subsection{Line graphs and Gaussian parents}
\label{subsec:majorana-parents}

We begin with line graphs, which admit a natural quadratic Majorana realisation. In this representation, the relevant expectation values are encoded directly in a covariance matrix, making Gaussian states particularly convenient seed states. We show that Gaussian seeds are in fact sufficient to generate the entire compatibility region. This provides a common framework for the cycle and path families discussed later.

Let $R=(V,E)$ be a finite simple graph with $V=[N]:=\{1,\ldots,N\}$ and $E\neq\varnothing$. Its vertices will label Majorana operators, while its edges will label the observables to be measured. Let $M=2\lceil N/2\rceil$ and choose Hermitian Majorana operators $\Gamma_1,\ldots,\Gamma_M$ satisfying $\{\Gamma_u,\Gamma_v\}=2\delta_{uv}\id$~\cite{TerhalDiVincenzo2002}. We work on the full Fock space of $M/2$ fermionic modes, including both parity sectors. When $N$ is odd, $\Gamma_{N+1}$ is an auxiliary Majorana unused by the observables.

For each edge $e=\{u,v\}\in E$, with $u<v$, define $A_e=\mathrm{i}\Gamma_u\Gamma_v$, and let
\[
 \mathscr A:=\{A_e:e\in E\}.
\]
Thus the edges of $R$ index the observables in $\mathscr A$. Each $A_e$ is a Hermitian unitary and is represented by a Pauli string under the Jordan--Wigner transformation.

Recall that the line graph $L(R)$ has vertex set $E$, with two vertices $e,f\in E$ adjacent exactly when the corresponding edges of $R$ share an endpoint. The quadratic Majorana construction then gives the following correspondence~\cite{ChapmanFlammia2020,ChapmanElmanMann2023}.

\begin{lemma}
\label{lem:line-graph-realisation}
The anti-commutativity graph of $\mathscr A$ is $L(R)$.
\end{lemma}

\begin{proof}
For distinct $e,f\in E$, the Majorana relations give
\[
 A_eA_f=(-1)^{|e\cap f|}A_fA_e.
\]
Since $R$ is simple, $|e\cap f|=1$ exactly when $e$ and $f$ share an endpoint. Hence $A_e$ and $A_f$ anti-commute exactly when $e$ and $f$ are adjacent in $L(R)$.
\end{proof}

The graph $R$ therefore specifies the quadratic observables, whereas $L(R)$ records their anti-commutation relations. Sharpnesses and weights are indexed by $E$, so $s\in[0,1]^E$ and $w\in\Delta(E)$. By representation independence, the resulting compatibility region applies to every Pauli realisation of $L(R)$.

The covariance matrix
\[
 Q_{uv}:=\frac{\mathrm{i}}2\Tr\bigl(\rho[\Gamma_u,\Gamma_v]\bigr)
\]
records the expectation values relevant to the quadratic Majorana observables. In particular, for $u\neq v$, the Majorana anti-commutation relations give $Q_{uv}=\Tr(\rho\,\mathrm{i}\Gamma_u\Gamma_v)$. For any state, $Q$ is real and skew-symmetric with $\|Q\|_\infty\leq1$. For a pure Gaussian state, moreover, $Q$ determines the state completely and satisfies $Q^2=-\id$. Conversely, every real skew-symmetric matrix satisfying this condition specifies a unique pure Gaussian state $\rho_Q$~\cite{HacklBianchi2021}.

For an edge $e=\{u,v\}$ with $u<v$, $\Tr(\rho_QA_e)=Q_{uv}$, so the corresponding Pauli-orbit parent has sharpness vector
\[
 c_{\mathscr A}(\rho_Q)
 =\bigl(|Q_{uv}|\bigr)_{\{u,v\}\in E}.
\]
Thus the covariance matrix specifies the Gaussian seed, while its edge entries determine the sharpnesses. We can construct the corresponding parent directly on the full Fock space, without first passing to a canonical irreducible representation.

For this purpose, use all Majorana monomials $\Gamma^x=\Gamma_1^{x_1}\cdots\Gamma_M^{x_M}$, with $x\in\F_2^M$, and write $D=2^{M/2}$. Up to phases, these $D^2$ operators form the full Pauli basis, so the conjugation averages of Appendix~\ref{app:clifford} give the orbit POVM
\begin{equation}
 \E_\rho(x)=\frac1D\Gamma^x\rho(\Gamma^x)^\dagger,
 \qquad x\in\F_2^M,
 \label{eq:majorana-full-fock-parent}
\end{equation}
for any seed state $\rho$. For an edge $e=\{u,v\}$, conjugation gives $\Gamma^xA_e(\Gamma^x)^\dagger=(-1)^{x_u+x_v}A_e$. Reporting $f_e(x)=\alpha_e(-1)^{x_u+x_v}$, with $\alpha_e\Tr(\rho A_e)=|\Tr(\rho A_e)|$ and $\alpha_e\in\{\pm1\}$, therefore gives sharpness $|\Tr(\rho A_e)|$ by Eq.~\eqref{eq:orbit-marginal}. For the Gaussian seed $\rho_Q$, this is $|Q_{uv}|$. The orbit includes odd-degree monomials, which exchange parity sectors and ensure normalisation on the full space even when the seed has definite parity.

It remains to choose a seed that maximises a prescribed weighted sum of sharpnesses. By Proposition~\ref{prop:boundary-variational}, this reduces to the spectral optimisation of signed Hamiltonians. For a fixed signing $a\in\{\pm1\}^E$, let $S_{w,a}$ be the real skew-symmetric $N\times N$ matrix with upper-triangular entries $w_ea_e$ on edges $e=\{u,v\}$ and zero on nonedges. The corresponding signed Hamiltonian is
\begin{equation}
 H_{w,a}
 =\frac{\mathrm{i}}2\sum_{u,v=1}^{N}
 (S_{w,a})_{uv}\Gamma_u\Gamma_v.
 \label{eq:quadratic-majorana-H}
\end{equation}
Writing $S=S_{w,a}$, a state with covariance $Q$ satisfies
\[
 \Tr(\rho H_{w,a})
 =\frac12\Tr(Q^{\mathsf T}S)
 \leq\frac12\|Q\|_\infty\|S\|_1
 \leq\frac12\|S\|_1,
\]
where we used $\|Q\|_\infty\leq1$ and trace-norm/operator-norm duality (padding $S$ by a zero row and column when $N$ is odd).

When $S$ is nonsingular, the bound is attained by the pure Gaussian covariance
\begin{equation}
 Q=S(-S^2)^{-1/2}\,,
 \label{eq:majorana-polar}
\end{equation}
as shown in Appendix~\ref{app:majorana-polar}. This is the orthogonal factor in the polar decomposition of $S$: it satisfies $Q^2=-\id$ and $Q^{\mathsf T}S=(-S^2)^{1/2}$, and hence gives $\lambda_{\max}(H_{w,a})=\frac12\|S\|_1$.

For singular $S$, the polar factor on $(\ker S)^\perp$ can be extended over $\ker S$ to a pure Gaussian covariance $Q$ satisfying $Q^2=-\id$. When $N$ is odd, we first adjoin the unused auxiliary Majorana. Appendix~\ref{app:majorana-polar} gives the construction. Thus, for every signing, the spectral optimum is attained by a pure Gaussian state.

\begin{proposition}
\label{prop:line-graph-gaussian}
Let $\mathscr A=\{A_e:e\in E\}$ be the quadratic Majorana family associated with $R$, so that its anti-commutativity graph is $L(R)$. Then
\begin{equation}
 h_{L(R)}(w)
 =\frac12\max_{a\in\{\pm1\}^E}\|S_{w,a}\|_1,
 \qquad w\in\Delta(E).
 \label{eq:line-graph-support}
\end{equation}
For any maximising signing $a_\star$, the corresponding pure Gaussian covariance $Q$ gives a seed with profile $s_{\{u,v\}}=|Q_{uv}|$ satisfying $w\cdot s=h_{L(R)}(w)$. Its orbit POVM is therefore an optimal joint measurement for $w$.
\end{proposition}

\begin{proof}
For each signing $a$, the construction above gives $\lambda_{\max}(H_{w,a})=\frac12\|S_{w,a}\|_1$, attained by a pure Gaussian state. Maximising over $a$ and applying Proposition~\ref{prop:boundary-variational} gives the support formula; the Gaussian seed for a maximising signing generates the stated parent.
\end{proof}

Equation~\eqref{eq:line-graph-support} extends the skew-energy approach of~\cite{McNulty2025Graph}. Recall that the skew energy of an orientation is the trace norm of its skew-adjacency matrix, so for uniform edge weights Eq.~\eqref{eq:line-graph-support} recovers the maximum skew-energy bound used there. Here the edge weights can also vary, and minimising over them gives the exact threshold $\eta(L(R))$. Thus uniform edge weights suffice when $L(R)$ is vertex-transitive, whereas optimising the weights can give a stronger bound for other root graphs.

For a $\Delta$-regular root graph $R$, the $\Delta$ edges incident to any vertex form a clique in $L(R)$, so $\eta(L(R))\leq1/\sqrt{\Delta}$. It was shown in~\cite{McNulty2025Graph} that equality can hold only if $R$ has an orientation whose skew-adjacency matrix $S$ satisfies $SS^{\mathsf T}=\Delta\id$. Proposition~\ref{prop:line-graph-gaussian} shows that this condition is also sufficient, and gives an explicit optimal Gaussian seed. The details are given in Appendix~\ref{app:majorana}.

\begin{corollary}
\label{cor:line-graph-saturation}
Let $R$ be $\Delta$-regular. Then $\eta(L(R))=1/\sqrt{\Delta}$ if and only if $R$ admits an orientation whose skew-adjacency matrix $S$ satisfies $SS^{\mathsf T}=\Delta\id$. In this case $Q=S/\sqrt{\Delta}$ is the covariance of a balanced pure Gaussian seed, and its orbit POVM attains the optimal uniform sharpness.
\end{corollary}

More generally, the weighted support values determine the full compatibility region. Since every support value is attained by a pure Gaussian seed, Gaussian seeds suffice to generate every compatible sharpness vector, not just the uniform optimum.

\begin{corollary}
\label{cor:line-graph-gaussian-region}
A vector $s\in[0,1]^E$ belongs to $\cC_{L(R)}$ if and only if there are finitely many real skew-symmetric matrices $Q^{(j)}\in\mathbb R^{M\times M}$ satisfying $(Q^{(j)})^2=-\id$, and probabilities $p_j\geq0$, $\sum_jp_j=1$, such that
\begin{equation}
 s_{\{u,v\}}
 \leq\sum_jp_j|Q^{(j)}_{uv}|,
 \qquad \{u,v\}\in E.
 \label{eq:line-graph-gaussian-region}
\end{equation}
A parent is obtained by mixing the orbit POVMs of the Gaussian seeds $\rho_{Q^{(j)}}$ and applying classical output noise when necessary.
\end{corollary}

\begin{proof}
By Theorem~\ref{thm:state-space}, the vectors satisfying Eq.~\eqref{eq:line-graph-gaussian-region} form a compact, convex subset of $\cC_{L(R)}$ that is closed under coordinatewise decrease. Proposition~\ref{prop:line-graph-gaussian} shows that this subset has the same nonnegative support values. The support characterisation~\eqref{eq:support-halfspaces} therefore makes the two regions equal.
\end{proof}

For a pure Gaussian seed, $Q^2=-\id$ implies $\sum_{u<v\leq M}|Q_{uv}|^2=\frac M2$. When $N$ is odd, the correlations with the auxiliary Majorana have total squared magnitude one, so in both cases $\sum_{u<v\leq N}|Q_{uv}|^2 =\left\lfloor\frac N2\right\rfloor$. Restricting the sum to edges can only decrease it. The squared Euclidean norm is convex and cannot increase under coordinatewise decrease of a nonnegative vector, so Corollary~\ref{cor:line-graph-gaussian-region} gives the necessary compatibility condition
\begin{equation}
 \sum_{e\in E}s_e^2
 \leq\left\lfloor\frac N2\right\rfloor.
 \label{eq:majorana-quadratic-bound}
\end{equation}

For a pure Gaussian seed with covariance $Q$, its profile saturates Eq.~\eqref{eq:majorana-quadratic-bound} if and only if $Q_{uv}=0$ on every nonedge of $R$ among the first $N$ Majoranas. In particular, when $R$ is complete there are no nonedges, so every pure Gaussian profile saturates the bound. We use this observation later for the complete quadratic Majorana family (cf. Sec.~\ref{subsec:complete-quadratic}). In general, the bound is only necessary for compatibility.

\subsection{Cycles}
\label{subsec:cycles}

We now consider families whose anti-commutativity graph is the cycle $C_n$, with $n\geq3$. Since $L(C_n)\cong C_n$, such a family is obtained from the line graph construction by taking $R=C_n$. Explicitly, $\mathscr A=\{A_1,\ldots,A_n\}$, where $A_j=\mathrm{i}\Gamma_j\Gamma_{j+1}$ for $1\leq j<n$ and $A_n=\mathrm{i}\Gamma_n\Gamma_1$.

For this family, the weighted support function reduces to two sign choices, and the optimal uniform sharpness is attained by a single pure Gaussian seed. We retain the full Fock space convention of Sec.~\ref{subsec:majorana-parents}, including an unused auxiliary Majorana when $n$ is odd.

The signing optimisation in Eq.~\eqref{eq:line-graph-support} simplifies because changes of sign of the individual Majoranas can move all edge signs to the closing edge. Explicitly, for $a\in\{\pm1\}^n$, conjugating the signed skew-adjacency matrix by the diagonal orthogonal matrix $D_a=\operatorname{diag}(1,a_1,a_1a_2,\ldots,a_1\cdots a_{n-1})$ changes the edge signs to $(1,\ldots,1,\tau)$, where $\tau=\prod_{j=1}^n a_j$. The trace norm is unchanged, so only the two choices $\tau=\pm1$ need be considered.

For $w\in\Delta([n])$, let $S_w^{(\tau)}$ be the skew-symmetric matrix defined by
\[
 (S_w^{(\tau)})_{j,j+1}=w_j\quad(1\leq j<n),
 \qquad
 (S_w^{(\tau)})_{n,1}=\tau w_n,
\]
with opposite entries fixed by skew symmetry and all remaining entries zero. Proposition~\ref{prop:line-graph-gaussian} gives
\begin{equation}
 h_{C_n}(w)
 =\frac12\max_{\tau=\pm1}\|S_w^{(\tau)}\|_1.
 \label{eq:cycle-support}
\end{equation}
Together with Eq.~\eqref{eq:support-halfspaces}, this determines the full nonuniform compatibility region. Each support value is attained by a parent generated by a pure Gaussian state obtained from an optimal signed matrix by the covariance construction above.

For uniform sharpness, vertex transitivity and Theorem~\ref{thm:vertex-transitive} give $\eta(C_n)=h_{C_n}(\bar w)$, where $\bar w=\mathbf1/n$. Evaluating the trace norms at these weights recovers the known threshold~\cite{McNulty2025Graph},
\begin{equation}
 \eta(C_n)=
 \begin{cases}
 \displaystyle\frac1n\cot\!\left(\frac{\pi}{2n}\right),
 &n\text{ odd},\\[2mm]
 \displaystyle\frac2n\csc\!\left(\frac{\pi}{n}\right),
 &n\text{ even}.
 \end{cases}
 \label{eq:cycle-uniform}
\end{equation}

The same support value gives the tight inequality $\sum_j s_j\leq n\eta(C_n)$ for every compatible vector.

For cycles, a parent attaining the optimal uniform sharpness can be specified explicitly by the covariance matrix of a single pure Gaussian seed. For even $n$, the maximising choice is $\tau=-1$, so we set $S=S_{\bar w}^{(-1)}$. This matrix is nonsingular, and the seed covariance is $Q_\star=S(-S^2)^{-1/2}$.

For odd $n$, both choices of $\tau$ are optimal; we take $\tau=+1$ and set $S=S_{\bar w}^{(+1)}$. Its kernel is spanned by $u=\mathbf1/\sqrt n$. Let $Q_0$ be the polar factor obtained from the same formula on the nonzero modes, with $Q_0u=0$. Including the auxiliary Majorana, the full seed covariance is
\[
 Q_\star=
 \begin{pmatrix}
 Q_0 & u\\
 -u^{\mathsf T} & 0
 \end{pmatrix}.
\]
This completion satisfies $Q_\star^2=-\id$ and leaves the correlations among the original $n$ Majoranas unchanged.

In both cases, $Q_\star$ specifies a pure Gaussian state $\rho_\star$ attaining the support value $h_{C_n}(\bar w)=\eta(C_n)$. The cyclic symmetry of the chosen signed matrix makes the cycle entries of its covariance equal in magnitude. Their average magnitude is therefore also their common value, giving
\begin{equation}
 |\Tr(\rho_\star A_j)|=\eta(C_n),
 \qquad j=1,\ldots,n.
 \label{eq:cycle-balanced-seed}
\end{equation}
Thus the seed's orbit POVM gives every observable the optimal common sharpness, without mixing different seed orbits or adding output noise.

Appendix~\ref{app:cycles} verifies these covariance and symmetry statements and shows that the construction also gives a single-orbit optimal parent in a canonical irreducible representation.

\subsection{Opening cycles: paths}
\label{subsec:paths}

We now consider families whose anti-commutativity graph is the path $P_n$, with $n\geq2$. Since $L(P_{n+1})\cong P_n$, such a family is obtained from the line graph construction by taking $R=P_{n+1}$. Explicitly, $\mathscr A=\{A_1,\ldots,A_n\}$, where $A_j=\mathrm{i}\Gamma_j\Gamma_{j+1}$ for $j=1,\ldots,n$.

This family can also be obtained from the cycle family for $C_{n+1}$ by removing its closing observable. At the level of the parent measurement, this requires only omitting the corresponding classical readout, so the seed state and parent POVM can remain unchanged. This relation allows the path support function to be obtained directly from that of the cycle. Setting that observable's weight to zero makes the two signed cycle matrices coincide with the $(n+1)\times(n+1)$ skew-symmetric matrix $\Omega_w$, defined by $(\Omega_w)_{j,j+1}=w_j$, $(\Omega_w)_{j+1,j}=-w_j$, and zero entries elsewhere. Hence, for $n\geq2$ and $w\in\Delta([n])$,
\begin{equation}
 \begin{aligned}
 h_{P_n}(w)
 &=h_{C_{n+1}}(w_1,\ldots,w_n,0)\\
 &=\frac12\|\Omega_w\|_1.
 \end{aligned}
 \label{eq:path-cycle-support-restriction}
\end{equation}
Thus paths require no optimisation over signings. For each choice of weights, Proposition~\ref{prop:line-graph-gaussian} supplies a pure Gaussian seed that attains the corresponding maximum. Varying the weights determines the full compatibility region through Eq.~\eqref{eq:support-halfspaces}.

For uniform sharpness, the question is whether opening a cycle allows a larger common value. Paths with at least three vertices are not vertex-transitive, so the uniform-weight argument used for cycles no longer applies. Nevertheless, removing one or two consecutive observables from an even cycle leaves its optimal uniform sharpness unchanged. The following resolves Conjecture~1 of~\cite{McNulty2025Graph} and also shows that an optimal parent can be obtained simply by restricting an optimal parent for the even cycle.

\begin{proposition}
\label{prop:paths-cycles}
For every $m\geq1$, the cycles and paths $C_{2m+2}$, $P_{2m+1}$, and $P_{2m}$ have the same optimal uniform sharpness:
\begin{equation}
 \begin{aligned}
 \eta(C_{2m+2})
 &=\eta(P_{2m+1})=\eta(P_{2m})\\
 &=\frac1{m+1}\csc\!\left(\frac{\pi}{2m+2}\right).
 \end{aligned}
 \label{eq:path-cycle-even}
\end{equation}
Every parent that attains this common sharpness for $C_{2m+2}$ also attains it for $P_{2m+1}$ and $P_{2m}$.
\end{proposition}

\begin{proof}
Deleting one or two consecutive observables preserves the sharpness of every retained marginal, so $\eta(C_{2m+2})\leq\eta(P_{2m+1})\leq\eta(P_{2m})$. It therefore suffices to prove a matching upper bound for the shorter path.

Set $\theta=\pi/(2m+2)$. The path trace-norm calculation in Appendix~\ref{app:paths} gives, for every $s\in\cC_{P_{2m}}$,
\begin{equation}
 \sum_{j=1}^{2m}\sin(j\theta)\sin((j+1)\theta)\,s_j
 \leq\cot\theta.
 \label{eq:path-sharpness-inequality}
\end{equation}
The coefficients sum to $(m+1)\cos\theta$, so setting $s_j=\eta$ gives $\eta(P_{2m})\leq1/((m+1)\sin\theta)=\eta(C_{2m+2})$, where the last equality is Eq.~\eqref{eq:cycle-uniform}. All three thresholds therefore coincide. Since restriction preserves the retained sharpnesses, it also preserves optimality for these two paths.
\end{proof}

The proof also identifies weights whose supporting inequality is saturated at the common optimal sharpness in Proposition~\ref{prop:paths-cycles}. Normalising the coefficients in Eq.~\eqref{eq:path-sharpness-inequality} gives an optimal weight vector $w^\star$ for $P_{2m}$. These weights are symmetric under reversal of the path, but are nonuniform for $m\geq2$.

The same inequality also applies to $P_{2m+1}$ and $C_{2m+2}$: one simply assigns zero weight to the additional observables, giving $(w^\star,0)$ and $(w^\star,0,0)$, respectively. The upper bound obtained from $P_{2m}$ is therefore sufficient to prove optimality for all three families.

In particular, the covariance $Q_\star$ constructed for $C_{2m+2}$ in Sec.~\ref{subsec:cycles} already specifies an optimal seed for both paths. The same state and orbit POVM  remain unchanged; only the first $2m+1$ or $2m$ readouts are retained.

For a two-qubit example, take
\[
 \begin{aligned}
 A_1&=X\otimes\id, & A_2&=Z\otimes\id,\\
 A_3&=X\otimes X,   & A_4&=\id\otimes Z,\\
 A_5&=\id\otimes X,& A_6&=Z\otimes Z.
 \end{aligned}
\]
Their anti-commutativity graph is $C_6$, and the state
\begin{equation}
 |\psi_\star\rangle
 =\frac{3|00\rangle+|01\rangle+|10\rangle+|11\rangle}
 {2\sqrt3}
 \label{eq:C6-balanced-seed}
\end{equation}
satisfies $\langle\psi_\star|A_j|\psi_\star\rangle=2/3$ for every $j$. Its Pauli orbit therefore gives parents attaining the optimal uniform sharpness for $C_6$, $P_5$ and $P_4$.

The weighted description also reveals the tradeoffs away from uniform sharpness. For $P_4$, consider the reflection-symmetric slice $s=(x,y,y,x)$. By reflection symmetry, it is enough to consider reflection-symmetric weights. Optimising the resulting support inequalities gives the exact compatibility condition
\[
 \bigl(\max\{2x-1,0\}\bigr)^2+2y^2\leq1.
\]
Its coordinatewise maximal points form the arc
\begin{equation}
 (2x-1)^2+2y^2=1,
 \qquad
 \frac12\leq x\leq1,\quad y\geq0.
 \label{eq:P4-boundary-arc}
\end{equation}
The rest of the compatible region (see Fig.~\ref{fig:P4-symmetric-slice}) is obtained by decreasing one or both sharpnesses from a point on the arc; see Appendix~\ref{app:P4-slice}.

The uniform optimum is $x=y=2/3$. The weights $w^\star=(1,2,2,1)/6$ give the bound $x+2y\leq2$, whose boundary line is tangent to the arc at this point. Setting $x=y=\eta$ gives $3\eta\leq2$, directly recovering the optimal common sharpness. Other weight choices select the remaining points on the arc.

Opening an \emph{odd} cycle behaves differently. Removing one observable from $C_{2m+1}$ gives $P_{2m}$, but Eqs.~\eqref{eq:cycle-uniform} and~\eqref{eq:path-cycle-even} imply $\eta(C_{2m+1})<\eta(P_{2m})$. A parent attaining the optimal uniform sharpness for the odd cycle remains a valid parent for the path, but its marginals do not attain the path optimum.

This also explains why every observable must receive positive weight in a bound that establishes the odd-cycle optimum. Setting any weight to zero leaves a path problem, whose optimal common sharpness is strictly larger; such a bound cannot give the smaller cycle threshold. Appendix~\ref{app:paths} gives the details.

An interesting phenomenon arises when an even path is closed without changing the number of observables. The graphs $P_{2m}$ and $C_{2m}$ differ by one additional edge, yet $\eta(P_{2m})<\eta(C_{2m})$ for $m\geq2$. Thus adding an anti-commutation relation can increase, rather than decrease, the optimal uniform sharpness~\cite{McNulty2025Graph}.

\begin{figure}[t]
\centering
\resizebox{\columnwidth}{!}{%
\begin{tikzpicture}[
  x=6.8cm,y=4.9cm,font=\small,>=Latex,
  line cap=round,line join=round
]
\definecolor{curveblue}{RGB}{0,114,178}
\definecolor{tangentorange}{RGB}{213,94,0}

% Path diagram: endpoint and internal sharpnesses.
\node[anchor=east] at (.13,1.25) {$P_4:$};
\draw[black!60,line width=.7pt]
  (.24,1.25)--(.84,1.25);

\foreach \xx/\lab in {.24/x,.44/y,.64/y,.84/x}{
  \draw[draw=curveblue,fill=white,line width=.8pt]
    (\xx,1.25) circle[radius=2.6pt];
  \node[below=4pt] at (\xx,1.25) {$\lab$};
}

% Compatible region.
% Parametrisation keeps the arc smooth at x=1.
\fill[curveblue!10]
  (0,0)--(1,0)
  --plot[domain=0:90,samples=100,variable=\t]
    ({(1+cos(\t))/2},{sin(\t)/sqrt(2)})
  --(0,{1/sqrt(2)})--cycle;

% Horizontal boundary segment.
\draw[black!45,line width=.65pt]
  (0,{1/sqrt(2)})--(.5,{1/sqrt(2)});

% Arc of optimal tradeoffs.
\draw[curveblue,line width=1.2pt]
  plot[domain=0:90,samples=100,variable=\t]
    ({(1+cos(\t))/2},{sin(\t)/sqrt(2)});

% Supporting line.
\draw[tangentorange,dashed,line width=.9pt]
  (0,1)--(1,.5);

\node[
  text=tangentorange,
  font=\scriptsize,
  anchor=south
] at (.25,.94) {$x+2y=2$};

% Uniform optimum, with its label above the curves.
\coordinate (optimum) at ({2/3},{2/3});

\draw[
  -{Latex[length=1.5mm,width=1mm]},
  tangentorange,
  line width=.5pt,
  shorten >=3pt
] (.76,.85)--(optimum);

\node[
  text=tangentorange,
  font=\scriptsize,
  align=center,
  anchor=south
] at (.81,.86) {uniform optimum\\$x=y=\frac23$};

\filldraw[
  fill=tangentorange,
  draw=white,
  line width=.45pt
] (optimum) circle[radius=2.1pt];

\node[text=black!65,font=\scriptsize]
  at (.29,.32) {compatible};

% Axes.
\draw[->,line width=.7pt]
  (0,0)--(1.065,0) node[right] {$x$};
\draw[->,line width=.7pt]
  (0,0)--(0,1.065) node[above] {$y$};

\node[below left=2pt,font=\scriptsize]
  at (0,0) {$0$};

% Horizontal-axis ticks.
\foreach \xx/\lab in {.5/{\frac12},1/{1}}{
  \draw[line width=.5pt]
    (\xx,.012)--(\xx,-.012);
  \node[below=3pt,font=\scriptsize]
    at (\xx,0) {$\lab$};
}

% Vertical-axis ticks.
\foreach \yy/\lab in {.70710678/{\frac1{\sqrt2}},1/{1}}{
  \draw[line width=.5pt]
    (.009,\yy)--(-.009,\yy);
  \node[left=4pt,font=\scriptsize]
    at (0,\yy) {$\lab$};
}

\end{tikzpicture}%
}

\caption{Sharpness tradeoffs for $P_4$ with $s=(x,y,y,x)$. The two endpoints have sharpness $x$ and the two internal vertices have sharpness $y$. The shaded region is the compatible region in this reflection-symmetric slice. The arc, $(2x-1)^2+2y^2=1$ for $1/2\leq x\leq1$ and $y\geq0$, shows the optimal tradeoffs: increasing one sharpness requires decreasing the other. The dashed line $x+2y=2$, corresponding to $w^\star=(1,2,2,1)/6$, is tangent to the arc at the optimal common sharpness $x=y=2/3$.}
\label{fig:P4-symmetric-slice}
\end{figure}
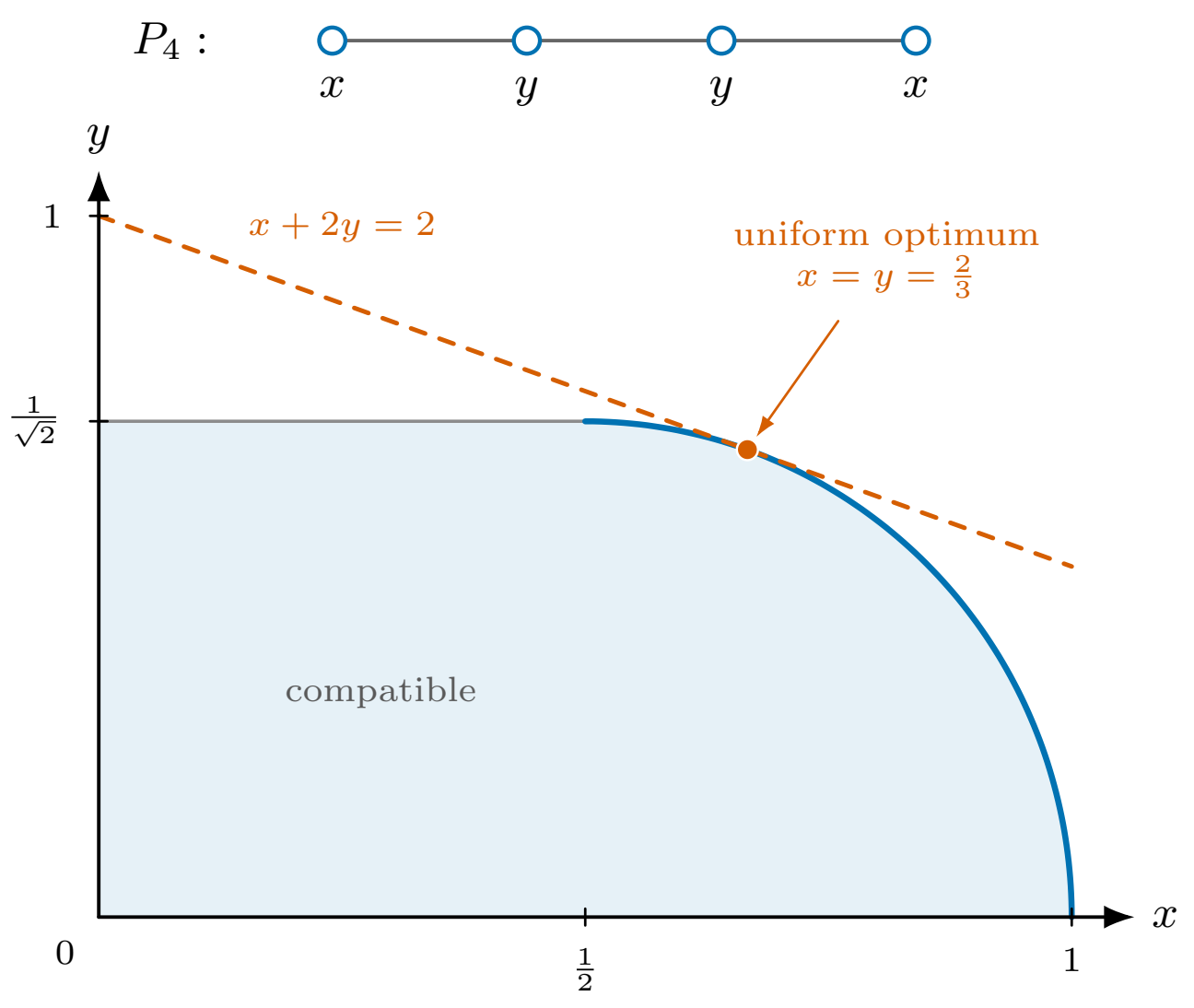

\subsection{Complete quadratic Majorana families}
\label{subsec:complete-quadratic}

We now consider the line graph of $R=K_N$. Thus, for $N$ Majorana operators, we have one quadratic observable for every unordered pair $\{u,v\}$. This gives the complete quadratic family of $\binom{N}{2}$ observables,
\[
 \cA_{N,2}
 =\{A_{uv}=\mathrm{i}\Gamma_u\Gamma_v:1\leq u<v\leq N\}.
\]
Its anti-commutativity graph is $L(K_N)=J(N,2)$, the Johnson graph.

For this family, the edge sum in Eq.~\eqref{eq:majorana-quadratic-bound} is the full sum $\sum_{1\leq u<v\leq N}|Q_{uv}|^2$. Every pure Gaussian profile therefore saturates the bound, $\sum_{1\leq u<v\leq N}|Q_{uv}|^2=\left\lfloor\frac N2\right\rfloor$. By Proposition~\ref{prop:line-graph-gaussian}, $h_{J(N,2)}(w)$ is the maximum of $\sum_{u<v}w_{uv}|Q_{uv}|$ over pure Gaussian covariances $Q$. Cauchy--Schwarz therefore gives, for $w\in\Delta(E(K_N))$,
\begin{equation}
 h_{J(N,2)}(w)
 \leq\sqrt{\left\lfloor\frac N2\right\rfloor}\,\|w\|_2,
 \label{eq:complete-quadratic-weighted-bound}
\end{equation}
with equality if and only if there exists a pure Gaussian covariance $Q$ satisfying
\begin{equation}
 |Q_{uv}|
 =\sqrt{\left\lfloor\frac N2\right\rfloor}
 \frac{w_{uv}}{\|w\|_2},
 \qquad 1\leq u<v\leq N.
 \label{eq:complete-quadratic-weighted-equality}
\end{equation}
Thus every pure Gaussian seed is optimal for some weighted sharpness objective: choosing the weights proportional to its absolute quadratic correlations makes its profile saturate Eq.~\eqref{eq:complete-quadratic-weighted-bound}.

We now turn to uniform sharpness. Since $J(N,2)$ is vertex-transitive, Theorem~\ref{thm:vertex-transitive} gives $\eta(J(N,2))=h_{J(N,2)}(\bar w)$, where $\bar w=\mathbf1/\binom{N}{2}$. Combining this with the line graph support formula~\eqref{eq:line-graph-support} gives
\begin{equation}
 \eta\bigl(J(N,2)\bigr)
 =\frac1{N(N-1)}\max_{S\in\cT_N}\|S\|_1,
 \label{eq:tournament-tracenorm}
\end{equation}
where $\cT_N$ is the set of real skew-symmetric $N\times N$ matrices with off-diagonal entries in $\{\pm1\}$. These are the skew-adjacency matrices of tournaments~\cite{Ito2017SkewEnergy}.

By Proposition~\ref{prop:line-graph-gaussian}, a maximising matrix gives a pure Gaussian seed $\rho_Q$ whose sharpness vector $c_{\mathscr A}(\rho_Q)$ attains the maximum in Eq.~\eqref{eq:vertex-transitive}. Hence
\[
 \frac{1}{\binom{N}{2}}
 \sum_{u<v}|\Tr(\rho_QA_{uv})|
 =\eta(J(N,2)).
\]
The individual sharpnesses need not, however, be equal. Averaging the parents obtained by permuting the Majorana labels, as in Theorem~\ref{thm:vertex-transitive}, makes them all equal to this average and therefore gives a parent attaining the optimal uniform sharpness.

Equation~\eqref{eq:tournament-tracenorm} gives the exact uniform optimum, whereas Eq.~\eqref{eq:complete-quadratic-weighted-bound} gives a simpler upper bound. Set $M=2\lceil N/2\rceil$, so that $M=N$ for even $N$ and $M=N+1$ for odd $N$. At uniform weights, the latter bound gives $\eta(J(N,2))\leq1/\sqrt{M-1}$. Equality requires all absolute quadratic correlations of an attaining Gaussian seed to be equal, by Eq.~\eqref{eq:complete-quadratic-weighted-equality}. The seed is then already balanced, so its orbit parent has uniform sharpness without symmetry averaging.

Saturation is characterised by \emph{skew-conference matrices}: matrices $S\in\cT_M$ satisfying $SS^{\mathsf T}=(M-1)\id$~\cite{KoukouvinosStylianou2008}. The existence of such a matrix is equivalent to the existence of a skew-Hadamard matrix of the same order, via $H_{\mathrm{SH}}=\id+S$. For even $N$, the following recovers the known saturation criterion~\cite{McNulty2025Graph} (see also~\cite[Theorem~2(i)]{McNulty2025Fermionic}); it also gives the corresponding criterion for odd $N$ and, in both cases, an optimal parent measurement from a Gaussian seed.

\begin{proposition}
\label{prop:skew-conference-parent}
Let $N\geq2$ and $M=2\lceil N/2\rceil$, i.e., $M=N$ for even $N$ and $M=N+1$ for odd $N$. Then $\eta(J(N,2))\leq1/\sqrt{M-1}$, with equality if and only if a skew-conference matrix of order $M$ exists. For every such matrix $S$, the covariance $Q_S=S/\sqrt{M-1}$ specifies a pure Gaussian seed $\rho_S$ whose orbit parent for $\cA_{N,2}$ has uniform sharpness
\begin{equation}
 s_{uv}=\eta(J(N,2))=\frac1{\sqrt{M-1}},
 \qquad 1\leq u<v\leq N.
 \label{eq:skew-conference-sharpness}
\end{equation}
\end{proposition}

\begin{proof}
The bound follows from Eq.~\eqref{eq:complete-quadratic-weighted-bound} at uniform weights. If equality holds, Eq.~\eqref{eq:complete-quadratic-weighted-equality} supplies a pure Gaussian covariance $Q$ with $|Q_{uv}|=1/\sqrt{M-1}$ for every $1\leq u<v\leq N$.

For even $N$, $M=N$, and $S=\sqrt{N-1}\,Q$ has off-diagonal entries in $\{\pm1\}$ and satisfies $SS^{\mathsf T}=(N-1)\id$. Hence $S$ is a skew-conference matrix.

For odd $N$, $M=N+1$. Since $Q^{\mathsf T}Q=\id$, the $N-1$ measured entries in each of the first $N$ rows contribute $(N-1)/N$ to its squared norm. The remaining entry, involving the auxiliary Majorana, therefore has magnitude $1/\sqrt N$. Thus every off-diagonal entry of the full covariance has magnitude $1/\sqrt N$, and $S=\sqrt N\,Q$ is a skew-conference matrix of order $N+1$.

Conversely, any skew-conference matrix $S$ of order $M$ gives a pure Gaussian covariance $Q_S=S/\sqrt{M-1}$. Its state satisfies $|\Tr(\rho_SA_{uv})|=1/\sqrt{M-1}$ for every measured pair. For odd $N$, the final Majorana is simply unused. Hence the orbit parent has uniform sharpness $1/\sqrt{M-1}$ and saturates the bound.
\end{proof}

The skew-conference seed also has a simple reduced-magic value. Since the independent sets of $J(N,2)$ are matchings of $K_N$, and a maximum matching has $\lfloor N/2\rfloor$ edges, vertex transitivity gives
\[
 \eta_{\mathrm{stab}}(J(N,2))
 =\frac{\lfloor N/2\rfloor}{\binom{N}{2}}
 =\frac1{M-1}.
\]
On the full Fock space, every nonempty commuting subset of the quadratic family is a matching whose product is a nontrivial Majorana monomial, and is therefore not scalar. Thus the family has no active dependencies. The seed's uniform sharpness $1/\sqrt{M-1}$, the definition of $R_G$, and Proposition~\ref{prop:reduced-magic-advantage} give
\[
 \begin{aligned}
 \operatorname{RoM}_{\cA_{N,2}}(\rho_S)
 &=R_{J(N,2)}(\rho_S)\\
 &=\frac{\eta(J(N,2))}{\eta_{\mathrm{stab}}(J(N,2))}
 =\sqrt{M-1}.
 \end{aligned}
\]
This statement concerns the full Fock-space realisation; restriction to a fixed parity sector can introduce active dependencies.

\subsection{Complete Pauli families and SIC parents}
\label{subsec:complete-pauli}

We finally consider all nonidentity Pauli observables on $n$ qubits. Let $d=2^n$ and choose Hermitian representatives $\{P_x:x\in\F_2^{2n}\}$ of the Pauli group modulo phase, with $P_0=\id$. The complete family is
\begin{equation}
 \cA_n
 :=
 \{P_x:x\in\F_2^{2n}\setminus\{0\}\},
 \label{eq:complete-pauli-family}
\end{equation}
containing $d^2-1$ observables.

For $x=(p,q)$ and $y=(p',q')$, the binary symplectic form $\omega(x,y)=p\cdot q'+q\cdot p'$ determines commutation: $P_xP_y=(-1)^{\omega(x,y)}P_yP_x$~\cite{DehaeneDeMoor2003}. Thus the anti-commutativity graph is the symplectic graph $G_n:=\mathrm{Sp}(2n,2)$, with nonzero Pauli labels as vertices and adjacency given by $\omega(x,y)=1$.

The Pauli operators form an orthogonal operator basis, $\Tr(P_xP_y)=d\,\delta_{xy}$. Expanding $\rho$ in this basis therefore gives $\Tr(\rho^2)=\frac1d\sum_x|\Tr(\rho P_x)|^2$. Since $P_0=\id$ and $\Tr(\rho)=1$,
\[
 \sum_{x\neq0}|\Tr(\rho P_x)|^2
 =d\Tr(\rho^2)-1\leq d-1.
\]
As in the Majorana case, averaging profiles and reducing their coordinates preserves this bound. Hence every $s\in\cC_{G_n}$ satisfies
\begin{equation}
 \sum_{x\neq0}s_x^2\leq d-1.
 \label{eq:complete-pauli-sharpness-bound}
\end{equation}
Every pure state profile saturates this bound and therefore lies on the Pareto boundary of $\cC_{G_n}$. Cauchy--Schwarz gives
\begin{equation}
 h_{G_n}(w)\leq\sqrt{d-1}\,\|w\|_2,
 \label{eq:complete-pauli-weighted-bound}
\end{equation}
with equality if and only if there exists a pure state $|\psi\rangle$ whose absolute Pauli expectations are proportional to $w$. Thus every pure state is optimal for a suitable weighted sharpness objective, attained by its Pauli-orbit parent.

We now turn to uniform sharpness. Clifford conjugations act transitively on the nonidentity Paulis, so $G_n$ is vertex-transitive and
\begin{equation}
 \eta(G_n)
 =
 \frac1{d^2-1}
 \max_\rho\sum_{x\neq0}|\Tr(\rho P_x)|.
 \label{eq:complete-pauli-eta}
\end{equation}
Setting $s_x=\eta$ in Eq.~\eqref{eq:complete-pauli-sharpness-bound} gives $\eta(G_n)\leq1/\sqrt{d+1}$. Saturating this bound requires a pure state whose nonidentity Pauli expectations all have magnitude $1/\sqrt{d+1}$. This is exactly the condition for its Pauli orbit to form a SIC.

\begin{proposition}
\label{prop:complete-pauli}
For the complete $n$-qubit Pauli family,
\begin{equation}
 \eta(G_n)\leq\frac1{\sqrt{d+1}}.
 \label{eq:complete-pauli-bound}
\end{equation}
Equality holds if and only if a Pauli-covariant SIC exists; equivalently, if and only if a parent attaining the optimal uniform sharpness can be chosen as a single Pauli orbit. If $|\psi\rangle$ is a SIC fiducial, then
\begin{equation}
 \E_\psi(x)
 =
 \frac1dP_x|\psi\rangle\langle\psi|P_x^\dagger,
 \qquad x\in\F_2^{2n},
 \label{eq:pauli-sic-parent}
\end{equation}
is an optimal parent with uniform sharpness $1/\sqrt{d+1}$.
\end{proposition}

\begin{proof}
By Eq.~\eqref{eq:complete-pauli-eta}, Cauchy--Schwarz and Pauli orthogonality give
\[
 \eta(G_n)
 \leq \frac{1}{d^2-1}
 \sqrt{(d^2-1)(d-1)}
 =\frac1{\sqrt{d+1}}.
\]
Equality holds if and only if a maximising state can be chosen pure with $|\langle\psi|P_x|\psi\rangle|^2=\frac1{d+1}$ for $x\neq0$. Since $P_x^\dagger P_y$ is proportional to $P_{x+y}$, this is exactly the condition that the Pauli orbit of $|\psi\rangle$ form a SIC~\cite{RenesEtAl2004SIC}. The orbit construction of Theorem~\ref{thm:state-space} then gives Eq.~\eqref{eq:pauli-sic-parent}.

Finally, if a parent attaining the optimal uniform sharpness is generated by a single Pauli orbit, Corollary~\ref{cor:balanced-orbit} gives a seed $\rho$ with $|\Tr(\rho P_x)|=\eta(G_n)$ for every $x\neq0$. Hence $\rho$ maximises Eq.~\eqref{eq:complete-pauli-eta} and is pure by Appendix~\ref{app:complete-pauli}. Pauli orthogonality then gives $(d^2-1)\eta(G_n)^2=d-1$, so the bound is saturated and the seed obeys the SIC condition above.
\end{proof}

Pauli-covariant SICs of this type exist exactly for $n=1$ and $n=3$~\cite{GodsilRoy2009}, giving the tetrahedral and Hoggar SICs~\cite{RenesEtAl2004SIC,SzymusiakSlomczynski2016}. In particular, the Hoggar SIC gives an optimal $64$-outcome parent for all $63$ nonidentity three-qubit Paulis, with sharpness $1/3$.

When no Pauli-covariant SIC exists, no single Pauli orbit attains the optimal uniform sharpness. Nevertheless, take any state maximising Eq.~\eqref{eq:complete-pauli-eta} and form the orbit POVMs generated by its Clifford conjugates. Mixing these POVMs as in Theorem~\ref{thm:vertex-transitive} gives every marginal the sharpness $\eta(G_n)$.

\section{Discussion}\label{sec:discussion}
% =============================================================================

Joint measurability is generally formulated as a search over the effects of a parent POVM, where the number of effects grows exponentially with the number of measurements. For Pauli observables, the present results replace this by a state-space problem: absolute Pauli expectation values determine the full compatibility region, and the same states generate explicit parent measurements through Pauli conjugations and classical post-processing. This region also determines the optimal worst-case total-variation error tradeoffs for jointly measurable binary approximations, and the full nonuniform tradeoffs under standard noise models.

This correspondence has several consequences. The full compatibility problem depends only on the anti-commutativity graph, rather than on the particular Pauli realisation. The parent effects need not be treated as independent optimisation variables, but can be organised into symmetry-related orbits generated by a bounded number of seed states. Optimal sharpness tradeoffs are determined by spectral problems for signed Pauli Hamiltonians, whose ground states provide seeds for attaining parents. 

The framework also reveals the role of magic in accessing the full compatibility region. Stabilizer seeds generate exactly the stable-set polytope of the anti-commutativity graph, so nonstabilizer resources are necessary to attain the full region. We quantified the maximal sharpness advantage and related it to the reduced robustness of magic. For several structured families, including highly symmetric ones, the framework yields explicit descriptions of the compatibility region and corresponding optimal joint measurements.

Our results also make it clear where the remaining difficulty lies when constructing optimal joint measurements. To construct a parent at a specified point on the Pareto boundary, one must find ground states of the relevant signed Pauli Hamiltonians and choose an ensemble whose averaged profile equals the desired sharpness vector. This suggests several natural questions, including how many seed orbits are genuinely required for a general graph, when a single orbit suffices, and which graph families admit tractable spectral descriptions of their compatibility regions. For estimation tasks, a further question is how to choose the seed ensemble, and hence the parent measurement, to balance implementation costs against statistical errors.

Finally, it is useful to distinguish what properties of Pauli observables are determined purely by their anti-commutativity graph. The full joint numerical range can depend on algebraic relations among commuting observables that the graph does not capture~\cite{Xu2024JointNumericalRange}. These relations also play an essential role in Pauli contextuality~\cite{Abramsky17,ChoudharyBarbosa2025,MullerGiorgetti2025} and in the reduced robustness of magic~\cite{LiuEtAl2026Nonstabilizerness}, where they constrain the allowed signed Pauli expectation values. By contrast, quantities built from absolute or squared expectation values can exhibit stronger graph invariance~\cite{deGois2023Uncertainty,XuEtAl2025Hbar}. Our results establish this invariance for the full compatibility region and the tradeoff regions under the noise models considered here.

\section*{Acknowledgements}

\begin{acknowledgments}
We acknowledge support from INFN through the project “QUANTUM” and from the Italian funding within the “Budget MUR - Dipartimenti di Eccellenza 2023–2027” - Quantum Sensing and Modelling for One-Health (QuaSiModO). We acknowledge support from Projects PN-RIC Q-SUD (Q-SUD B99H26000380007) and PN-RIC QUANTAS (SYNERGIA B99H26000410007). PF acknowledges support from the Italian National Group of Mathematical Physics (GNFM-INdAM). This work was supported in part by a grant of access to OpenAI models through the ChatGPT for Academic Researchers program. AI assistance was used in preparing figures, checking and developing proofs, and copyediting.
\end{acknowledgments}

\bibliography{references}

\appendix

\setlength{\abovedisplayskip}{6pt plus 2pt minus 2pt}
\setlength{\belowdisplayskip}{6pt plus 2pt minus 2pt}
\setlength{\abovedisplayshortskip}{0pt plus 2pt}
\setlength{\belowdisplayshortskip}{6pt plus 2pt minus 2pt}
\renewcommand{\thesection}{\Alph{section}}
\renewcommand{\theequation}{\Alph{section}\arabic{equation}}
\renewcommand{\thelemma}{\Alph{section}.\arabic{lemma}}
\setcounter{section}{0}

% =============================================================================
\refstepcounter{section}
\setcounter{equation}{0}
\setcounter{lemma}{0}
\section*{Appendix \thesection: Graph-Clifford representations and Pauli conjugation symmetries}
\label{app:clifford}
% =============================================================================

We describe the irreducible representations of the graph-Clifford relations and derive the Pauli-conjugation identities used in the orbit construction of Sec.~\ref{subsec:state-space-characterisation}. We begin by recalling the decomposition of the graph-Clifford algebra into a direct sum of full matrix algebras~\cite{Khovanova2010}, which determines these irreducible representations; see also~\cite{Cuypers2021}. Appendix~\ref{app:representations} uses this representation structure to show that the state-space condition in Eq.~\eqref{eq:main-theorem} is independent of the realisation and to transfer parent measurements between realisations.

\subsection{Canonical representation and sign-change group}
\label{app:canonical-sign-change}

Let $G=(V,E)$ be an anti-commutativity graph with adjacency matrix $B_G$ over $\F_2$. We use the binary rank $r$ and nullity $\nu_2(G)=|V|-r$ defined in Eq.~\eqref{eq:rank-nullity}. Consider the universal complex graph-Clifford algebra $\mathfrak C(G)$, generated by abstract self-adjoint unitaries $A_v$ subject only to the graph-Clifford relations of Sec.~\ref{subsec:pauli-symmetries}. Here ``universal'' means that we impose no additional algebraic relations that might hold in a particular physical realisation. The ordered monomials $A^z$ therefore satisfy
\[
 A^zA^{z'}
 =(-1)^{z^{\mathsf T}B_Gz'}A^{z'}A^z,
 \qquad z,z'\in\F_2^V.
\]
Thus the binary form $z^{\mathsf T}B_Gz'$ records whether two monomials commute or anticommute.

Since $B_G$ is alternating over $\F_2$, i.e., symmetric with zero diagonal, its rank $r$ is even. The normal form for an alternating bilinear form gives an invertible binary matrix $T$ such that
\[
 T^{\mathsf T}B_GT
 =
 \left(
 \bigoplus_{j=1}^{r/2}
 \begin{pmatrix}
  0&1\\
  1&0
 \end{pmatrix}
 \right)
 \oplus 0_{\nu_2(G)}.
\]
Each column of $T$ labels a monomial, which we take as a new generator. Each $2\times2$ block corresponds to an anticommuting pair, while generators belonging to different blocks commute. The $\nu_2(G)$ columns corresponding to the zero block lie in $\ker B_G$, so their monomials commute with every generator and are central.

Every monomial squares to $\pm\id$, so multiplying it by a suitable phase makes it a self-adjoint involution. Denote the resulting anticommuting pairs by $P_j,Q_j$, $j=1,\ldots,r/2$, and the central involutions by $C_\ell$, $\ell=1,\ldots,\nu_2(G)$. Because the columns of $T$ form a basis of $\F_2^V$, every original generator $A_v$ can be recovered, up to a fixed phase, as a product of the new generators. The change of generators is therefore invertible and gives an equivalent set of generators for $\mathfrak C(G)$.

Each pair $P_j,Q_j$ satisfies the same relations as the single-qubit Pauli operators $X,Z$ and generates $M_2(\mathbb C)$. Since different pairs commute, together they generate a tensor product of $r/2$ copies of $M_2(\mathbb C)$. In an irreducible representation, each central involution must be scalar, so $C_\ell=\lambda_\ell\id$ with $\lambda_\ell\in\{\pm1\}$. To see that every choice of these signs occurs, represent $P_j,Q_j$ by $X,Z$ on the $j$th of $r/2$ qubits and set $C_\ell=\lambda_\ell\id$. These operators satisfy the relations among the new generators, and reversing the change of generators gives a representation of the original graph-Clifford relations. Thus no constraint links the choices of the signs $\lambda_\ell$.

For each sign vector $\lambda\in\{\pm1\}^{\nu_2(G)}$, fixing these central values leaves the full matrix algebra on $r/2$ qubits. Consequently,
\[
 \mathfrak C(G)
 \cong
 \bigoplus_{\lambda\in\{\pm1\}^{\nu_2(G)}}
 M_{2^{r/2}}(\mathbb C).
\]
In particular, every irreducible realisation has dimension $d_G=2^{r/2}$. Two irreducible realisations are unitarily equivalent if and only if their central signs agree, or equivalently if all central monomials have the same scalar values in both realisations. The inverse change of generators also realises the original generators as signed Pauli strings on $r/2$ qubits.

For the remainder of this appendix, we fix a canonical irreducible representation and use the operators $U_x$ defined in Sec.~\ref{subsec:pauli-symmetries} and the characters $\chi_v$ of Eq.~\eqref{eq:character}. If $k\in\ker B_G$, then $A^k$ commutes with every generator and is therefore scalar by irreducibility. Changing the representative of $x\in\cK_G$ consequently multiplies $U_x$ only by a phase. Moreover, $U_xU_y=\mu(x,y)U_{x+y}$ with $|\mu(x,y)|=1$, so multiplication follows the group law of $\cK_G$ up to a phase. Thus the $U_x$ form a projective representation of $\cK_G$. Since these phases cancel under conjugation, the maps $X\mapsto U_xXU_x^\dagger$ define a group action on the full operator space, independent of the chosen representatives. Their action on the generators follows from the monomial commutation rule and is given in Eq.~\eqref{eq:sign-change-action}.

\subsection{Monomial basis and conjugation averages}
\label{app:monomial-characters}

\begin{lemma}\label{lem:B-basis}
The monomials $\{U_x:x\in\cK_G\}$ form an orthogonal unitary basis of the operator space of the canonical irreducible representation:
\begin{equation}
 \Tr(U_x^\dagger U_y)=d_G\,\delta_{x,y}.
 \label{eq:B-orthogonal}
\end{equation}
Their conjugation action is
\begin{equation}
 U_xU_yU_x^\dagger
 =(-1)^{\widetilde x^{\mathsf T}B_G\widetilde y}U_y,
 \label{eq:B-eigenoperator}
\end{equation}
where $\widetilde x,\widetilde y\in\F_2^V$ represent $x,y\in\cK_G$. The exponent is independent of these choices.
\end{lemma}

\begin{proof}
For $z\in\F_2^V\setminus\ker B_G$, choose $v\in V$ such that $(B_Gz)_v=1$. Then $A^z$ anticommutes with $A_v$, so $\Tr(A^z)=\Tr(A_vA^zA_v)=-\Tr(A^z)$. Hence $\Tr(A^z)=0$.

If $x\neq y$ in $\cK_G$, then $\widetilde x+\widetilde y\notin\ker B_G$, and $U_x^\dagger U_y$ is, up to phase, the monomial $A^{\widetilde x+\widetilde y}$. Its trace therefore vanishes. For $x=y$, unitarity gives $\Tr(U_x^\dagger U_x)=d_G$. This proves Eq.~\eqref{eq:B-orthogonal}. Since $|\cK_G|=d_G^2$, these orthogonal monomials form a basis of the operator space.

Equation~\eqref{eq:B-eigenoperator} follows from the monomial commutation rule in Appendix~\ref{app:canonical-sign-change}. Adding an element of $\ker B_G$ to either representative leaves the exponent unchanged.
\end{proof}

The vertex characters $\chi_v$ need not be distinct. If $\chi_u=\chi_v$, then $A_uA_v$ is unchanged by every conjugation $U_x$ and hence commutes with every generator. It is therefore scalar by irreducibility, and self-adjointness and unitarity give $A_u=\pm A_v$. An isolated vertex similarly satisfies $A_v=\pm\id$, since it commutes with every generator.

\begin{lemma}\label{lem:B-projections}
The character identities in Eq.~\eqref{eq:character-projection} hold for every operator $X$ on the canonical irreducible space:
\[
 \begin{aligned}
 \frac{1}{|\cK_G|}
 \sum_{x\in\cK_G}U_xXU_x^\dagger
 &=\frac{\Tr(X)}{d_G}\id,\\
 \frac{1}{|\cK_G|}
 \sum_{x\in\cK_G}\chi_v(x)U_xXU_x^\dagger
 &=\frac{\Tr(A_vX)}{d_G}A_v.
 \end{aligned}
\]
\end{lemma}

\begin{proof}
Conjugation by any $U_y$ permutes the summands of the first average, since $U_yU_x$ differs from $U_{x+y}$ only by a phase. The average therefore commutes with every $U_y$. Each generator $A_v$ is a phase multiple of one of these monomials, so the average commutes with every generator and is scalar by irreducibility. Taking its trace gives the first identity.

For the second identity, Eq.~\eqref{eq:sign-change-action} gives $U_xA_v=\chi_v(x)A_vU_x$. Together with $A_v^2=\id$, this yields
\[
 \begin{aligned}
 &\frac{1}{|\cK_G|}
 \sum_{x\in\cK_G}\chi_v(x)U_xXU_x^\dagger\\
 &\qquad =
 A_v\left(
 \frac{1}{|\cK_G|}
 \sum_{x\in\cK_G}U_x(A_vX)U_x^\dagger
 \right)
 =\frac{\Tr(A_vX)}{d_G}A_v,
 \end{aligned}
\]
where the last equality applies the first identity to $A_vX$.
\end{proof}

% =============================================================================
\refstepcounter{section}
\setcounter{equation}{0}
\setcounter{lemma}{0}
\section*{Appendix \thesection: Independence of the physical representation}
\label{app:representations}
% =============================================================================

We show that the state-space condition in Eq.~\eqref{eq:main-theorem} and the compatibility region defined in Eq.~\eqref{eq:compatibility-region} are independent of the chosen finite-dimensional realisation of the graph-Clifford relations. The key fact, stated in Lemma~\ref{lem:C-normalform}, is that every such realisation decomposes, up to a unitary change of basis, into copies of a fixed canonical irreducible family with some generator signs changed. We use this decomposition to compare ensembles of states in different representations and to construct parent measurements with the same sharpnesses.

\begin{lemma}\label{lem:C-normalform}
Fix a canonical irreducible realisation $\{A_v\}_{v\in V}$ of $G$ on $\mathbb C^{d_G}$. For any finite-dimensional realisation $\{A'_v\}_{v\in V}$ of the same graph-Clifford relations on $\mathcal H'$, there exist a finite-dimensional Hilbert space $\mathcal K$, commuting self-adjoint unitaries $\{D_v\}_{v\in V}$ on $\mathcal K$, and a unitary $W:\mathcal H'\longrightarrow\mathbb C^{d_G}\otimes\mathcal K$ such that
\begin{equation}
 W A'_v W^\dagger=A_v\otimes D_v,
 \qquad v\in V.
 \label{eq:C-generalrep}
\end{equation}
\end{lemma}

\begin{proof}
Theorem~2 of~\cite{Xu2024JointNumericalRange} states that, after a unitary change of basis, every realisation is a fixed standard irreducible family tensored with commuting self-adjoint unitaries. Applying this decomposition to the fixed canonical irreducible family leaves a one-dimensional multiplicity space, so the additional factors are scalar signs. Multiplying the corresponding factors for the physical realisation by these signs gives Eq.~\eqref{eq:C-generalrep} relative to the chosen family $\{A_v\}$.

In this decomposition, the physical realisation is a direct sum of canonical copies, with possibly different generator signs in different sectors. The operators $D_v$ record these signs; they are commuting self-adjoint unitaries because they act as $\pm\id$ on each sector.
\end{proof}

The factor $\mathcal K$ records the sector labels and their multiplicities within the original physical Hilbert space. We take $\{P_z\}$ to be the nonzero joint spectral projections of the commuting operators $\{D_v\}_{v\in V}$, so that
\begin{equation}
 \sum_zP_z=\id_{\mathcal K},
 \qquad
 D_vP_z=d_v(z)P_z,
 \qquad d_v(z)\in\{\pm1\}.
 \label{eq:C-central-signs}
\end{equation}
Thus, in sector $z$, the observables act as $d_v(z)A_v$ on the canonical factor and as the identity on the multiplicity factor.

The next lemma adapts the state-decomposition argument of~\cite[Appendix~C]{Xu2024JointNumericalRange} to absolute expectation-value profiles.

\begin{lemma}\label{lem:C-statevectors}
For every state $\rho'$ on $\mathcal H'$, there is an ensemble $\{q_z,\rho_z\}$ of states on the canonical irreducible space such that
\begin{equation}
 \bigl|\Tr(\rho' A'_v)\bigr|
 \leq
 \sum_z q_z\bigl|\Tr(\rho_zA_v)\bigr|,
 \qquad v\in V.
 \label{eq:C-profile-reduction}
\end{equation}
Conversely, every absolute expectation-value profile in the canonical representation is attained by a state on $\mathcal H'$.
\end{lemma}

\begin{proof}
Set $\rho:=W\rho'W^\dagger$, a state on $\mathbb C^{d_G}\otimes\mathcal K$, and define $q_z:=\Tr(\rho(\id\otimes P_z))$. For each $z$ with $q_z>0$, let
\begin{equation}
 \rho_z
 :=\frac{1}{q_z}
 \Tr_2\left(
  (\id\otimes P_z)\rho(\id\otimes P_z)
 \right),
 \label{eq:C-rhoz}
\end{equation}
where $\Tr_2$ denotes the partial trace over $\mathcal K$. Each $\rho_z$ is a state, and $\sum_zq_z=1$; labels with $q_z=0$ may be omitted.

The observables $A_v\otimes D_v$ are block diagonal with respect to the projections $\id\otimes P_z$, so coherences between different sectors do not contribute to their expectations. Using Eqs.~\eqref{eq:C-generalrep} and \eqref{eq:C-central-signs}, we therefore obtain
\begin{equation}
 \Tr(\rho' A'_v)
 =\sum_z q_z d_v(z)\Tr(\rho_zA_v).
 \label{eq:C-profile-decomp}
\end{equation}
The triangle inequality gives Eq.~\eqref{eq:C-profile-reduction}.

For the converse, let $\sigma$ be any canonical state. Choose a state $\tau_z$ supported in the range of a nonzero $P_z$ and set $\sigma':=W^\dagger(\sigma\otimes\tau_z)W$. Then
\[
 \Tr(\sigma'A'_v)
 =\Tr(\sigma A_v)\Tr(\tau_zD_v)
 =d_v(z)\Tr(\sigma A_v).
\]
Taking absolute values shows that $\sigma'$ has the same absolute expectation-value profile as $\sigma$.
\end{proof}

Applying the first part of Lemma~\ref{lem:C-statevectors} to each state of an ensemble, and multiplying its original weight by the corresponding $q_z$, gives a canonical ensemble whose averaged absolute expectations are at least as large. For each canonical state, the converse construction gives a state on $\mathcal H'$ with the same absolute expectation values. Thus the inequalities in Eq.~\eqref{eq:main-theorem} are feasible for the same sharpness vectors in every realisation. Equivalently, averaging profiles and then reducing individual coordinates gives the same set, although the profiles attainable by individual states may differ.

The next lemma extends the representation-independence argument of~\cite[Proposition~2]{McNulty2025Graph} from the uniform threshold to arbitrary sharpness vectors, with explicit maps between parent measurements.

\begin{lemma}\label{lem:C-parent-transfer}
Let $\{A'_v\}_{v\in V}$ be any finite-dimensional realisation of the same graph-Clifford relations as the canonical family $\{A_v\}_{v\in V}$. A sharpness vector $s\in[0,1]^V$ admits a parent measurement for $\{A_v\}_{v\in V}$ if and only if it admits a parent measurement for $\{A'_v\}_{v\in V}$.
\end{lemma}

\begin{proof}
Suppose first that $\{\E(a)\}_{a\in\{\pm1\}^V}$ is a parent in the canonical representation satisfying Eq.~\eqref{eq:parent-condition}. In sector $z$, the generator signs $d_v(z)$ can be implemented by reversing the corresponding outcome labels. Define $d(z)a:=\bigl(d_v(z)a_v\bigr)_{v\in V}$ and set
\begin{equation}
 \widetilde{\E}(a)
 :=\sum_z \E\bigl(d(z)a\bigr)\otimes P_z.
 \label{eq:C-lift}
\end{equation}
These effects are positive. For each $z$, the map $a\mapsto d(z)a$ is a permutation of $\{\pm1\}^V$, so the effects sum to $\id\otimes\id_{\mathcal K}$. Relabelling $b=d(z)a$ gives
\begin{align}
 \sum_a a_v\widetilde{\E}(a)
 &=\sum_z d_v(z)
   \left(\sum_b b_v\E(b)\right)\otimes P_z \nonumber\\
 &=s_vA_v\otimes D_v.
 \label{eq:C-liftproof}
\end{align}
Conjugating these effects by $W^\dagger$ gives a parent in the physical realisation.

Conversely, after conjugating a physical parent by $W$, let $\{\E'(a)\}_{a\in\{\pm1\}^V}$ denote its effects on $\mathbb C^{d_G}\otimes\mathcal K$. They satisfy
\begin{equation}
 \sum_a a_v\E'(a)=s_vA_v\otimes D_v.
 \label{eq:C-arbparent}
\end{equation}
We compress this parent to one canonical copy in any nonzero sector and reverse the same outcome signs. Choose a unit vector $\zeta_z$ in the range of a nonzero $P_z$ and define the isometry $V_z\psi:=\psi\otimes\zeta_z$. Set
\begin{equation}
 \E_z(a)
 :=V_z^\dagger\E'\bigl(d(z)a\bigr)V_z.
 \label{eq:C-partial}
\end{equation}
These effects are positive and sum to $V_z^\dagger V_z=\id$. Relabelling $a\mapsto d(z)a$ and using $D_v\zeta_z=d_v(z)\zeta_z$ gives
\begin{align}
 \sum_a a_v\E_z(a)
 &=s_vd_v(z)V_z^\dagger(A_v\otimes D_v)V_z \nonumber\\
 &=s_vA_v.
 \label{eq:C-sectorparent}
\end{align}
Thus $\{\E_z(a)\}$ is a parent in the canonical representation.
\end{proof}

Comparing every realisation with the fixed canonical family shows that the compatibility region depends only on the anti-commutativity graph $G$, as stated in Corollary~\ref{cor:graph-invariance}.

% =============================================================================
\refstepcounter{section}
\setcounter{equation}{0}
\setcounter{lemma}{0}
\section*{Appendix \thesection: Orbit bounds and binary nullity}
\label{app:orbit-bounds}
\label{app:stateproof}
% =============================================================================

The Carath\'eodory bound $N_{\mathrm{orb}}\leq |V|+1$ in Corollary~\ref{cor:orbit-bound} follows from Theorem~\ref{thm:state-space}, as explained in the main text. Here we prove the nullity bound by combining seed states whose expectation signs can be aligned by Pauli conjugation. Once the signs agree, mixing the states preserves their averaged absolute expectation values.

\subsection{The nullity bound}

Fix a canonical irreducible realisation $\cA=\{A_v\}_{v\in V}$ on $\mathbb C^{d_G}$. For $s\in\cC_G$, Theorem~\ref{thm:state-space} supplies a finite ensemble $\{p_j,\rho_j\}$ satisfying $s\leq\sum_jp_jc_{\cA}(\rho_j)$. For each seed, choose $z_j\in\F_2^V$ such that
\[
 \Tr(\rho_jA_v)
 =(-1)^{(z_j)_v}c_{\cA}(\rho_j)_v,
 \qquad v\in V,
\]
with either sign allowed when the expectation vanishes.

By Eq.~\eqref{eq:sign-change-action}, conjugating a seed by $U_x$ changes its sign vector from $z_j$ to $z_j+B_G\widetilde x$, without changing its absolute profile. The available sign changes form $S_G:=\operatorname{im}B_G$. There are $|\F_2^V/S_G|=2^{\nu_2(G)}$ cosets, and signs within each coset can therefore be aligned.

For each coset $C$ containing some $z_j$ with $p_j>0$, choose a representative $z_C$. Whenever $z_j\in C$, choose $x_j\in\cK_G$ such that $B_G\widetilde x_j=z_j+z_C$ and set $\widetilde\rho_j:=U_{x_j}\rho_jU_{x_j}^\dagger$. Then
\[
 \Tr(\widetilde\rho_jA_v)
 =(-1)^{(z_C)_v}c_{\cA}(\rho_j)_v.
\]
Define
\[
 q_C:=\sum_{j:z_j\in C}p_j,
 \qquad
 \sigma_C:=\frac1{q_C}
 \sum_{j:z_j\in C}p_j\widetilde\rho_j.
\]
All expectations being averaged have the same sign in each coordinate. Hence
\[
 q_Cc_{\cA}(\sigma_C)
 =\sum_{j:z_j\in C}p_jc_{\cA}(\rho_j).
\]
Summing over the occupied cosets gives
\[
 \sum_Cq_Cc_{\cA}(\sigma_C)
 =\sum_jp_jc_{\cA}(\rho_j)\geq s.
\]
Thus one mixed seed per occupied coset suffices. Applying the parent construction in Eq.~\eqref{eq:ensemble-orbit-parent}, together with the Carath\'eodory bound, proves
\begin{equation}
 N_{\mathrm{orb}}
 \leq
 \min\bigl\{|V|+1,\,2^{\nu_2(G)}\bigr\}.
 \label{eq:D-orbit-bound}
\end{equation}
If $B_G$ is nonsingular over $\F_2$, all sign vectors lie in one coset and a single orbit suffices.

Each orbit POVM has $|\cK_G|=d_G^2$ labelled outcomes. Consequently, every compatible sharpness vector admits a parent in the canonical representation with
\begin{equation}
 |\Omega|
 \leq
 d_G^2\min\bigl\{|V|+1,\,2^{\nu_2(G)}\bigr\}.
 \label{eq:D-outcome-bound}
\end{equation}
Here $\Omega$ records the seed and orbit labels. Classical post-processing, including output noise when needed, gives the required binary marginals without increasing the number of POVM outcomes.

The same cosets reduce the spectral optimisation in Eq.~\eqref{eq:support-spectral}. Writing $a_v(z)=(-1)^{z_v}$, we have
\[
 U_xH_{w,a(z)}U_x^\dagger
 =H_{w,a(z+B_G\widetilde x)}.
\]
It therefore suffices to check one signing per coset.

\subsection{Refinement by complex conjugation}

Choose a basis in which the canonical generators are Pauli strings, and let $*$ denote entrywise complex conjugation. Then $A_v^*=(-1)^{\kappa_v}A_v$ for some $\kappa\in\F_2^V$, and
\[
 \Tr(\rho^*A_v)
 =(-1)^{\kappa_v}\Tr(\rho A_v).
\]
Thus complex conjugation changes a seed's sign vector from $z$ to $z+\kappa$ while preserving its absolute profile.

If $\kappa\notin S_G$, this transformation pairs distinct cosets $C$ and $C+\kappa$. Conjugating the seeds in one coset of each pair places their sign vectors in the other coset. Pauli conjugation then makes the expectation values have the same sign for each observable. Mixing these states gives one seed for each pair of cosets. Consequently,
\begin{equation}
 N_{\mathrm{orb}}
 \leq
 \min\bigl\{|V|+1,\,2^{\nu_2(G)-1}\bigr\}.
 \label{eq:D-refined-orbit-bound}
\end{equation}
The factor $2^{\nu_2(G)}$ in Eq.~\eqref{eq:D-outcome-bound} can likewise be replaced by $2^{\nu_2(G)-1}$.

A sufficient graph condition for this improvement is the existence of $k\in\ker B_G$ whose support induces an odd number of edges. Indeed, $A^k$ is scalar by irreducibility, so $(A^k)^*=(A^k)^\dagger$. Reversing its factors contributes one minus sign per induced edge, whereas conjugating the factors individually contributes $(-1)^{\kappa\cdot k}$. Hence
\[
 (-1)^{\kappa\cdot k}A^k
 =(A^k)^*
 =(A^k)^\dagger
 =-A^k.
\]
Thus $\kappa\cdot k=1$, and $S_G=(\ker B_G)^\perp$ implies $\kappa\notin S_G$.

For an odd cycle $C_{2m+1}$, $m\geq1$, the kernel equation $B_Gk=0$ gives $k_{j+2}=k_j$, with indices taken cyclically. The odd cycle length forces all coordinates to be equal, so the kernel is spanned by the all-ones vector. Its support induces $2m+1$ edges. Thus $\nu_2(C_{2m+1})=1$, and Eq.~\eqref{eq:D-refined-orbit-bound} gives $N_{\mathrm{orb}}\leq1$, proving the odd-cycle assertion in Corollary~\ref{cor:orbit-bound}.

The paired signing classes also have identical spectra, since
\[
 H_{w,a(z+\kappa)}=H_{w,a(z)}^*.
\]
Under the same graph condition, at most $2^{\nu_2(G)-1}$ representatives therefore suffice in the spectral optimisation.

The combined seeds may be mixed; decomposing them into pure states need not preserve these bounds on the number of orbit POVMs. Complex conjugation is used only to specify a new seed, not as an operation on the unknown input state.

% =============================================================================
\refstepcounter{section}
\setcounter{equation}{0}
\setcounter{lemma}{0}
\section*{Appendix \thesection: Automorphisms and symmetry reduction}
\label{app:symmetry}
% =============================================================================

Fix a canonical irreducible family $\cA=\{A_v\}_{v\in V}$ on $\mathbb C^{d_G}$, with $|V|\geq1$, and let $\mathfrak G=\operatorname{Aut}(G)$. We first show that graph automorphisms permute the attainable absolute expectation-value profiles. We then use this symmetry to reduce the weight minimisation in Eq.~\eqref{eq:eta-support}.

\begin{lemma}\label{lem:E-symmetry}
For every $\pi\in\mathfrak G$, there exists a unitary $W_\pi$ such that, for every state $\rho$, the state $\rho^\pi:=W_\pi\rho W_\pi^\dagger$ satisfies
\begin{equation}
 c_{\cA}(\rho^\pi)_v
 =
 c_{\cA}(\rho)_{\pi^{-1}(v)},
 \qquad v\in V.
 \label{eq:E-profile-permutation}
\end{equation}
\end{lemma}

\begin{proof}
The permuted family $\{A_{\pi(v)}\}_{v\in V}$ is an irreducible realisation of the same graph-Clifford relations. Since both families act on $\mathbb C^{d_G}$, the multiplicity space in Lemma~\ref{lem:C-normalform} is one-dimensional. Hence there exist a unitary $W_\pi$ and signs $\sigma_{\pi,v}\in\{\pm1\}$ such that
\begin{equation}
 W_\pi A_vW_\pi^\dagger
 =
 \sigma_{\pi,v}A_{\pi(v)}.
 \label{eq:E-autlift}
\end{equation}
It follows that $\Tr(\rho^\pi A_{\pi(v)}) =\sigma_{\pi,v}\Tr(\rho A_v)$. Taking absolute values and relabelling $v$ gives Eq.~\eqref{eq:E-profile-permutation}.

The unitaries may be chosen separately for each $\pi$; no group representation property of $\pi\mapsto W_\pi$ is required.
\end{proof}

For $w\in\Delta(V)$, define $(\pi w)_v:=w_{\pi^{-1}(v)}$. Equation~\eqref{eq:E-profile-permutation} gives $(\pi w)\cdot c_{\cA}(\rho^\pi) =w\cdot c_{\cA}(\rho)$. As $\rho$ ranges over all states, so does $\rho^\pi$. Maximising over $\rho$ and using Eq.~\eqref{eq:support-state} therefore yields
\begin{equation}
 h_G(\pi w)=h_G(w).
 \label{eq:E-hinvariant}
\end{equation}

We now apply the standard averaging argument for convex optimisation~\cite[Appendix~B.1]{Nguyen2020Symmetries}. For a subgroup $\Lambda\leq\mathfrak G$, define $\bar w:=|\Lambda|^{-1}\sum_{\pi\in\Lambda}\pi w$. This vector belongs to $\Delta(V)$ and is fixed by every element of $\Lambda$, so it is constant on each vertex orbit of $\Lambda$. Convexity of the support function and Eq.~\eqref{eq:E-hinvariant} give
\begin{equation}
 h_G(\bar w)
 \leq
 \frac{1}{|\Lambda|}
 \sum_{\pi\in\Lambda}h_G(\pi w)
 =
 h_G(w).
 \label{eq:E-average}
\end{equation}
Averaging a minimiser therefore gives another minimiser. Thus the minimisation in Eq.~\eqref{eq:eta-support} may be restricted to weights that are constant on each vertex orbit of $\Lambda$.

If $G$ is vertex-transitive, taking $\Lambda=\mathfrak G$ gives $\bar w=\mathbf1/|V|$, so uniform weights are optimal. The corresponding parent construction is given in Theorem~\ref{thm:vertex-transitive}.

% =============================================================================
\refstepcounter{section}
\setcounter{equation}{0}
\setcounter{lemma}{0}
\section*{Appendix \thesection: Reduced magic and sharpness advantage}
\label{app:magic}

\setlength{\abovedisplayskip}{4pt plus 2pt minus 1pt}
\setlength{\belowdisplayskip}{4pt plus 2pt minus 1pt}
\setlength{\abovedisplayshortskip}{0pt plus 2pt}
\setlength{\belowdisplayshortskip}{4pt plus 2pt minus 1pt}
% =============================================================================

We use the characterisation of the reduced stabilizer polytope in~\cite[Theorem~3]{VarelaEtAl2026Magic} and~\cite{LiuEtAl2026Nonstabilizerness} to prove the equality in Proposition~\ref{prop:reduced-magic-advantage}. The distinction is that joint measurability depends on absolute Pauli expectation values, whereas reduced magic retains their signs.

For the signed correlation vector $b_{\cA}(\rho)=\bigl(\Tr(\rho A_v)\bigr)_{v\in V}$, define the reduced stabilizer polytope
\begin{equation}
 \mathcal P_{\mathrm{stab}}(\cA)
 :=\operatorname{conv}\{b_{\cA}(\sigma):
 \sigma\text{ a pure stabilizer state}\}.
 \label{eq:magic-reduced-stabilizer-polytope}
\end{equation}
In the decomposition defining $\operatorname{RoM}_{\cA}$ in Sec.~\ref{subsec:sv-reduced-magic}, the coefficients are real and sum to one, but may be negative. One may equivalently use points of $\mathcal P_{\mathrm{stab}}(\cA)$: expanding each point into pure stabilizer correlations does not increase the total absolute weight.

For a pure stabilizer state, the support of $b_{\cA}(\sigma)$ is an independent set of $G$, but Pauli product relations can constrain its signs. Discarding these sign constraints gives the symmetric graph relaxation~\cite{LiuEtAl2026Nonstabilizerness}
\begin{equation}
 \widehat{\mathcal P}_G
 :=\{b\in[-1,1]^V:|b|\in\operatorname{STAB}(G)\},
 \label{eq:magic-sign-relaxed-polytope}
\end{equation}
where $|b|$ is taken componentwise. When $\cA$ has no active dependencies, meaning that no nonempty mutually commuting subset has product proportional to $\id$, the relaxation is exact (see~\cite[Theorem~3]{VarelaEtAl2026Magic} and~\cite{LiuEtAl2026Nonstabilizerness}):
\begin{equation}
 \mathcal P_{\mathrm{stab}}(\cA)=\widehat{\mathcal P}_G.
 \label{eq:magic-relaxation-exact}
\end{equation}

Under this assumption, set $t=R_G(\rho)\geq1$ and $p=b_{\cA}(\rho)/t$. By Eq.~\eqref{eq:seed-sharpness-advantage}, $|p|=c_{\cA}(\rho)/t\in\operatorname{STAB}(G)$, so Eq.~\eqref{eq:magic-relaxation-exact} gives $p,-p\in\mathcal P_{\mathrm{stab}}(\cA)$. The identity
\begin{equation}
 b_{\cA}(\rho)
 =\frac{t+1}{2}\,p+\frac{1-t}{2}\,(-p)
 \label{eq:magic-two-point-decomposition}
\end{equation}
has coefficients whose sum is one and whose absolute values sum to $t$. Expanding $p$ and $-p$ into pure stabilizer mixtures therefore gives $\operatorname{RoM}_{\cA}(\rho)\leq t=R_G(\rho)$. The reverse inequality is Eq.~\eqref{eq:rom-radial-bound}, so if $\cA$ has no active dependencies,
\begin{equation}
 \operatorname{RoM}_{\cA}(\rho)=R_G(\rho)\,.
 \label{eq:magic-exact-rom-radial}
\end{equation}
Replacing $\mathcal P_{\mathrm{stab}}(\cA)$ by $\widehat{\mathcal P}_G$ in this optimisation gives minimum total absolute weight $R_G(\rho)$, even when active dependencies are present.

% =============================================================================
\refstepcounter{section}
\setcounter{equation}{0}
\setcounter{lemma}{0}
\section*{Appendix \thesection: Quadratic Majorana Hamiltonians and Gaussian seeds}
\label{app:majorana-polar}
\label{app:majorana}

We collect the standard facts about quadratic Majorana Hamiltonians and fermionic Gaussian states needed in Sec.~\ref{sec:structured-parents}. Further details can be found, for example, in \cite{HacklBianchi2021,Bravyi2005,McNulty2025Fermionic}. We first bound the energy using the covariance matrix, then diagonalise the Hamiltonian and construct a pure Gaussian state attaining the bound. We also apply this construction to the saturation criterion for regular root graphs in Corollary~\ref{cor:line-graph-saturation}.

\subsection{Spectral optima and Gaussian seeds}

Let $M=2\lceil N/2\rceil$ and choose Hermitian Majorana operators $\Gamma_1,\ldots,\Gamma_M$ satisfying $\{\Gamma_j,\Gamma_k\}=2\delta_{jk}\id$. We work on the full Fock space of $M/2$ fermionic modes, of dimension $2^{M/2}$, without restricting to a fixed fermion-parity sector. When $N$ is odd, $\Gamma_{N+1}$ is an unused auxiliary Majorana. For a real skew-symmetric $N\times N$ matrix $S$, consider
\begin{equation}
 H(S)=\frac{\mathrm{i}}{2}
 \sum_{j,k=1}^{N}S_{jk}\Gamma_j\Gamma_k.
 \label{eq:app-majorana-H}
\end{equation}
Set $\widehat S=S\oplus0_{M-N}$. The added zero row and column, when present, change neither the Hamiltonian nor the trace norm: $H(\widehat S)=H(S)$ and $\|\widehat S\|_1=\|S\|_1$.

For any state $\rho$, define its covariance matrix on all $M$ Majoranas by
\begin{equation}
 (Q_\rho)_{jk}
 =\frac{\mathrm{i}}{2}
 \Tr\left(\rho[\Gamma_j,\Gamma_k]\right).
 \label{eq:majorana-covariance-definition}
\end{equation}
This matrix is real and skew-symmetric, with zero diagonal and $(Q_\rho)_{jk}=\mathrm{i}\Tr(\rho\Gamma_j\Gamma_k)$ for $j\neq k$. It satisfies $\|Q_\rho\|_\infty\leq1$. Indeed, for $x\in\mathbb R^M$, the operator $\Gamma(x):=\sum_jx_j\Gamma_j$ obeys $\Gamma(x)^2=\|x\|_2^2\id$, so $\|\Gamma(x)\|_\infty=\|x\|_2$. The commutator norm bound therefore gives
\begin{equation}
 \begin{aligned}
 |x^{\mathsf T}Q_\rho y|
 &=\frac12\left|
 \Tr\left(\rho\,\mathrm{i}[\Gamma(x),\Gamma(y)]\right)
 \right|\\
 &\leq\|\Gamma(x)\|_\infty\|\Gamma(y)\|_\infty
 =\|x\|_2\|y\|_2.
 \end{aligned}
 \label{eq:magic-covariance-contraction}
\end{equation}
Using the covariance representation of the quadratic Hamiltonian and the inequality $|\Tr(X^{\mathsf T}Y)|\leq \|X\|_\infty\|Y\|_1$, we obtain
\begin{equation}
 \begin{aligned}
 \Tr\bigl(\rho H(S)\bigr)
 &=\frac12\Tr(Q_\rho^{\mathsf T}\widehat S)\\
 &\leq\frac12\|Q_\rho\|_\infty\|\widehat S\|_1
 \leq\frac12\|S\|_1.
 \end{aligned}
 \label{eq:majorana-energy-covariance}
\end{equation}

To construct a state saturating the bound, we use the real skew-symmetric normal form. There is a real orthogonal matrix $O\in O(M)$, satisfying $O^{\mathsf T}O=\id_M$, such that
\begin{equation}
 \widehat S
 =O\left[\bigoplus_{k=1}^{M/2}\sigma_kJ_2\right]
 O^{\mathsf T},
 \qquad
 J_2=\begin{pmatrix}0&1\\-1&0\end{pmatrix},
 \qquad \sigma_k\geq0.
 \label{eq:skew-canonical-form}
\end{equation}

Define the transformed Majorana operators by
\[
 \widetilde\Gamma_j
 :=\sum_{\ell=1}^{M}O_{\ell j}\Gamma_\ell.
\]
They remain Hermitian and satisfy $\{\widetilde\Gamma_j,\widetilde\Gamma_k\} =2(O^{\mathsf T}O)_{jk}\id=2\delta_{jk}\id$. Substituting $\Gamma_\ell=\sum_jO_{\ell j}\widetilde\Gamma_j$ into the Hamiltonian gives
\begin{equation}
 \begin{aligned}
 H(S)
 &=\frac{\mathrm{i}}{2}\sum_{j,k=1}^{M}
 (O^{\mathsf T}\widehat S O)_{jk}
 \widetilde\Gamma_j\widetilde\Gamma_k\\
 &=\sum_{k=1}^{M/2}\sigma_k\,
 \mathrm{i}\widetilde\Gamma_{2k-1}\widetilde\Gamma_{2k}.
 \end{aligned}
 \label{eq:majorana-normal-modes}
\end{equation}
These quadratic terms commute because they involve disjoint Majorana pairs.

Their eigenvalues have the usual occupation-number interpretation. The fermionic annihilation operators $f_k:=(\widetilde\Gamma_{2k-1} +\mathrm{i}\widetilde\Gamma_{2k})/2$ have occupation operators $n_k:=f_k^\dagger f_k$, with $\mathrm{i}\widetilde\Gamma_{2k-1}\widetilde\Gamma_{2k} =2n_k-\id$. Thus Eq.~\eqref{eq:majorana-normal-modes} is a sum of independent terms $\sigma_k(2n_k-\id)$. On the full Fock space, the occupations can be chosen independently from $\{0,1\}$. Choosing every occupation to be one, including those with $\sigma_k=0$, gives the rank-one projector
\begin{equation}
 \rho_\star
 =\prod_{k=1}^{M/2}
 \frac{\id+\mathrm{i}\widetilde\Gamma_{2k-1}
 \widetilde\Gamma_{2k}}{2}.
 \label{eq:majorana-optimal-state}
\end{equation}
This occupation-number state is a pure Gaussian state~\cite{Bravyi2005}. In the transformed basis, its covariance is $\bigoplus_kJ_2$. Returning to the original basis gives
\begin{equation}
 Q_\star
 =O\left[\bigoplus_{k=1}^{M/2}J_2\right]O^{\mathsf T}.
 \label{eq:majorana-optimal-covariance}
\end{equation}
Writing $|\widehat S|=(\widehat S^{\mathsf T}\widehat S)^{1/2}$, the block forms in Eqs.~\eqref{eq:skew-canonical-form} and~\eqref{eq:majorana-optimal-covariance} give $Q_\star^2=-\id_M$ and $Q_\star^{\mathsf T}\widehat S=|\widehat S|$. Thus $\rho_\star$ attains the bound in Eq.~\eqref{eq:majorana-energy-covariance}. Each $\sigma_k$ occurs twice among the singular values of $\widehat S$, so
\begin{equation}
 \Tr\bigl(\rho_\star H(S)\bigr)
 =\lambda_{\max}\bigl(H(S)\bigr)\\
 =\sum_{k=1}^{M/2}\sigma_k
 =\frac12\|S\|_1.
 \label{eq:majorana-max-eigenvalue}
\end{equation}
In particular, $\rho_\star$ is a ground state of $H(-S)=-H(S)$. Replacing every occupation by its complement reverses the energy, so the spectrum is symmetric about zero and $\|H(S)\|_\infty=\lambda_{\max}(H(S))$. When $S$ is nonsingular, $N=M$ and all $\sigma_k>0$. Maximal energy then requires every occupation to be one, making $\rho_\star$ unique.

The covariance can also be expressed through the polar decomposition. For nonsingular $S$, its block form gives
\begin{equation}
 Q_\star=S|S|^{-1}=S(-S^2)^{-1/2},
 \label{eq:majorana-polar-appendix}
\end{equation}
so $S=Q_\star|S|$. For singular $\widehat S$, this formula determines the covariance only on the orthogonal complement of the kernel. Let $\mathcal Z:=\ker\widehat S\subseteq\mathbb R^M$ and let $P_{\mathcal Z}$ be the orthogonal projector onto $\mathcal Z$. Define $\widehat Q_0$ to equal $\widehat S(-\widehat S^2)^{-1/2}$ on $\mathcal Z^\perp$ and to vanish on $\mathcal Z$. Then
\begin{equation}
 \widehat Q_0^2=-(\id_M-P_{\mathcal Z}),
 \qquad
 \widehat Q_0^{\mathsf T}\widehat S=|\widehat S|.
 \label{eq:majorana-partial-polar-identities}
\end{equation}
The kernel has even dimension, since $\widehat S$ has even size and even rank. Choose a real skew-symmetric operator $Q_{\mathcal Z}$ on $\mathcal Z$ satisfying $Q_{\mathcal Z}^2=-\id_{\mathcal Z}$, and extend it by zero on $\mathcal Z^\perp$. The completion $Q_\star=\widehat Q_0+Q_{\mathcal Z}$ satisfies $Q_\star^2=-\id_M$ and $Q_\star^{\mathsf T}\widehat S=|\widehat S|$. Every real skew-symmetric covariance with square $-\id$ specifies a unique pure Gaussian state~\cite{Bravyi2005}, so this again gives a state saturating the bound in Eq.~\eqref{eq:majorana-energy-covariance}. The freedom in $Q_{\mathcal Z}$ corresponds to choosing a pure Gaussian state for the zero-energy fermionic modes. It leaves the energy unchanged but can change individual quadratic correlations.

For odd $N$ with $\ker S$ one-dimensional, the completion can preserve the entire covariance block on the original Majoranas. This is the case used for uniformly weighted odd cycles in Appendix~\ref{app:cycles}. Let $u$ be a unit vector spanning $\ker S$, and let $Q_0$ equal $S(-S^2)^{-1/2}$ on $u^\perp$ and vanish on $u$. The kernel of $\widehat S=S\oplus0$ is then spanned by $(u,0)$ and the auxiliary coordinate. Pairing these two directions gives
\begin{equation}
 \widehat Q
 =\begin{pmatrix}Q_0&u\\-u^{\mathsf T}&0\end{pmatrix}.
 \label{eq:majorana-odd-completion}
\end{equation}
Since $Q_0u=0$ and $Q_0^2=-(\id_N-uu^{\mathsf T})$, direct multiplication gives $\widehat Q^2=-\id_{N+1}$ and $\widehat Q^{\mathsf T}\widehat S=|S|\oplus0$. It therefore specifies a pure Gaussian state $\widehat\rho_\star$ attaining the largest eigenvalue of $H(S)$. Its upper-left block is $Q_0$, so the original quadratic correlations are preserved, namely, $\mathrm{i}\Tr(\widehat\rho_\star\Gamma_j\Gamma_k) =(Q_0)_{jk}$, $1\leq j\neq k\leq N$.

\subsection{Saturation for regular root graphs}

We now prove Corollary~\ref{cor:line-graph-saturation}. Recall that, for a finite simple $\Delta$-regular graph $R=([N],E)$ with $\Delta\geq1$, the corollary states that $\eta(L(R))=1/\sqrt\Delta$ if and only if $R$ admits an orientation whose skew-adjacency matrix satisfies
\[
 SS^{\mathsf T}=\Delta\id.
\]
In this case, $Q=S/\sqrt\Delta$ is the covariance of a balanced pure Gaussian seed whose orbit parent attains the optimal uniform sharpness. We also show that the construction gives a single-orbit parent in a canonical irreducible representation.

\begin{proof}
The $\Delta$ edges incident at any vertex correspond to mutually anti-commuting observables, giving $\eta(L(R))\leq1/\sqrt\Delta$. The necessity of $SS^{\mathsf T}=\Delta\id$ for equality follows from~\cite{McNulty2025Graph}.

Conversely, suppose an orientation has a skew-adjacency matrix $S$ satisfying this condition. Then $S$ is nonsingular, so $N$ is even, and $|S|=\sqrt\Delta\,\id$. Thus
\[
 Q=\frac{S}{\sqrt\Delta}=S|S|^{-1}
\]
is the polar covariance in Eq.~\eqref{eq:majorana-polar-appendix}. It specifies a pure Gaussian state $\rho$ attaining the largest eigenvalue of $H(S)$. Since $S_{uv}\in\{\pm1\}$ on every edge,
\[
 |\Tr(\rho A_e)|=\frac1{\sqrt\Delta},
 \qquad e\in E.
\]
The orbit POVM in Eq.~\eqref{eq:majorana-full-fock-parent} therefore attains the upper bound.

It remains to obtain a single orbit in a canonical representation. Since $S$ is nonsingular, all coefficients $\sigma_k$ in Eq.~\eqref{eq:majorana-normal-modes} are positive, so the maximal-energy state $\rho$ is unique. Decompose the full space representation of the selected observables into irreducible summands. The Hamiltonian, being a linear combination of these observables, is block diagonal in this decomposition. Its unique maximal-energy eigenvector must lie in one summand: otherwise its nonzero components in different summands would give linearly independent eigenvectors with the same maximal energy. Restricting the observables to this summand preserves the balanced profile. The orbit construction of Theorem~\ref{thm:state-space} then gives an optimal single-orbit parent in this canonical representation.
\end{proof}

For the seed constructed above, the full Fock space parent in Eq.~\eqref{eq:majorana-full-fock-parent} is already optimal. The final part of the proof establishes the additional conclusion that a single canonical orbit attains the same sharpnesses. We restrict the seed and observables to an irreducible invariant subspace and then construct a new parent using Theorem~\ref{thm:state-space}. The single-orbit form is therefore established by this construction, rather than by simply restricting the original full-space POVM.

For cycles, a suitable subspace can instead be identified using \emph{fermion parity}. The full Fock space splits into two parity sectors, spanned by occupation-number states with even and odd total occupation, respectively. Quadratic Majorana observables preserve both sectors, and each pure Gaussian seed is supported in one of them. A parity sector need not itself be irreducible. In Appendix~\ref{app:cycles}, we verify that the sector containing the constructed seed is irreducible.

% =============================================================================
\refstepcounter{section}
\setcounter{equation}{0}
\setcounter{lemma}{0}
\section*{Appendix \thesection: Cycles}
\label{app:cycles}
% =============================================================================

We verify that the Gaussian seeds specified in Sec.~\ref{subsec:cycles} are balanced, so a single orbit attains the optimal uniform sharpness $\eta(C_n)$. We then show that the balanced seed may be taken in a canonical irreducible representation.

Let $\bar w=n^{-1}(1,\ldots,1)$ and $S_n^{(\tau)}=nS_{\bar w}^{(\tau)}$. At these weights, the maximum in Eq.~\eqref{eq:cycle-support} is attained by $\tau=-1$ for even $n$ and by either sign for odd $n$~\cite{McNulty2025Graph}. We use $\tau=+1$ in the latter case.

Suppose first that $n$ is even and set $S=S_n^{(-1)}$. This matrix is nonsingular, so the covariance $Q$ in Eq.~\eqref{eq:majorana-polar} defines a pure Gaussian state $\rho_\star$ attaining the largest eigenvalue of $H(S)$.

To show that the absolute expectations are equal, let $T$ be the signed cyclic shift with $T_{j,j+1}=1$ for $j<n$ and $T_{n,1}=-1$. Then $S=T-T^{\mathsf T}$ and $TST^{\mathsf T}=S$. The formula for $Q$ preserves this symmetry, giving $TQT^{\mathsf T}=Q$. Comparing entries gives $Q_{j,j+1}=Q_{j+1,j+2}$ for $j<n-1$ and $Q_{n-1,n}=-Q_{n,1}$. Thus $Q_{j,j+1}=q$ for all $j<n$ and $Q_{n,1}=-q$, for some real $q$. Each term in $H(S)$ therefore has expectation $q$, giving
\begin{equation}
 \begin{aligned}
 nq
 &=\Tr\bigl(\rho_\star H(S)\bigr)
 =\frac12\Tr(Q^{\mathsf T}S)\\
 &=\frac12\|S\|_1
 =n\eta(C_n).
 \end{aligned}
 \label{eq:app-cycle-balanced-even}
\end{equation}
Hence $q=\eta(C_n)>0$.

For odd $n$, take $S=S_n^{(+1)}$. Its kernel is spanned by $u=n^{-1/2}(1,\ldots,1)^{\mathsf T}$. Let $Q_0$ equal $S(-S^2)^{-1/2}$ on $(\ker S)^\perp$ and vanish on the kernel. The cyclic shift with $T_{n,1}=1$ preserves $S$ and hence $Q_0$. The same entrywise comparison gives $(Q_0)_{j,j+1}=(Q_0)_{n,1}=q$ for all $j<n$. Since $Q_0^{\mathsf T}S=|S|$, repeating the calculation in Eq.~\eqref{eq:app-cycle-balanced-even} gives $q=\eta(C_n)$.

The completion in Eq.~\eqref{eq:majorana-odd-completion} pairs the kernel direction $u$ with the auxiliary Majorana. It gives a pure Gaussian covariance without changing the block on the original $n$ Majoranas. Thus, for both even and odd $n$, the constructed state satisfies the balanced expectations in Eq.~\eqref{eq:cycle-balanced-seed}.

To obtain a single canonical orbit, it remains to show that the parity sector containing the seed is irreducible. Including the auxiliary Majorana when $n$ is odd, each parity sector has dimension $2^{\lceil n/2\rceil-1}$. The kernel equation $B_{C_n}k=0$ over $\F_2$ gives $k_{j+2}=k_j$, with indices taken cyclically, so $\nu_2(C_n)=2$ for even $n$ and $\nu_2(C_n)=1$ for odd $n$. Thus each parity sector has dimension $d_{C_n}=2^{(n-\nu_2(C_n))/2}$. Since every irreducible summand has this dimension by Appendix~\ref{app:clifford}, each parity sector is itself irreducible. Restricting the observables to the sector containing the seed preserves its expectation values, so Theorem~\ref{thm:state-space} gives an optimal single-orbit parent in this canonical representation.

\refstepcounter{section}
\setcounter{equation}{0}
\setcounter{lemma}{0}
\section*{Appendix \thesection: Paths}
\label{app:paths}
% =============================================================================

We prove Proposition~\ref{prop:paths-cycles}, which establishes Conjecture~1 of~\cite{McNulty2025Graph}, and then derive the reflection-symmetric $P_4$ boundary slice. We use the path realisation of Sec.~\ref{subsec:paths} and the weighted support formula in Eq.~\eqref{eq:path-cycle-support-restriction}. The remaining step is to exhibit weights giving a matching upper bound. Throughout, we use the full Fock space convention of Appendix~\ref{app:majorana-polar}.

\subsection{Proof of the path threshold}
\label{app:opening-even-cycle}

We show that deleting one or two consecutive observables from an even cycle leaves the optimal uniform sharpness unchanged. Write
\begin{equation}
 \ell=m+1\geq2,
 \qquad
 \theta=\frac{\pi}{2\ell},
 \qquad
 \eta_0=\frac1{\ell\sin\theta}.
 \label{eq:app-path-parameters}
\end{equation}
By Eq.~\eqref{eq:cycle-uniform}, $\eta(C_{2\ell})=\eta_0$.

A parent for the cycle also measures any subfamily. Deleting one observable gives $P_{2\ell-1}$, and deleting an endpoint of this path gives $P_{2\ell-2}$. At each step, any uniform sharpness achievable before the deletion remains achievable afterwards. Hence
\begin{equation}
 \eta_0
 =
 \eta(C_{2\ell})
 \leq
 \eta(P_{2\ell-1})
 \leq
 \eta(P_{2\ell-2}).
 \label{eq:app-path-primal-bound}
\end{equation}

Thus both paths can attain $\eta_0$; the question is whether removing observables allows a larger uniform sharpness. To show that it does not, it suffices to bound the threshold of the shorter path by $\eta_0$. We will choose weights for $P_{2\ell-2}$ whose support value is at most $\eta_0$. Since the weights sum to one, the corresponding support inequality bounds every achievable uniform sharpness by $\eta_0$, forcing equality throughout Eq.~\eqref{eq:app-path-primal-bound}.

Let $\beta_j:=\sin(j\theta)$ and choose
\begin{equation}
 w_j^{\star}
 =
 \frac{\beta_j\beta_{j+1}}{\ell\cos\theta},
 \qquad j=1,\ldots,2\ell-2.
 \label{eq:app-path-weights}
\end{equation}
Using the recurrence $\beta_{r-1}+\beta_{r+1}=2\cos\theta\,\beta_r$ and the identity $\sum_{k=1}^{\ell-1}\beta_{2k}^2=\ell/2$, we obtain
\begin{equation}
 \begin{aligned}
 \sum_{j=1}^{2\ell-2}\beta_j\beta_{j+1}
 &=
 \sum_{k=1}^{\ell-1}\beta_{2k}(\beta_{2k-1}+\beta_{2k+1})\\
 &=
 2\cos\theta\sum_{k=1}^{\ell-1}\beta_{2k}^2
 =
 \ell\cos\theta.
 \end{aligned}
 \label{eq:app-path-weight-normalisation}
\end{equation}
Thus $w^{\star}\in\Delta([2\ell-2])$. These weights are positive, symmetric about the centre of the path, and nonuniform for $\ell\geq3$.

Reorder the Majorana indices into odd and even sites, and multiply the $k$th basis vector within each block by $(-1)^{k-1}$. This orthogonal change of basis gives
\begin{equation}
 \Omega_w\simeq
 \begin{pmatrix}
 0&X_w\\
 -X_w^{\mathsf T}&0
 \end{pmatrix},
 \label{eq:app-path-block}
\end{equation}
where $X_w\in\mathbb R^{\ell\times(\ell-1)}$ is bidiagonal, with
\begin{equation}
 \begin{gathered}
 (X_w)_{k,k}=w_{2k-1},\qquad (X_w)_{k+1,k}=w_{2k},
 \end{gathered}
 \label{eq:app-path-bidiagonal}
\end{equation}
for $k=1,\ldots,\ell-1$. Each singular value of $X_w$ occurs twice in the skew block matrix, so Eq.~\eqref{eq:path-cycle-support-restriction} becomes $h_{P_{2\ell-2}}(w)=\|X_w\|_1$.

For the weights in Eq.~\eqref{eq:app-path-weights}, $X_{w^{\star}}=\frac1{\ell\cos\theta}X_{\ell}$, where the only nonzero entries of $X_{\ell}$ are
\begin{equation}
 (X_{\ell})_{k,k}=\beta_{2k-1}\beta_{2k},
 \qquad
 (X_{\ell})_{k+1,k}=\beta_{2k}\beta_{2k+1}.
 \label{eq:app-path-CN}
\end{equation}
Let $M_{\ell}$ be the $\ell\times(\ell-1)$ matrix with ones on the diagonal and subdiagonal and zeros elsewhere. Then $X_{\ell}=M_{\ell}\odot(uv^{\mathsf T})$, where $\odot$ denotes entrywise multiplication, $u_j=\sin((2j-1)\theta)$ for $j=1,\ldots,\ell$, and $v_k=\sin(2k\theta)$ for $k=1,\ldots,\ell-1$. Both squared Euclidean norms equal $\ell/2$.

Entrywise multiplication by $M_\ell$ retains the diagonal and subdiagonal entries of a matrix and sets all other entries to zero. Applying~\cite[Theorem~4.2]{Levene2014} to $M_\ell^{\mathsf T}$ with $n=\ell-1$, and using that transposition preserves the operator norm, gives
\[
 \|M_\ell\odot Y\|_\infty
 \leq\frac{2}{\ell}\cot\theta\,\|Y\|_\infty.
\]
To obtain a bound on $\|X_\ell\|_1$, we use the variational formula for the trace norm: for a real matrix $X$, this is the maximum of $|\Tr(Y^{\mathsf T}X)|$ over real matrices $Y$ of the same size with $\|Y\|_\infty\leq1$. Since $X_\ell=M_\ell\odot(uv^{\mathsf T})$, expanding the trace entrywise gives
\[
 \begin{aligned}
 \|X_\ell\|_1
 &=\max_{\|Y\|_\infty\leq1}
   |\Tr(Y^{\mathsf T}X_\ell)|\\
 &=\max_{\|Y\|_\infty\leq1}
   |u^{\mathsf T}(M_\ell\odot Y)v|\\
 &\leq\frac{2}{\ell}\cot\theta\,
   \|u\|_2\|v\|_2
 =\cot\theta,
 \end{aligned}
\]
where the last equality uses $\|u\|_2^2=\|v\|_2^2=\ell/2$.

By Eq.~\eqref{eq:eta-support}, this estimate gives
\begin{equation}
 \begin{aligned}
 \eta(P_{2\ell-2})
 &\leq h_{P_{2\ell-2}}(w^{\star})\\
 &\leq\frac{\cot\theta}{\ell\cos\theta}
 =\eta_0.
 \end{aligned}
 \label{eq:app-path-dual}
\end{equation}
Combining this with Eq.~\eqref{eq:app-path-primal-bound} forces equality throughout:
\begin{equation}
 \eta(P_{2\ell-2})
 =
 \eta(P_{2\ell-1})
 =
 \eta(C_{2\ell})
 =
 \frac1{\ell\sin\theta}.
 \label{eq:app-path-cycle-equality}
\end{equation}
The inequalities also imply $\|X_{\ell}\|_1=\cot\theta$. Substituting $\ell=m+1$ proves Eq.~\eqref{eq:path-cycle-even}. Multiplying the supporting inequality $w^{\star}\cdot s\leq\eta_0$ by $\ell\cos\theta$ gives Eq.~\eqref{eq:path-sharpness-inequality}, with the optimal normalised weights in Eq.~\eqref{eq:app-path-weights}.

Restricting the cycle parents to either path attains this common threshold, completing the proof of Proposition~\ref{prop:paths-cycles}.

For $m\geq1$, let $x=\pi/[2(2m+1)]$ and $y=\pi/[2(m+1)]$. Equations~\eqref{eq:cycle-uniform} and~\eqref{eq:app-path-cycle-equality} give
\[
 \eta(C_{2m+1})
 =\frac2\pi\,x\cot x
 <\frac2\pi
 <\frac2\pi\,\frac{y}{\sin y}
 =\eta(P_{2m}),
\]
using $\tan x>x$ and $\sin y<y$. Since setting any cycle weight to zero leaves the path support problem in Eq.~\eqref{eq:path-cycle-support-restriction}, every optimal odd-cycle weighting has strictly positive entries.

\subsection{A boundary slice of \texorpdfstring{$P_4$}{P4}}
\label{app:P4-slice}

We derive the compatibility condition for $s=(x,y,y,x)$, where the two endpoints have sharpness $x$ and the two internal vertices have sharpness $y$, with $0\leq x,y\leq1$. By Eq.~\eqref{eq:support-halfspaces}, such a vector is compatible if and only if $w\cdot s\leq h_{P_4}(w)$ for every $w\in\Delta([4])$. We first reduce the weights by reflection symmetry, then evaluate the resulting bounds.

Let $\mathsf R$ reverse the vertex order. For any $w\in\Delta([4])$, the average $\bar w=(w+\mathsf Rw)/2$ satisfies $\bar w\cdot s=w\cdot s$. Equation~\eqref{eq:E-hinvariant} gives $h_{P_4}(\mathsf Rw)=h_{P_4}(w)$, so convexity implies $h_{P_4}(\bar w)\leq h_{P_4}(w)$. Thus averaging the weights leaves the left-hand side of the inequality unchanged and can only decrease the right-hand side. Any violated bound therefore remains violated after averaging. It follows that we need only consider weights of the form
\begin{equation}
 w=\frac{(a,b,b,a)}{2(a+b)},
 \qquad a,b\geq0,\quad a+b>0.
 \label{eq:app-P4-symmetric-weights}
\end{equation}

To evaluate Eq.~\eqref{eq:path-cycle-support-restriction} for these weights, reorder the five Majorana indices as $(1,3,5;2,4)$. In this order, the matrix $2(a+b)\Omega_w$ is
\begin{equation}
 \begin{pmatrix}0&B\\-B^{\mathsf T}&0\end{pmatrix},
 \qquad
 B=\begin{pmatrix}a&0\\-b&b\\0&-a\end{pmatrix}.
 \label{eq:app-P4-block}
\end{equation}
The eigenvalues of $B^{\mathsf T}B$ are $a^2$ and $a^2+2b^2$, so $\|B\|_1=a+\sqrt{a^2+2b^2}$. The skew-symmetric block matrix has each singular value of $B$ twice, together with one zero. Its trace norm is therefore $2\|B\|_1$. Dividing by the normalisation factor $2(a+b)$ and using $h_{P_4}(w)=\tfrac12\|\Omega_w\|_1$ gives
\begin{equation}
 h_{P_4}\!\left(\frac{(a,b,b,a)}{2(a+b)}\right)
 =\frac{a+\sqrt{a^2+2b^2}}{2(a+b)}.
 \label{eq:app-P4-symmetric-support}
\end{equation}
Substituting this expression and $w\cdot s=(ax+by)/(a+b)$ into Eq.~\eqref{eq:support-halfspaces} shows that $s$ is compatible exactly when
\begin{equation}
 a(2x-1)+2by\leq\sqrt{a^2+2b^2}
 \quad\text{for all }a,b\geq0.
 \label{eq:app-P4-support-inequalities}
\end{equation}

Both sides scale by the same factor when $(a,b)$ is rescaled, so it suffices to consider $a^2+2b^2=1$. If $2x-1<0$, the left-hand side is maximised by taking $a=0$ and $b=1/\sqrt2$, giving $\sqrt2y$. If $2x-1\geq0$, Cauchy--Schwarz applied to $(a,\sqrt2b)$ and $(2x-1,\sqrt2y)$ gives the maximum $\sqrt{(2x-1)^2+2y^2}$. This value is attained by choosing the two vectors proportional; when the second vector is zero, the maximum is zero for every choice of weights. Thus compatibility is equivalent to
\begin{equation}
 \bigl(\max\{2x-1,0\}\bigr)^2+2y^2\leq1,
 \qquad x,y\geq0.
 \label{eq:app-P4-slice-region}
\end{equation}
For $0\leq x\leq1/2$, this condition requires only $y\leq1/\sqrt2$; for $x\geq1/2$, it becomes $(2x-1)^2+2y^2\leq1$. Points with $x<1/2$ cannot be coordinatewise maximal, since $x$ can be increased to $1/2$ without reducing $y$. The coordinatewise maximal points therefore form the arc in Eq.~\eqref{eq:P4-boundary-arc}.

% =============================================================================
\refstepcounter{section}
\setcounter{equation}{0}
\setcounter{lemma}{0}
\section*{Appendix \thesection: The complete Pauli family}
\label{app:complete-pauli}
% =============================================================================

We now prove that every state maximising
\begin{equation}
 \mathcal L(\rho):=\sum_{x\neq0}|\Tr(\rho P_x)|,
 \label{eq:complete-pauli-l1}
\end{equation}
is pure. Together with Corollary~\ref{cor:balanced-orbit}, which characterises optimal single-orbit parents by seeds having equal absolute Pauli expectations, this gives the SIC characterisation in Proposition~\ref{prop:complete-pauli} of Sec.~\ref{subsec:complete-pauli}.

\Needspace{10\baselineskip}
\begin{lemma}\label{lem:pauli-L1-purity}
Every maximiser of $\mathcal L$ is pure.
\end{lemma}

\begin{proof}
Let $\mathcal L_\star:=\max_\sigma \mathcal L(\sigma)$, and let $\rho=\sum_{j=1}^k p_j\rho_j$ be the spectral decomposition of a maximiser, where $p_j>0$ and the $\rho_j$ are mutually orthogonal rank-one projectors. Convexity gives
\[
 \mathcal L_\star=\mathcal L(\rho)
 \leq\sum_jp_j\mathcal L(\rho_j)
 \leq\mathcal L_\star.
\]
Equality holds throughout. The first inequality is a sum of triangle inequalities, one for each Pauli expectation, so each must be saturated. Since every $p_j>0$, for each $x\neq0$ the nonzero expectations $\Tr(\rho_jP_x)$ all have the same sign.

Suppose now that $k\geq2$. Since the $d^2$ Pauli operators form an orthogonal basis on the $d$-dimensional Hilbert space, Pauli expansion gives
\[
 d\,\Tr(\rho_1\rho_2)
 =1+\sum_{x\neq0}
   \Tr(\rho_1P_x)\Tr(\rho_2P_x).
\]
Every term in the sum is nonnegative by the sign property above, so $d\,\Tr(\rho_1\rho_2)\geq1$. But $\rho_1$ and $\rho_2$ are orthogonal, and hence $\Tr(\rho_1\rho_2)=0$, giving a contradiction. Therefore $k=1$, so $\rho$ is pure.
\end{proof}

% =============================================================================
\end{document}